\documentclass[a4paper,aps,twocolumn,nobibnotes,rmp,superscriptaddress,longbibliography,floatfix]{revtex4}

\usepackage[utf8]{inputenc}
\usepackage{amsmath,amsthm,amsfonts,bbm}
\usepackage{braket}
\usepackage{xcolor}
\definecolor{dark-gray}{rgb}{.35,.55,.55}
\definecolor{dark-blue}{rgb}{.0,.0,.6}
\usepackage[colorlinks=true,linkcolor=dark-blue,citecolor=dark-blue,urlcolor=dark-blue]{hyperref}
\usepackage{graphicx}
\usepackage{epstopdf}
\usepackage[normalem]{ulem}

\newcommand{\ketbra}[2]{{\ket{#1}\!\bra{#2}}}
\newcommand{\proj}[1]{\ketbra{#1}{#1}}
\DeclareMathOperator{\tr}{tr}

\newcommand{\reals}{{\mathbb R}}
\newcommand{\complexes}{{\mathbb C}}

\newcommand{\hilbert}{{\mathcal H}}

\newcommand{\mean}[1]{{\langle #1 \rangle}}
\theoremstyle{definition}
\newtheorem*{theorem*}{Theorem}
\newcommand{\pp}{{\mathbf p}}
\newcommand{\ee}{{\mathbf e}}
\newcommand{\xx}{{\mathbf x}}
\newcommand{\uu}{{\mathbf u}}
\newcommand{\mm}{{\mathbf m}}
\newcommand{\cc}{{\mathbf c}}
\newcommand{\bb}{{\mathbf b}}
\iftrue
\newcommand{\ketbrac}[1]{\ensuremath{| #1 \rangle \!\langle #1 |}}
\else
\newcommand{\ketbrac}[1]{\proj{#1}}
\fi
\DeclareMathOperator{\Prob}{Prob}
\providecommand{\exv}[1]{\langle{#1}\rangle}

\newcommand\bigcdot\bullet

\usepackage{import}
\usepackage{xifthen}
\usepackage{pdfpages}
\usepackage{transparent}

\begin{document}
	

\title{Kochen-Specker Contextuality}


\author{Costantino Budroni}
\email{costantino.budroni@unipi.it}
\affiliation{Department of Physics 'E. Fermi' University of Pisa, Largo B. Pontecorvo 3, I-56127 Pisa, Italy}
\affiliation{Faculty of Physics, University of Vienna, Boltzmanngasse 5, 1090 Vienna, Austria}
\affiliation{Institute for Quantum Optics and Quantum Information (IQOQI), Austrian Academy of Sciences, Boltzmanngasse 3, 1090 Vienna, Austria}

\author{Ad\'an Cabello}
\affiliation{
Departamento de F\'{\i}sica Aplicada II,
Universidad de Sevilla,
41012 Sevilla,
Spain
}
\affiliation{
Instituto Carlos~I de F\'{\i}sica Te\'orica y Computacional, 
Universidad de Sevilla, 
41012 Sevilla, 
Spain
}
\author{Otfried G\"uhne}
\affiliation{
Naturwissenschaftlich-Technische Fakult\"at,
Universit\"at Siegen,
Walter-Flex-Stra{\ss}e 3,
57068 Siegen,
Germany
}
\author{Matthias Kleinmann}
\affiliation{
Naturwissenschaftlich-Technische Fakult\"at,
Universit\"at Siegen,
Walter-Flex-Stra{\ss}e 3,
57068 Siegen,
Germany
}
\author{Jan-{\AA}ke Larsson}
\affiliation{
Institutionen f\"or Systemteknik och Matematiska Institutionen,
Link\"opings Universitet,
58183 Link\"oping,
Sweden
}

\date{\today}
\begin{abstract}
	A central result in the foundations of quantum mechanics is the Kochen-Specker theorem. In short, it states that quantum mechanics is in conflict with classical models in which the result of a measurement does not depend on which other compatible measurements are jointly performed. Here compatible measurements are those that can be implemented simultaneously or, more generally, those that are jointly measurable. This conflict is generically called quantum contextuality. In this review, an introduction to this subject and its current status is presented. Several proofs of the Kochen-Specker theorem and different notions of contextuality are reviewed. How to experimentally test some of these notions is explained and  connections between contextuality and nonlocality or graph theory are discussed. Finally,  some applications of contextuality in quantum information processing are reviewed.

\end{abstract}

\maketitle


\tableofcontents


\section{Introduction}


In the realm of classical physics, it is possible to consistently assume the existence of values for intrinsic properties (such as the length) of a physical object, and that nondeterministic measurement outcomes are caused by imperfect preparation or measurement procedures. Quantum theory fundamentally challenges such a point of view: It admits situations in which any assignment of a value to the result of the measurement of a physical property must depend on the {\it measurement context}, namely, on what other properties are simultaneously measured with it. Quantum contextuality then, as the name suggests, refers to the impossibility of such context-independent classical description of the predictions of quantum theory, which originated in the work of \citet{Specker:1960D} and \citet{Kochen:1967JMM}.

Quantum contextuality is a phenomenon that combines many of the interesting aspects of quantum theory in a single framework, from {\it measurement incompatibility}, as the impossibility of performing simultaneous measurements of arbitrary observables, to {\it Bell nonlocality} and {\it entanglement}, when the system examined is composed of several spatially separated parts. The adopted perspective is that of observed statistics, which allows 	for an analysis of the experimental results that is independent of quantum mechanics. On the one hand, quantum contextuality generated an intense debate on the foundations of quantum mechanics and stimulated the search for physical principles explaining why quantum theory is the way it is. On the other hand, the nonclassical properties of contextual correlations have been directly connected to quantum information processing applications such as quantum computation.

The central role of quantum contextuality in quantum theory, from both a fundamental and an applied perspective, is what provided motivation this review. The difficulty is that there is broad variety of perspectives from which to approach quantum contextuality, ranging from physics to mathematics, computer science, and philosophy, to mention a few, and consequently a vast literature. It is impossible to review all the literature and, at the same time, it would not be useful for the reader. We are thus forced to make a selection of topics to be presented. 
We decided to focus on {\it Kochen-Specker contextuality} \cite{Specker:1960D,Kochen:1967JMM}, which we often refer to simply  as contextuality. A different notion of nonclassicality proposed in \citet{Spekkens:2005PRA} is only briefly covered to highlight the differences with the notion of contextuality reviewed here.

Our goal with this review is to provide an introduction to contextuality that covers all the most important topics. In particular, we address the following questions:
\begin{itemize}
	\item[(a)] What is the structure of noncontextual hidden-variable models?
	\item[(b)] What are the physical assumptions involved in the definition of contextuality and how to operationally define contexts?
	\item[(c)] How does one perform experimental tests? What are the assumptions, the loopholes, and the methods to deal with them?
	\item[(d)] What are the applications of quantum contextuality in quantum information processing?
\end{itemize}
These can be summarized as follows: What is quantum contextuality, how do we test it, what is it useful for, and what do we learn from it?

This review aims to reach a broad audience, from people with little or no experience with quantum contextuality, to experts working in the field, both on the theoretical and experimental side. As a consequence, it can be read in different ways and some parts can be skipped by readers with some experience in contextuality. 

The content of this review can be outlined as follows. Section~\ref{sec:nutshell} contains a brief introduction to the basic concepts involved in quantum contextuality, such as compatible measurements, contexts and noncontextuality inequalities. Section~\ref{section3} contains the statement and proof of the original Kochen-Specker theorem and further simplifications and related arguments. Section~\ref{section4} addresses the main question of this review: What is the mathematical structure of noncontextual hidden-variable theories, and how can we put these theories to test in an experiment? This includes questions such as the operational identification of contexts, the problem involving imperfect experimental realizations, and experimental tests performed thus far. In Sec.~\ref{sec:Advanced}, we present a collection of advanced topics associated with quantum contextuality, from the definition of the noncontextuality polytope and the computation of noncontextuality inequalities, to the relations between contextuality and graph theory and the connection between quantum contextuality and Bell nonlocality. Finally, in Sec.~\ref{sec:applications} information-theoretical applications of quantum contextuality, such as quantum computation and random number generation, are
discussed. The Appendix~\ref{app:history} puts 
the results presented in the review into a historical perspective.


\section{Quantum contextuality in a nutshell}\label{sec:nutshell}
In this section, we explain the essence of the Kochen-Specker theorem and the modern view on Kochen-Specker contextuality. This is intended to be a simple explanation for introducing the topic to readers with little or no experience in the topic of quantum contextuality. Many of the subtleties and open problems, particularly those connected to the definition of contexts and compatible measurements, are discussed in more detail in the subsequent sections.

\subsection{A first example}

Arguably the simplest example of Kochen-Specker contextuality, in which state preparation plays no role, is provided by the so-called Peres-Mermin (PM) square, \cite{Peres:1990PLA, Mermin:1990PRL, Peres:1991JPA, Peres:1992FP, Mermin:1993RMP, Peres:1993} a construction of nine measurements arranged in a square:
\begin{equation}
\left[\begin{matrix}
 A&B&C\\
 a&b&c\\
 \alpha&\beta&\gamma
\end{matrix}\right].
\label{eq:basicPM}
\end{equation}
Each measurement is dichotomic, i.e., it has only two possible outcomes, in this case labeled as $+1$ and $-1$. If we think in classical terms, there could 
be nine properties of an object and performing a measurement reveals 
whether the property is present ($+1$) or absent ($-1$). 

In the following, it is assumed that the three measurements 
in each column and row form a ``context,'' i.e., a set of 
measurements whose values could in principle be jointly 
measured. We write $ABC$ to denote the 
product of the values of the measurements $A$, $B$, and $C$ 
for a single object. Similarly we use $abc$, $Aa\alpha$, 
etc. In a classical model describing the object, each 
of the nine measurements has a definite value, regardless 
of which context the measurement is contained in. 
Such a value assignment is then said to be {\it noncontextual}.
Thus, for the set $\{ABC$, $abc$, $\alpha\beta\gamma$, $Aa\alpha$, $Bb\beta$,
 $Cc\gamma\}$ there can be only an even number of products with the assigned value $+1$.
This holds since assigning $+1$ to all measurements gives six positive products
 and changing the value assigned to any measurement changes the value of two of the
 products, since each measurement appears in two of them.

Defining the expectation value
\begin{equation}
 \exv{ABC}\equiv \Prob[ABC=+1] -\Prob[ABC=-1],
\end{equation}
we have thus shown the validity of the inequality \cite{Cabello:2008PRL}
\begin{equation}\label{s2:PMineq}
 \exv{\mathsf{PM}}\equiv \exv{ABC} + \exv{abc} + \exv{\alpha\beta\gamma} +
 \exv{Aa\alpha} +\exv{Bb\beta} - \exv{Cc\gamma} \le 4.
\end{equation}

The significance of this inequality comes from the fact that it can
 be violated by a quantum system.
The quantum example works for a system composed of two spin-1/2 particles.
If we denote the Pauli operators $\sigma_x$, $\sigma_y$, and $\sigma_z$, the observables are
\begin{equation}\label{s2:PMobs}
\left[\begin{matrix}
 A&B&C\\
 a&b&c\\
 \alpha&\beta&\gamma
\end{matrix}\right]
=
\left[\begin{matrix}
 \sigma_z \otimes \openone&
\openone \otimes \sigma_z&
\sigma_z \otimes \sigma_z\\
 \openone \otimes \sigma_x&
 \sigma_x \otimes \openone&
 \sigma_x \otimes \sigma_x\\
 \sigma_z \otimes \sigma_x&
 \sigma_x \otimes \sigma_z&
 \sigma_y \otimes \sigma_y
\end{matrix}\right].
\end{equation}
Notice that the observables within one row or one column mutually commute, which allows one to simultaneously measure them and make sense of the expectation value for the product of outcomes, e.g., $ABC$. One verifies that for a system in the state $\ket{\psi}$, such an expectation value is given by
 $\exv{ABC}= \exv{\psi| ABC |\psi}$.
In fact, for the value of the terms in $\exv{\mathsf{PM}}$ the state $\ket{\psi}$
 does not play any role, since $ABC=\openone$ and we thus have $\exv{ABC}= +1$, and likewise for all
 products except $Cc\gamma = -\openone$, which gives $\exv{Cc\gamma}= -1$.
Summing these we obtain $\exv{\mathsf{PM}}= 6$ in clear contradiction to
 Eq.~\eqref{s2:PMineq} \cite{Cabello:2008PRL}.
Since we derived
 Eq.~\eqref{s2:PMineq} under the assumption that it is possible to consistently
 assign a value to the nine observables of the object, the violation of Eq.~\eqref{s2:PMineq} implies either that there is no value assignment or that 
 the value-assignment must depend on which context the observable appears in. This phenomenon is known as {\it quantum contextuality}.

\subsection{A second look}

At this point, we preview why quantum contextuality is a more subtle topic than it may seem from the argument presented thus far. As an entry point, one might wonder why we chose the particular form of the previously mentioned inequality instead of a simpler form like
\begin{equation}\label{eq:pm_all}
 \exv{ABC\;abc\;\alpha\beta\gamma\;Aa\alpha\;Bb\beta\;Cc\gamma}= +1.
\end{equation}
The reason is that, in the quantum example, in order to violate inequality \eqref{s2:PMineq}, we have to choose the observables in such a way that they are not all jointly measurable; i.e., they do not all mutually commute as in the rhs of Eq.~\eqref{s2:PMobs}. In such a case, according to quantum mechanics there is no measurement able to reveal the value of those observables on the same object consistently. For example, there is no common eigenstate of the previously defined observables $A$ and $b$. Hence, to experimentally test the expression in Eq.~\eqref{eq:pm_all}, one would need to perform a joint measurement of incompatible observables.

Another common misconception, associated with the particular realization of the PM square as two-qubit observables in Eq.~\eqref{eq:pm_all},  is that the measurements of each observable in the last row and last column can be performed as two single-qubit local measurements (with four outcomes) instead of a global (dichotomic) measurement. By doing so, however, one has in the last row a measurement of six incompatible single-qubit observables, which cannot form a context.
The use of single-qubit measurements, therefore, is at variance with the assumption that the last row and last column form a context and hence removes the contradiction. Performing coherent global measurements on the two qubits can indeed be a crucial challenge in experiments, see also
Sec.~\ref{ssec:experiments}.

A third source of confusion, diametrically opposed to the previous
one, is to consider the measurement of each row and column as
a single global and fundamental measurement. In this view, one has six
measurements, corresponding to the three rows and columns,
that simulate the nine measurements $A, \dots, \gamma$. Each
of these six global measurements has four outcomes, corresponding
to the outcomes of the three simulated measurements under
the constraint that their product equals $+1$ (or $-1$ for
the last column). Thus, one may be surprised that none of the
$2^9$ joint assignments of outcomes to the measurements
$A, \dots, \gamma$ are logically possible. For the original 
formulation of the PM square, however, only the 
nine dichotomic measurements $A, \dots, \gamma$ are fundamental 
entities and an evaluation of, for instance, $\mean{Cc\gamma}$ 
entails a product of three numbers, which experimentally is 
by no means guaranteed to be $-1$, see also Fig.~\ref{fig-exp-2} 
in Sec.~\ref{ssec:experiments}. Interpreting the PM square 
in terms of the six previously defined global measurements, however, 
possible in the framework of Spekkens contextuality \cite{KrishnaNJP2017}, 
see also Section \ref{ssec:other_cont}.

Finally, returning to the discussion after Eq.~(\ref{eq:pm_all}), we note
 that incompatibility does not immediately rule out a classical 
description: One can imagine a classical theory where values of all physical 
properties are simultaneously defined, but the classical measurement 
procedure of a property introduces some disturbance in the system and modifies 
the value of other physical properties. We revisit this problem in 
Sec.~\ref{section4}.

There is a price that we have to pay to see a violation of the inequality in Eq.~\eqref{s2:PMineq}: The current status of research is that it is impossible to conceive a quantum experiment featuring contextual behavior without additional assumptions. The basic reason is that we accepted that there are sets of observables the value of which cannot be revealed on the same object. But how can we ensure that a specific measurement in two different contexts does reveal the value of the same physical property?
 
This question brings us to the notion of compatibility. Intuitively, this corresponds to some notion of 
simultaneous measurability and nondisturbance among quantum measurements. In textbook quantum mechanics, an 
observable corresponds to an Hermitian operator, i.e.,  $A=A^\dagger$, with outcomes  identified with its 
eigenvalues, i.e., $A=\sum_i \lambda_i P_i$ and the spectral projections $P_i$ identified with the measurement 
effects, i.e., ${\rm Prob}(\lambda_i)=\tr(\rho P_i)$. Two observables $A$ and $B$ are said to be compatible if they 
commute, i.e., $[A,B]=0$. This type of measurements is called projective, ideal, or 
sharp, depending on which property one wants to emphasize. Commutativity is a strong property that implies 
 several other properties for these measurements.
In fact, if $[A,B]=0$, then there is another observable $C$ such that the spectral projections of $A$ and $B$ are a 
coarse graining of those of $C$ and thus measuring $C$ allows one to infer the result of both $A$ and $B$, a 
property called {\it joint measurability}.  At this point, we remark that joint measurability is the minimal requirement to define some notion of a context. Nevertheless, we highlight here some other properties of commuting projective measurements that will turn out to provide useful intuition for an operational definition of contexts presented in Sec.~\ref{ssec:compatible} and used to deal with imperfect measurements in experimental tests of contextuality; See Sec.~\ref{ssec:exper_imp} and Sec.~\ref{ssec:experiments}). More precisely, notice that
from the state-update rule $\rho \mapsto P_i \rho P_i$, one 
can see that, if $[A,B]=0$, the outcomes of $A$ are not disturbed by a subsequent measurement of $B$ and are repeated by a later measurement of $A$, such as in the sequence $ABA$. This is the property of {\it outcome repeatability}.
Conversely, projective measurements that satisfy one of the previously mentioned properties are necessarily commuting~\cite{HW2010}.

This is no longer true in the case of generalized measurements, which may be nonprojective (or, nonideal, unsharp). In this case, notions such as commutativity, nondisturbance, and joint measurability are no longer equivalent \cite{HW2010}, and the 
term ``incompatibility'' usually denotes the lack of joint measurability \cite{HeinosaariJPA2016}. Different 
notions, corresponding to stronger or weaker assumptions, are also possible.  

For the moment, we do not enter into this problem. 
Consider the case of ideal measurements, where these ambiguities do not arise and which were the focus of most contextuality arguments until recent times (such as all arguments based on the examples presented in  Sec.~\ref{section3} and \ref{ssec:NC_ineq}).
Notice, however, that in this case the conclusions of the tests of contextuality must take into account imperfections, and  techniques hence need to be developed for analyzing the experimental data; see Sec.~\ref{ssec:exper_imp} and  \ref{ssec:experiments}).

A different notion of classicality for the case of nonideal measurements is presented in Sec.~\ref{ssec:other_cont}, namely, Spekkens contextuality \cite{Spekkens:2005PRA}.
We provide an account of Spekkens contextuality in Sec.~\ref{ssec:other_cont} in order to clarify the distinctions between his approach and the one presented here.

Finally, there are two other related research directions that introduce some notion of contextuality that we do not cover in this review. One was developed by \citet{Grangier2021}, \citet{Auffeves2020, Auffeves2022} and the other was developed by \citet{Griffiths2017,Griffiths2019,Griffiths2020}. Despite the similar terminology and motivation, i.e., the analysis of Bell- and Kochen-Specker-type arguments and experiments, these approaches discuss a different notion of contextuality with respect to the one presented in this review. We refer the interested reader to the corresponding literature.


\section{The Kochen-Specker theorem}\label{section3}


This section is devoted to the Kochen-Specker theorem, which can be considered the starting point of the research in quantum contextuality. It states that noncontextual models conflict with quantum theory. 
Understanding this theorem, as well as the several variants presented in the following, is important for understanding quantum
contextuality and the further developments in this research line.


\subsection{Kochen-Specker sets}\label{sec:KS-sets}
The example presented in Sec.~\ref{sec:nutshell} is based on the violation of an 
inequality that is satisfied by any noncontextual value assignment (or convex combinations thereof). 
This is a ``modern'' tool to witness quantum contextuality. In contrast, the 
original argument by \textcite{Kochen:1967JMM} was designed as a logical 
impossibility proof for value assignments. In the following, we  explain the 
original argument and some of its  simplifications. 

\begin{figure}[t]
\centerline{\includegraphics[width=0.5\textwidth]{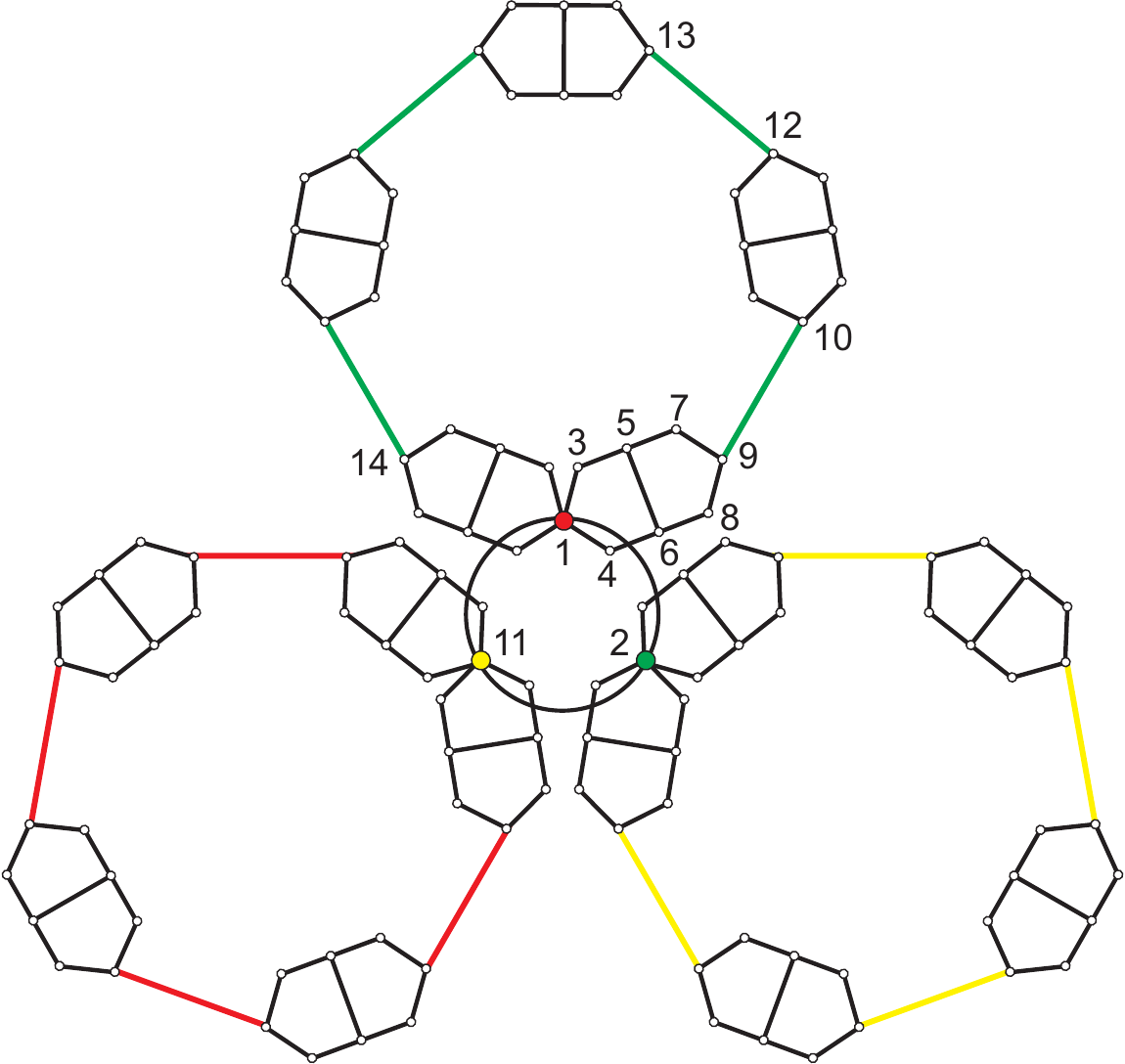}}%
\caption{\label{fig:KS117}The set in the original proof of KS has 117 vectors 
and 118 contexts. Each node represents a vector. Nodes in the same straight 
line or circumference represent mutually orthogonal vectors. The red node is 
orthogonal to all nodes connected to the red edge. We proceed similarly with the green and 
yellow nodes.
A proof of the KS theorem can be obtained as follows. One of the nodes $1$, $2$ 
and $11$ has to be true. The symmetry of the graph allows us to assume without 
loss of generality that it is node $1$. Therefore, node $9$ must be false, because of the ``bug'' subgraph (see Fig.~\ref{fig:bugs}) between node $1$ and node $9$. Thus, since nodes $2$, $9$, and $10$ are mutually orthogonal and node $2$ is connected to node $1$, node $10$ must be true. Applying the same argument, node $12$ must be false and node $13$ must be true since nodes $12,13$, and $2$ form a basis. When this is repeated twice, node $14$ must be true. However, nodes $1$ and $14$ cannot both  be  true. This concludes the proof. This figure improves the representation in the work of \textcite{Kochen:1967JMM}, where the 117 vectors are represented by 120 nodes using two nodes each for 3 of the 117 vectors.}
\end{figure}

The Kochen-Specker (KS) theorem \cite{Kochen:1967JMM} deals with assignments of truth values to potential measurement results. In quantum mechanics, such measurement results are, for ideal measurements, described by projectors. Each projector defines a subspace of the Hilbert space, namely, the subspace that is left invariant under the action of the projector. If this subspace is one dimensional, then this subspace is a ray spanned by a single vector. The KS theorem can be seen as a statement about the impossibility of certain assignments to sets of vectors or, equivalently to sets of rank-1 projectors.

We first fix the framework for the theorem. In a $d$-dimensional Hilbert space $\hilbert$ consider $d$ rank-1 projectors $P_1,P_2,\ldots,P_d$ associated with $d$ orthogonal vectors in $\hilbert$. They satisfy the following relations
\begin{itemize}
\item[({\bf O})] $P_i P_j=0$ for any $i\neq j$ (orthogonality).
\item[({\bf C})] $\sum_{i=1}^d P_i=\openone$ or, equivalently, using ({\bf O}), ${\prod_{i=1}^d(\openone - P_i)=0}$ (completeness).
\end{itemize}
Such relations can be interpreted in terms of yes-no questions (or true-false propositions) $Q_1,\dotsc,Q_d$ as follows:
\begin{itemize}
\item[($\mathbf{O'}$)] $Q_i$ and $Q_j$ are  exclusive; i.e., they cannot be  
simultaneously ``true'' for $i\neq j$.
\item[($\mathbf{C'}$)] $Q_1,\dotsc,Q_d$ cannot be simultaneously 
``false''; one of them has to be true.
\end{itemize}
Note that while the relation ($\mathbf{O'}$) was already used by \textcite{Kochen:1967JMM}, use of the name exclusive is modern, see \textcite{CSWPRL2014} and \textcite{AcinCMP2015}.

\begin{figure}[t]
\includegraphics[width=.4\linewidth]{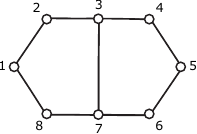}
\large{(a)}
\includegraphics[width=.4\linewidth]{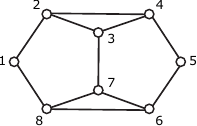}
\large{(b)}
\caption{\label{fig:bugs}%
Different representation of the same orthogonality relations. (a) Vectors are represented by nodes, contexts are represented by straight lines (or, more generally, by smooth lines). Three vectors in the same straight line are mutually orthogonal. This graph is also called a Greechie diagram; see \citet{Greechie:1971}; (b) Vectors are represented by nodes and orthogonal vectors are connected by edges. In particular, triples of mutually orthogonal vectors form triangles, e.g., nodes $2,3$, and $4$. An example of vectors realizing such relations is given by:
$v_1=( 1,-1, 1), v_2=(1, 1, 0), v_3=(0,0, 1), v_4=(1,-1, 0)$, $v_5=( 1, 1,-1), v_6=(1, 0, 1), v_7=(0,1, 0), 
v_8=(1, 0,-1)$.
This graph is called the bug \cite{Svozil_priv_comm} and it has the 
property that, for $d=3$, an assignment of true to node 1 implies a 
false assignment for node 5. In fact, if nodes 1 and 5 are both true, then nodes 2, 4, 6, and 
8 must be false (they are connected), which implies nodes 3 and 7 are true 
(they are the remaining nodes of two triples), which gives a contradiction 
since nodes 3 and 7 are connected. As can be seen in Fig.~\ref{fig:KS117}, they are 
the building block of the original proof of the KS theorem.}
\end{figure}

\begin{figure}[t]
	\centerline{\includegraphics[width=0.90\linewidth]{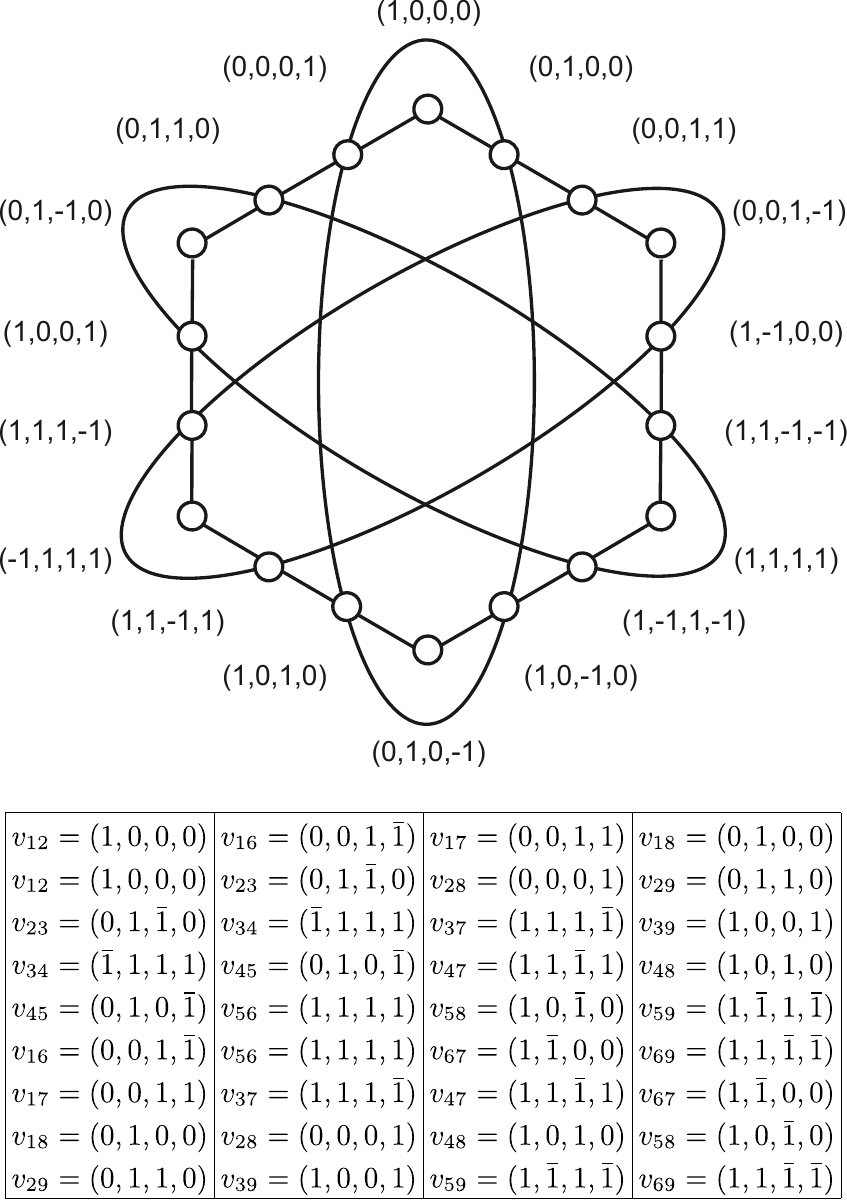}}%
	\caption{\label{fig:cab18} Graphical representation of the 18-vector KS set by \citet{Cabello:1996PLA}. Each node represents a vector. For simplicity, vectors are unnormalized. Each smooth line, i.e., every straight line or ellipse, represents a context. Vectors in each context are mutually orthogonal. Each vector appears in exactly two contexts e.g., $v_{12}$ appears both in context $1$ and $2$, etc. As a consequence, by assigning a noncontextual ``true'' to some vectors, one obtains an even number of true, whereas one should get exactly nine true propositions, one for each context. Therefore, the set is not KS colorable. Notice that there are additional relations of orthogonality not shown in the graph and not used to prove the contradiction. Vectors are also listed in the table, where each row represent a context. For simplicity, we denote $-1$ as $\bar{1}$.}
\end{figure}

\begin{figure}[t]
	\centering
	\centerline{\includegraphics[scale=0.44]{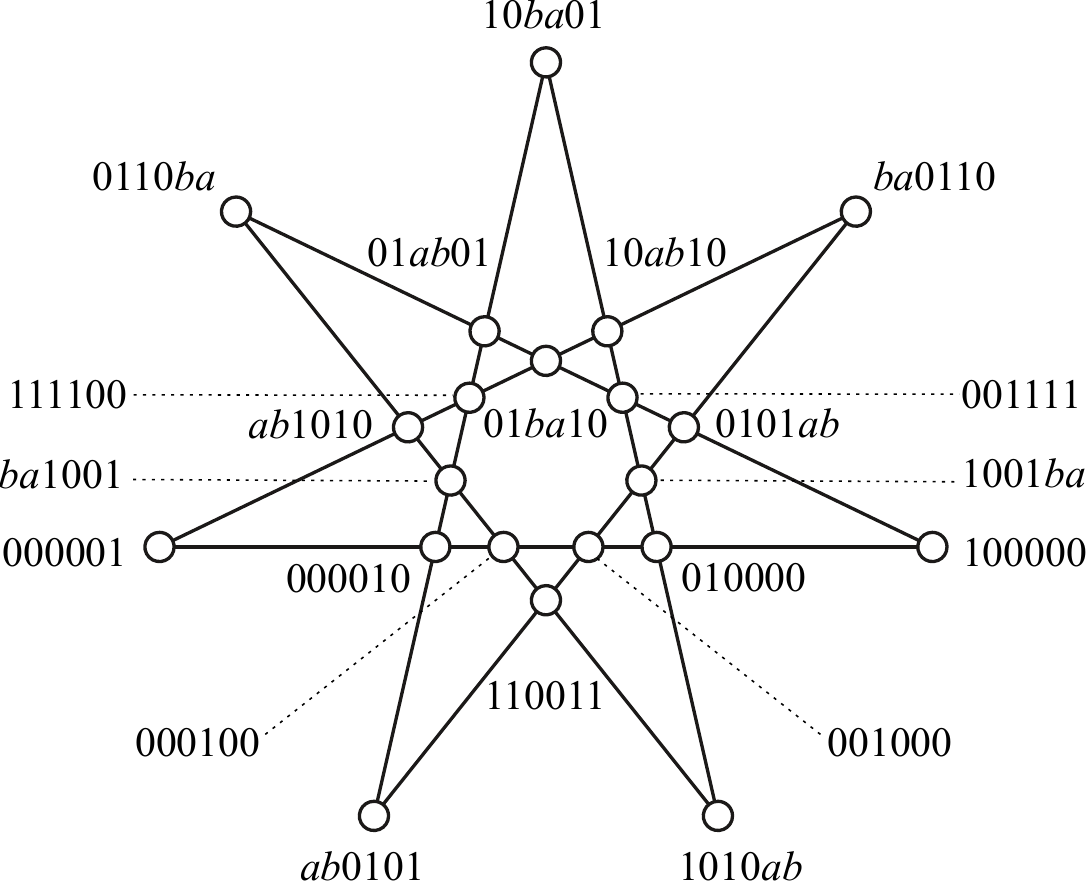}}
	\caption{\label{Fig:7_cont}Orthogonality relations between the vectors of the 21-vector, seven-context KS set given by \citet{LisonekPRA2014}. Vectors are represented by nodes and contexts by straight lines. $1010ab$ denotes the vector $(1,0,1,0,a,b)$, where $a=e^{2 \pi i/3}$ and $b=a^2$. For simplicity, normalization factors are omitted. Contexts contain mutually orthogonal vectors. The set has 21 vectors and each vector is in two contexts. To map one and only one of the vectors in each context to $1$ and preserve the latter property, one would need to associate $1$ with $21/2$ vectors, which is not an integer. This makes the mapping impossible and proves that the set is a KS set.}
\end{figure}

\begin{figure}[t]
	\centerline{\includegraphics[width=0.5\linewidth]{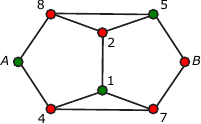}}%
	\caption{\label{fig:KS8-Yu-Oh} Subgraph of the Yu-Oh graph given by the vertices $\set{A,B,1,2,4,5,7,8}$, corresponding to a basic block of the original KS graph (the bug), with a valid coloring, i.e., green=true, red=false. As discussed in Sec.~\ref{sec:KS-sets} (Fig.~\ref{fig:bugs}), $A$ and $B$ must be exclusive events.} 
\end{figure}

For an arbitrary set of rank-1 projectors in dimension $d$, only certain subsets may obey conditions ({\bf O}) and ({\bf C}) and, analogously, for an arbitrary set of propositions and a fixed $d$, certain subsets can be subject to conditions ($\mathbf{O'}$) and ($\mathbf{C'}$). A set of $d$ mutually exclusive propositions is a context. We use two different graphical representations for the relations ($\mathbf{C'}$),($\mathbf{O'}$) in a set of propositions, as in Figs.~\ref{fig:KS117} and \ref{fig:bugs}. In one representation, sets of mutually exclusive propositions are nodes in the same straight or smooth line; see Fig.~\ref{fig:bugs}~(a) as well as Figs.~\ref{fig:KS117},~\ref{fig:cab18}, and \ref{Fig:7_cont}. In the other representation in Fig.~\ref{fig:bugs}~(b), edges simply connect exclusive propositions; see also  Figs.~\ref{fig:yo13int} and \ref{fig:KS8-Yu-Oh}. The constraints ($\mathbf{O'}$) and ($\mathbf{C'}$) can be translated into rules for coloring the vertices of the graph with two colors (e.g., green for true and red for false), namely, such that each two exclusive nodes cannot be both green, condition ($\mathbf{O'}$), and each set of $d$ mutually exclusive propositions must contain a green node, condition ($\mathbf{C'}$); see also Fig.~\ref{fig:KS8-Yu-Oh}. The problem of finding a coloring  with two colors according to the previously stated rules is referred to as the {\it KS colorability problem} \cite{Belinfante1973}. We see in Sec.~\ref{sssec:NCHVE} how the conditions ($\mathbf{O'}$) and ($\mathbf{C'}$) can be relaxed in modern approaches to contextuality.

Kochen and Specker provided a physical interpretation of certain rank-1 projectors in $d=3$ as spin operators for a spin-1 particle.
More precisely, the spin operator along one direction, say $S_x$ along the direction $x$ in the Euclidean three-dimensional space, has the eigenvalues ${-1,0,+1}$. Hence, the operator $S_x^2$ is a projector on the two-dimensional space corresponding to the eigenvalues $\pm 1$. Moreover, such spin operators have the property that for each triple of orthogonal directions, say  $x,y$, and $z$, the operators commute, i.e., $[S_x^2,S_y^2]=[S_x^2,S_z^2]=[S_y^2,S_z^2]=0$.
In this way, directions in the Euclidean space correspond to spin measurements and can be identified with directions in the Hilbert space. For a given direction $\vec v$, the one-dimensional projector $P_{\vec v}=\openone- S_{\vec v}^2$ can be interpreted as the measurement outcome $0$ of a spin measurement in direction $\vec v$. For this reason, in the context of the KS theorem one often considers vectors $\vec v\in \reals^3$ in the place of rank-1 projectors $P$ in a three-dimensional Hilbert space.


It is then straightforward to show that a joint measurement of the dichotomic observables $P_{\vec u},P_{\vec v}$, and $P_{\vec w}$ for three orthogonal vectors $\vec u$, $\vec v$, and $\vec w$ can be obtained as a single trichotomic measurement having as effects precisely $\{P_{\vec u},P_{\vec v},P_{\vec w}\}$, i.e., a measurement in a given orthogonal basis. With the usual convention for truth assignments, i.e., $1$ for true and $0$ for false, the previously mentioned assignments can be formulated as a map from a (finite) set of vectors $S\subset \mathbb{R}^3$ to $0$ and $1$ such that for any orthogonal basis contained in the set $S$ one and only one of the vectors is mapped to the value~1. 
The set $S$ consists of several bases, possibly with intersection; i.e., one vector may be part of several bases in $S$. Kochen and Specker then proved the following: 
\begin{theorem*}{\bf (Kochen-Specker, 1967)}
There is a finite set $S\subset \mathbb{R}^3$ such that there is no function $f:S\rightarrow \{0,1\}$ satisfying
\begin{equation}
f(\vec u) + f(\vec v) + f(\vec w) = 1
\end{equation}
for all triples $(\vec u, \vec v,\vec w)$  of mutually orthogonal vectors in $S$.
\end{theorem*}

In general, a KS set $\mathcal S$ in dimension $d$ is defined as a set $\mathcal S$ of vectors in a $d$-dimensional Hilbert space, with the property that there is no map $f:\mathcal S\rightarrow \{0,1\}$ satisfying $\sum_{\ket\psi\in \mathcal{B}} f(\ket\psi)=1$, for any subset $\mathcal{B}\subset \mathcal S$ of $d$ orthogonal vectors. Since any such set provides a proof of the KS theorem in dimension $d$, these sets are also called  a ``proof of the KS theorem.'' That there is no KS set for $d=2$ follows from the fact that one can construct explicit noncontextual assignments for all projectors in $\mathbb{C}^2$; see e.g., \cite{Kochen:1967JMM}.

The original proof of the KS theorem consists of a set of 117 vectors that realize the graph in Fig.~\ref{fig:KS117} for which it is impossible to assign values true and false such that two adjacent nodes cannot be both true [condition ($\mathbf{O'}$)] and each set of three mutually exclusive nodes must contain a value true [condition ($\mathbf{C'}$)]. For any two vectors that are orthogonal but do not participate in a basis one can readily add a vector to complete the pair to a basis. This enlarges the 117 vectors to 192 vectors and those 192 vectors form then the set $S$ in the KS theorem.

The original KS proof is long and complicated, given the high number of vectors necessary to obtain a contradiction. Some worked on the problem and simplified it by finding KS sets with an increasingly small number of vectors in different dimensions. For example, in dimension~3 \cite{Belinfante1973, BubFOP1996,deObaldia:1988FPH,Alda:1980AM, ConwayKochen,Peres:1988,Peres:1991JPA,Peres:1993}, in dimension~4 \cite{Penrose:2000, Zimba:1993SHPS, Peres:1991JPA, Kernaghan:1994JPA,Cabello:1996PLA}, in dimension~6 \cite{LisonekPRA2014}, and in dimension~8 \cite{Kernaghan:1995PLA,Toh:2013aCPL,Toh:2013bCPL}. Subsequent works have identified many other examples of KS sets in different dimensions \cite{Arends:2011,Cabello:1994EJP,Gould:2010FPH,Pavicic:2005JPA,Pavicic:2006, Waegell:2012JPA,RuugeJPA2012,Aravind:1998JPA, Waegell:2010JPA, Pavicic:2011JMP,Waegell:2011FPH:2, Waegell:2011FPH, Megill:2011PLA, Waegell:2011JPA, Waegell:2013PRA, Waegell:2017PRA, WaegellJPA2015}.
The method used by \citet{Kochen:1967JMM} can be extended to construct KS sets in any dimension $d>3$ \cite{Cabello:1996JPA}. Other methods for obtaining KS sets in $d>3$ were proposed in \citet{Zimba:1993SHPS,RuugeJPA2007,Cabello:2005PLA,MatsunoJPA2007,Pavicic:2005JPA}. KS sets with a continuum of vectors in dimension~3 were presented by \citet{Galindo:1975, Gill:1996JPA}. A way to further reduce the number of vectors was discussed by \citet{Cabello:1996PLAb}.
 

The smallest KS set in terms of vectors is the 18-vector (nine-context) set in dimension~4 introduced by~\citet{Cabello:1996PLA} and shown in Fig.~\ref{fig:cab18}. A proof of the minimality was presented by~\citet{Xu:2020PRL}. The impossibility of an assignment satisfying the conditions ({\bf O}) and ({\bf C}) is proven by a parity argument: since there are nine contexts, one must assign true exactly nine times. However, this is not possible, since each vector appears in two contexts.

The smallest KS set known in terms of contexts is the 21-vector seven-context set in dimension~6 introduced by~\citet{LisonekPRA2014} and shown in Fig.~\ref{Fig:7_cont}. This KS set has also been proven to be the one with the smallest number of contexts, thereby, allowing for a symmetric parity proof of the KS theorem \cite{LisonekPRA2014}.


The first step in the proof in \citet{Kochen:1967JMM} consists of identifying a set of eight vectors whose relations of orthogonality are represented by the eight-node graph in Fig.~\ref{fig:KS8-Yu-Oh}. Specker called this graph the bug \cite{Svozil_priv_comm}. It has the peculiarity that, whenever $A$ is true, then $B$ must be false. This is at the basis of several KS-type contradictions, such as the ones by \citet{Stairs:1983PS} and \citet{Clifton:1993AJP}. They provide a realization of the bug as orthogonality relations of a set of rank-1 projectors $P_A,\ldots, P_B$  such that $P_A=\ketbrac{\psi}$, with  $\mean{\psi|P_A|\psi}=1$ and  $\mean{\psi|P_B|\psi} > 0$, which contradict the KS assignment rules; see Fig~\ref{fig:bugs}. The bug is the simplest example \cite{Cabello:2018PRA} of other ``true-implies-false'' structures; see \citet{Cabello:1995JPA}. Hardy's proof \cite{HardyPRL1993} can be recast as a true-implies-false proof~\cite{Cabello:1996PLA} in which the initial truth corresponds to being in a particular entangled state. Similarly, one can construct proofs in which the initial and final propositions are product states \cite{CabelloPRA1997}.


\subsection{Generalized Kochen-Specker-type arguments}\label{ssec:sicnotks}
A different approach to the Kochen-Specker contradiction has been undertaken
using other types of algebraic relations instead of ({\bf O}) and ({\bf C}). 
Important examples are the PM magic square \cite{Peres:1990PLA, 
Mermin:1990PRL, Mermin:1993RMP}, the Mermin magic pentagram, and the scenario 
of Yu and Oh \cite{Yu:2012PRL, Bengtsson:2012PLA, Kleinmann:2012PRL}.


\subsubsection{The magic square and the magic pentagram}
\label{PM}


The PM square \cite{Peres:1990PLA, Mermin:1990PRL, 
Mermin:1993RMP} introduced in Sec.~\ref{sec:nutshell} is a proof of 
the Kochen-Specker theorem even though it does not explicitly use a KS set of 
vectors. The difference resides in the fact that instead of imposing  
({\bf O}) and ({\bf C}) relations on rank-1 projectors they impose analogous 
algebraic relations 
on $\pm 1$ observables. Consider, again, the square of observables
\begin{equation}\label{s3:PM}
\left[\begin{matrix}
 A&B&C\\
 a&b&c\\
 \alpha&\beta&\gamma
\end{matrix}\right]
=
\left[\begin{matrix}
 \sigma_z \otimes \openone&
\openone \otimes \sigma_z&
\sigma_z \otimes \sigma_z\\
 \openone \otimes \sigma_x&
 \sigma_x \otimes \openone&
 \sigma_x \otimes \sigma_x\\
 \sigma_z \otimes \sigma_x&
 \sigma_x \otimes \sigma_z&
 \sigma_y \otimes \sigma_y
\end{matrix}\right].
\end{equation}
Each row and column contains a set of commuting observables. In addition, we have the product of observables along the rows and column being $+\openone$, with the exception of the last column, where it is $-\openone$. The logical relations ($\mathbf{O'}$) and ($\mathbf{C'}$) are  substituted here with the algebraic relations
\begin{equation}\label{eq:alcondPM}
\begin{split}
v(A)v(B)v(C) =  v(a)v(b)v(c) = \\
= \cdots  = - v(C)v(c)v(\gamma) = +1, 
\end{split}
\end{equation}
where with $v(A)$ we denoted the value $\pm 1$ assigned to the measurements $A$, etc. It is then clear that Eq.~\eqref{eq:alcondPM} can never be satisfied, since it would imply
\begin{equation}
\begin{split}
[v(A)v(B)v(C)]  [v(a)v(b)v(c)]  \cdots  [v(C)v(c)v(\gamma)]\\
 = 1 \times 1\times \ldots \times (-1)= -1.
\end{split}
\end{equation}
But, on the other hand, as in Eq.~\eqref{eq:pm_all},
\begin{equation}
\begin{split}
v(A)v(B)v(C)   v(a)v(b)v(c)  cldots  v(C)v(c)v(\gamma) \\
=  v(A)^2v(B)^2v(C)^2 \cdots v(\gamma)^2 = 1,
\end{split}
\end{equation}
which gives a contradiction. The magic square can be converted into 
a standard proof of the KS theorem with vectors \cite{Peres:1991JPA}.


\begin{figure}[t]%
	\centerline{\includegraphics[width=\linewidth]{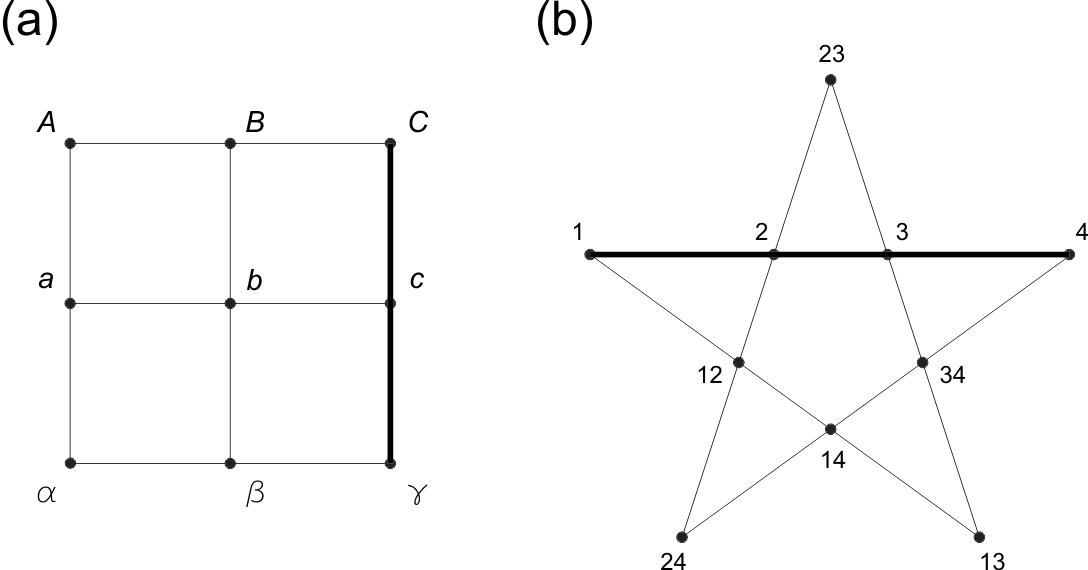}}
	\caption{\label{fig:magicsets} (a) PM magic square and (b) Mermin's magic pentagram. Each dot represents an observable with possible outcome $-1$ or $1$. Each line contains mutually compatible observables. For each line, the product of the corresponding observables is the identity, except for the bold lines, where it is minus the identity. A possible choice of observables satisfying the conditions in (a) is given in Eq.~\eqref{s3:PM}.  A possible choice of observables satisfying the conditions in (b) is the following: $O_1=\sigma^{(1)}_z \otimes \sigma^{(2)}_x \otimes \sigma^{(3)}_x$, $O_2=\sigma^{(1)}_x \otimes \sigma^{(2)}_z \otimes \sigma^{(3)}_x$, $O_3=\sigma^{(1)}_x \otimes \sigma^{(2)}_x \otimes \sigma^{(3)}_z$, $O_4=\sigma^{(1)}_z \otimes \sigma^{(2)}_z \otimes \sigma^{(3)}_z$, $O_{12}=\sigma^{(3)}_x$, $O_{13}=\sigma^{(2)}_x$, $O_{14}=\sigma^{(1)}_z$, $O_{24}=\sigma^{(2)}_z$, and $O_{34}=\sigma^{(3)}_z$.
}%
\end{figure}


There is a similar compact proof of the KS theorem with Pauli operators for three qubits found by \citet{Mermin:1990PRL,Mermin:1993RMP}. It is based on ten observables that can be arranged as shown in Fig.~\ref{fig:magicsets} (b), a construction that is sometimes called the magic pentagram. 

The PM magic square and Mermin's magic pentagram have the minimum 
number of Pauli observables required for proving the KS theorem for two and 
three qubits, respectively. There are several similar proofs of the KS theorem 
with Pauli observables for more than three qubits 
\cite{Saniga:2012QIC,Planat:2012EPJ,Planat:2013,Waegell:2013PRA,Waegell:2013PLA,Waegell:2014PRA}. 
A result by \textcite{Arkhipov:2012XXX} showed that all critical 
(i.e., the contradiction disappears by removing one observable) parity proofs 
(i.e., based on a parity argument, as described before, see Sec.~\ref{sec:KS-sets}) of the KS theorem for 
more than three qubits with Pauli observables, where each observable is in 
exactly two contexts, can be reduced to the magic square or the magic 
pentagram. This is not true if each observable is in an even number of contexts larger than two \cite{Trandafir:2022PRL}.


\subsubsection{Yu and Oh's set}\label{sssec:Yu-Oh_KS}


\begin{figure}[t]%
\centerline{\includegraphics[width=.92\linewidth]{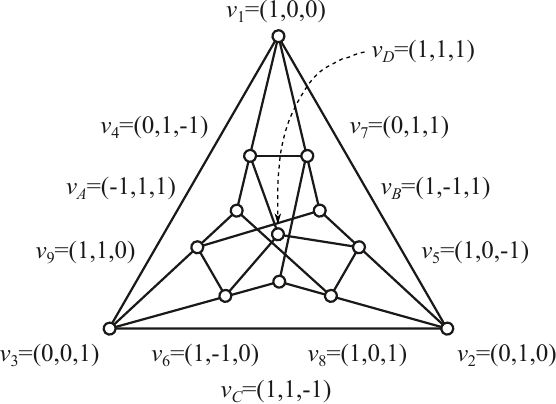}}
\caption{\label{fig:yo13int} Graph of orthogonality between the vectors of the 
Yu and Oh set \cite{Yu:2012PRL}. Adjacent nodes represent orthogonal vectors. 
For simplicity, vectors are unnormalized. See the text for further details.}
\end{figure}



Yu and Oh's argument \cite{Yu:2012PRL} does not provide a proof of the 
Kochen-Specker theorem. However, it fits in the more general framework of {\it 
state-independent contextuality} (SI-C)\footnote{The acronym ``SIC'' could generate some confusion, as it is often used in the quantum foundations and information community to denote
{\it symmetric informationally complete} sets. That is not the case here. Following the notation of \citet{CKB2015}, we use SI-C to denote state-independent contextuality and distinguish it from symmetric informationally complete.}, namely, it is a contextuality argument 
that depends not on the choice of a particular quantum state, but rather on 
the properties of the observables alone. Like KS proofs, Yu and Oh's 
argument is based on a set of rank-1 projectors. However, contrary to KS 
proofs, the corresponding vectors admit a value assignment that is consistent with the 
conditions ($\mathbf{O'}$) and ($\mathbf{C'}$). The contextuality argument arises from the 
fact that every probability distribution that is consistent with those assignments, 
i.e., that comes from a convex mixture of them, is in contradiction with the 
probabilities that can be obtained  using the 13 projectors, for all quantum 
states.

The basic elements are 13 vectors in $\mathbb{C}^3$, listed in Fig.~\ref{fig:yo13int}, and the corresponding set of projectors $\ketbrac{v}$.
The orthogonality relations of such vectors are depicted in Fig.~\ref{fig:yo13int}.
The projectors associated with nodes $A,B,C$, and $D$ sum up to a multiple of the identity, namely,
\begin{equation}
\ketbrac{v_A}+\ketbrac{v_B}+\ketbrac{v_C}+\ketbrac{v_D}=\frac{4}{3}\openone.
\end{equation}
Thus, for any quantum state the sum of their probabilities is ${4}/{3}> 1$. On the other hand,  the orthogonality relations among the vectors $\{v_i\}$, which correspond to the exclusivity of the respective propositions, imply that the propositions associated with nodes $A,B,C$, and $D$ are also exclusive, i.e., they cannot be simultaneously true. This exclusivity implies that the sum of probabilities ${\rm Prob}(A)+{\rm Prob}(B)+{\rm Prob}(C)+{\rm Prob}(D)\leq 1$ in any noncontextual hidden-variable model.

This can be easily proven by identifying subgraphs containing two of the vertices $\set{A,B,C,D}$ and two of the triangles $\set{1,4,7}$, $\set{2,5,8}$, $\set{3,6,9}$ as the basic blocks of the original KS proof, i.e., the bug, depicted in Fig.~\ref{fig:bugs}; for instance, the subgraph $\set{A,B,1,2,4,5,7,8}$ depicted in Fig.~\ref{fig:KS8-Yu-Oh}, implies that $B$ and $C$ are exclusive. By symmetry the same argument applies to any two vertices in  $\set{A,B,C,D}$.

To summarize, even if the nodes $A,B,C$, and $D$ are not connected in the graph in Fig.~\ref{fig:yo13int}, their relations with other compatible elements imply that such elements correspond to exclusive propositions. Thus the sum of their probabilities is bounded by 1, whereas is quantum mechanics (QM) such a bound can be violated. We 
see in Sec.~\ref{sssec:state-indep} how one can demonstrate this contradiction via a state-independent violation of a noncontextuality inequality. It has been proven that the Yu-Oh set is the SI-C set with the smallest number of vectors in any dimension \cite{Cabello:2016JPA}. 



\subsection{A special instance of Gleason's theorem}\label{ssec:Gleason}
In the following, we outline the connection between Gleason's theorem 
\cite{Gleason:1957JMM} and the Kochen-Specker theorem. It is helpful to recall 
the result underlying Gleason's theorem.
\begin{theorem*}{\cite{Gleason:1957JMM}} Let $f:S^2\rightarrow \mathbb{R}$ be a 
non-negative function on the real sphere $S^2\subset \mathbb{R}^3$, such that 
all orthonormal bases $(\vec u,\vec v,\vec w)$ in $S^2$ obey
\begin{equation}\label{eq:ffadd}
 f(\vec u)+f(\vec v)+f(\vec w)=1.
\end{equation}
Thus, there is a positive semidefinite matrix $R$ with $\tr(R)=1$, such 
that
\begin{equation}\label{eq:dmff}
f(\vec v) = \vec v^\top R \vec v.
\end{equation}
\end{theorem*}

It is evident that an assignment of values $0,1$ according to ($\mathbf{O'}$) to 
all the orthonormal bases in $\complexes^3$ satisfies the assumptions of the 
theorem; hence, it must be given by a density matrix, according to 
Eq.~\eqref{eq:dmff}. On the other hand, there is no density matrix providing 
$0,1$ assignments to all orthonormal bases in $\mathbb{R}^3$  (hence the contradiction \cite{Kochen:1967JMM}). In contrast, Kochen 
and Specker obtain a contradiction using only a finite set of vectors.

A similar argument, connecting Gleason's theorem to the impossibility of a noncontextual hidden-variable assignment, was provided by \citet{Bell:1966RMP}. He showed that given a function $f$ satisfying Eq.~\eqref{eq:ffadd}, two vectors $\vec{v}$ and $\vec{w}$ such that $f(\vec v)=1$ and $f(\vec w)=0$ cannot be arbitrary close. This, in turn, is in contradiction to the possibility of assigning $0,1$ values to all orthonormal bases while obeying the rules in Eq.~\eqref{eq:ffadd}, since there would be arbitrary close pairs with different assignments. Such a minimal angle between vectors has been quantified as $\tan^{-1}(1/2)\approx 0.464$, and the argument was further refined by \citet{Mermin:1993RMP}, who noticed that the previous reasoning  can be easily extended to an argument that uses only a finite set of vectors, such as the original KS argument.


\section{Contextuality as a property of nature}\label{section4}


The Kochen-Specker theorem, originally presented as a 
logical impossibility proof, did not involve any 
statistical argument but rather was based on perfect assignments 
of $0$ (false) or $1$ (true) to a set of quantum propositions. 
This caused a debate on the role of finite precision measurements (see Sec.~\ref{nullif}) that also stimulated the development of statistical 
versions of the Kochen-Specker contradiction. The results 
of this effort were {\it noncontextuality inequalities}, 
which, under certain assumptions, are able to experimentally 
detect the phenomenon of quantum contextuality. In the following, 
we introduce the basic notions and open problems associated with 
noncontextuality inequalities and contextuality tests. We present the 
definition of noncontextual hidden variable models and 
noncontextuality inequalities in Sec.~\ref{ssec:NC_ineq}, 
 the operational definition of contexts in Sec.~\ref{ssec:compatible}, 
the problem of noise and imperfections in Sec.~\ref{ssec:exper_imp}, 
and finally experimental tests of contextuality in Sec.~\ref{ssec:experiments}.
In Sec.~\ref{ssec:other_cont}, we review a different notion of contextuality
introduced by \citet{Spekkens:2005PRA}.


\subsection{Noncontextuality inequalities}\label{ssec:NC_ineq}

Noncontextuality inequalities provide bounds obeyed by noncontextual hidden-variable models, in analogy to Bell inequalities that provide bounds for local-hidden variable models~\cite{Bell:1964PHY,Brunner:2014RMP}. The first proposals of Kochen-Specker-type inequalities were made by \citet{Simon:PRL2001} and \citet{Larsson:2002EPL}, but these require stronger assumptions than later noncontextuality inequalities~\cite{Klyachko:2008PRL,Cabello:2008PRL}, as we later discuss.

We start by introducing the mathematical formulation of noncontextual hidden variable models. We then discuss basic examples of noncontextuality inequalities
such as the  the Klyachko-Can-Binicio{\u{g}}lu-Shumovsky (KCBS)  \cite{Klyachko:2008PRL} and the Yu-Oh inequalities \cite{Yu:2012PRL}, exhibiting, respectively, state-dependent and state-independent quantum violations. Finally, we compare these constructions with other related approaches to noncontextual hidden-variable models.

\subsubsection{Mathematical structure of noncontextual hidden-variable models}\label{sssec:NCHV}

Different definitions of noncontextual hidden-variable (NCHV) models are present in the literature that despite their substantial equivalence, use different terminology and different mathematical structures, from the marginal problem definition in the work of KCBS~\cite{Klyachko:2008PRL,Chaves:2012PRA,FritzIEEE13}, the equivalent definition of the noncontextuality polytope \cite{Kleinmann:2012PRL}, to the sheaf-theoretical approach~\cite{Abramsky:2011NJP}, the graph-theoretical approach by~\citet{CSWPRL2014}, and the hypergraph-theoretical approach by~\citet{AcinCMP2015} \cite[see also the book by][]{Amaral2018Book}.
Here we adopt what we consider a minimal mathematical structure based on the noncontextuality polytope and the marginal problem characterization of NCHV. Further properties of NCHV models related to graphs and hypergraphs are discussed in Sec.~\ref{sec:Advanced}.

Given a set of observables $\mathcal{G}=\set{A_1,\ldots,A_n}$, a collection of contexts 
is a subset $\mathcal{M}$ of the power set of $\mathcal{G}$, i.e., 
$\mathcal{M}\subset 2^{\mathcal{G}}$; $\mathcal{M}$ is sometimes called the 
{\it marginal scenario} \cite{Chaves:2012PRA}.
The idea behind this name is that the observed data from measurements in 
each context arise as a marginal of a global probability distribution on 
all observables.  For each context $\{A_{i}\}_{i\in \mathcal{C}} \in \mathcal{M}$, i.e., with  $\mathcal{C}\subset\{1,\ldots,n\}$, we  have a distribution $p_{\mathcal{C}}$ of the outcomes over it. A necessary but not sufficient condition for the existence of a global distribution is for these marginals to be locally consistent. In other words, for each $\mathcal{C}$ and $\mathcal{C}'$ we have that
\begin{equation}\label{eq:p_nondist}
p_{\mathcal{C}| \mathcal{C}\cap \mathcal{C}'}= p_{\mathcal{C}'| \mathcal{C}\cap \mathcal{C}'},
\end{equation} 
where ${| \mathcal{C}\cap \mathcal{C}'}$ denotes the restriction of the distribution to observables in the intersection of the two contexts, obtained simply by marginalization, i.e., by summing over the variables not in $\mathcal{C}\cap \mathcal{C}'$. This consistency condition of the marginals is sometimes called the sheaf condition \cite{Abramsky:2011NJP} and, in Bell scenarios, it is equivalent to the nonsignaling condition  of \citet{Popescu:1994FPH}. In a contextuality scenario this condition on probability distributions has also been called {\it nondisturbance} \cite{RamanathanPRL2012}.

In a NCHV model, we assume the existence of a hidden 
variable that determines the outcomes of each observable regardless of 
the context.  For each context, given by the observables $\{A_{i}\}_{i\in \mathcal{C}}$ and $\mathcal{C}\subset\{1,\ldots,n\}$ and the outcomes $\{a_{i}\}_{i\in \mathcal{C}}$, this corresponds to
 \begin{equation}\label{eq:nchvdef}
 p_\mathcal{C}(\{a_i\}_{i\in \mathcal{C}}) = \sum_\lambda p(\lambda) \prod_{i\in \mathcal{C}}p(a_{i} | \lambda),
 \end{equation}
 with $p(\lambda)\geq 0, \sum_\lambda p(\lambda)=1$, $p(a_{i} | \lambda)\geq 0$, $\sum_{a_{i}} p(a_{i} | \lambda) =1$, for $i \in \mathcal{C}$. Notice that the outcomes $a_s$ are arbitrary at this level. In most cases, we consider either $a_s\in\{0,1\}$ or $a_s\in \{-1,1\}$.
 
Equation~\eqref{eq:nchvdef} implies that each outcome  depends only on the hidden variable $\lambda$, not on the specific context in which the observable is measured. Given the factorization properties of the distribution in Eq.~\eqref{eq:nchvdef}, i.e., in the marginal problem approach, it can  easily be proven that the response functions $p(a_s | \lambda)$ that are not deterministic (i.e., $\neq 0,1$) can always be transformed into deterministic functions of a new hidden variable $\lambda'$.  This is due to the fact that all probability measures on a finite set of events (finite set of measurements and outcomes) are a convex mixtures of $\{0,1\}$-valued measures, i.e., deterministic assignments. At the same time, using that $\{0,1\}$-valued measures are multiplicative, i.e., $\delta(a_i, a_j)=\delta(a_i)\delta(a_j)$, one finds that a global probability distribution over all variables, which can be written as a convex mixture of global deterministic assignments, factorizes in a way similar to Eq.~\eqref{eq:nchvdef}.

By further developing this intuition, one can show that  Eq.~\eqref{eq:nchvdef} is equivalent to the existence of a global probability distribution over all observables 
 $A_1,\ldots,A_n$, such that $P(\{a_i\}_{i\in \mathcal{C}})$ is obtained by summing over all possible 
outcomes for the other observables, namely, 
\begin{equation}\label{eq:fineth}
p_\mathcal{C}(\{a_i\}_{i\in \mathcal{C}}) = \sum_{a_s: s\notin \mathcal{C}} p_\mathcal{G}(a_1,\ldots,a_n).
\end{equation}
This general argument is known in the literature as Fine's theorem \cite{Fine:1982PRL}, although Fine stated it only in the case of the \citet*[CHSH][]{Clauser:1969PRL} Bell scenario and a complete proof appeared as a straightforward  corollary of the {\it theorem of common causes} by \citet{Suppes:1981SYN}, published one year earlier. Even though it must have been quite well known, as it was used more or less explicitly in several works on Bell inequalities \cite[see, e.g.,][the paragraph below Eq.~(2)]{Froissart:1981}, a proof in the language of  contextuality has appeared in the literature only relatively recently, in connection with the sheaf-theoretical framework of  \citet[]{Abramsky:2011NJP}.

Starting from the previously mentioned definition, noncontextuality inequalities can be derived with the same methods used for Bell inequalities, namely with the {\it correlation polytope} method~\cite{Pitowsky:1989}. Like Bell inequalities, noncontextuality inequalities are satisfied by NCHV models, and their violation by data collected in a quantum experiment demonstrates quantum contextuality. We discuss general methods to derive them in Sec.~\ref{sssec:polytope} and Sec.~\ref{sssec:LP}.

\subsubsection{State-dependent contextuality}\label{sssec:state-dep}

\begin{figure}[t]
\begin{center}
\includegraphics[width=0.9\linewidth]{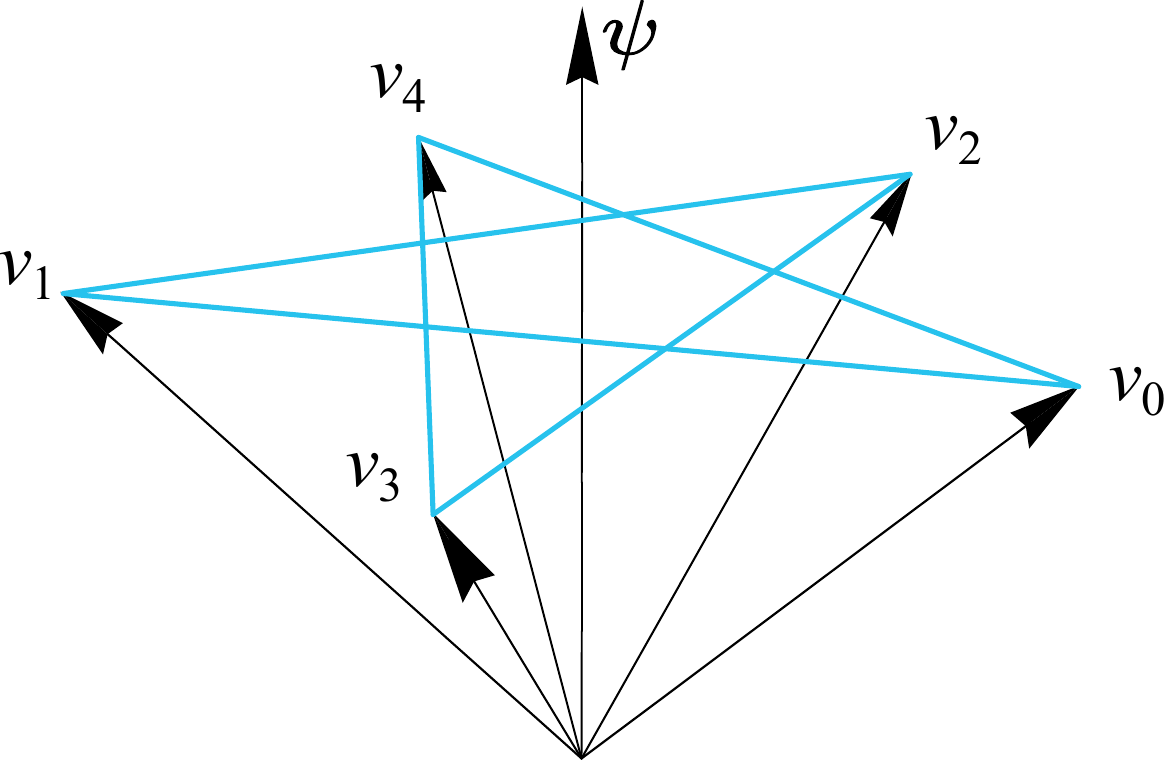}
\caption{\label{fig:klyset} The set of vectors $v_j\in \mathbb{R}^3$ giving the dichotomic observables $A_j=2\ketbrac{v_j}-\openone$ providing the maximum violation of the KCBS inequality form a regular pentagram, with orthogonal vectors connected by a blue line. The state $\psi$ is directed along its symmetry axis. A realization is $\ket{\psi}=(1,0,0)^\top$ and $\ket{v_k}=(\cos \theta, \sin \theta \cos \varphi_k, \sin \theta \sin \varphi_k)^\top$, with $\cos \theta=1/\sqrt[4]{5}$, $\varphi_k=2 \pi k/5$, and $k=(2 j +1) \mod 5$.}
\end{center}
\end{figure}

We can now proceed to discuss noncontextuality inequalities in the state-dependent scenario. 
The minimal dimension to witness contextuality is $d=3$ and the KCBS scenario \cite{Klyachko:2008PRL} is the simplest scenario where qutrits produce contextuality.  The scenario is defined by five  measurements $A_0,\ldots,A_4$, with outcomes $a_i\in\{-1,1\}$ such that $A_i$, and $A_{i+1}$, with sum modulo $5$, are compatible [i.e., they have a marginal scenario $\mathcal{M}=\set{(A_i,A_{i+1})}_{i=0}^4$]; see Fig.~\ref{fig:klyset}. KCBS  proposed the following inequality as being valid for NCHV models
\begin{equation}\label{eq:kcbs}
\mean{A_0 A_1} + \mean{A_1 A_2} + \mean{A_2A_3}
+ \mean{A_3A_4} + \mean{A_4A_0}\geq -3,
\end{equation}
where $\mean{A_i A_j}:= \sum_{a_i,a_j} a_i a_j\ p(a_i,a_j)$. According to the discussion in Sec.~\ref{sssec:NCHV} and by convexity arguments, the noncontextual bound $-3$ can be proven by trying all possible $\pm 1$ noncontextual assignments to the observables $A_i$. All other noncontextuality inequalities for this scenario can be obtained by relabeling the outcomes, i.e., by mapping $A_i\mapsto -A_i$. For instance, with the transformations $A_1 \mapsto -A_1$ and $A_3 \mapsto -A_3$, we obtain the inequality
\begin{equation}\label{eq:kcbs_alt}
\mean{A_0 A_1} + \mean{A_1 A_2} + \mean{A_2A_3}
+ \mean{A_3A_4} - \mean{A_4A_0}\leq 3.
\end{equation}

In contrast to Bell inequalities, here there is no bipartition of the set of observables such that 
every observable in one part is compatible with every observable of the other. 
Consequently, Eq. (\ref{eq:kcbs}) cannot be interpreted as a Bell inequality: The measurements 
must be performed on a single system. 

On a three-level system Eq. (\ref{eq:kcbs}) can be violated up to 
$5-4\sqrt{5} \approx -3.94$ 
with the state ${\ket{ \psi } =(1,0,0)}$ and measurement settings 
$A_j=2\ketbrac{v_j} -\openone$ and
$\ket{v_j}= \left(\cos\theta,\sin\theta\cos[j \pi 4/5],\sin\theta\sin[j \pi 4/5]\right)$ with 
$\cos^2\theta =\cos(\pi/5)/[1+\cos(\pi/5)]$, see Fig.~\ref{fig:klyset}. One can straightforwardly verify that $\mean{v_i|v_{i+1}}=0$ and, thus, $[A_i,A_{i+1}]=0$, where the sum is $\mod 5$. The KCBS inequality has been violated in several experiments, see Sec.~\ref{ssec:experiments} for more details.

The KCBS inequality, together with the inequality by \citet*[CHSH][]{Clauser:1969PRL}, is part of a general family of noncontextuality inequalities associated with compatibility structures forming an $n$-cycle, for $n=5$ and $4$ respectively. As a result of Vorob'ev's theorem~\cite{Vorobev1962} (see also the discussion in Sec.~\ref{sssec:graph_hyp}) cycles of length strictly greater than $3$ are necessary to witness contextuality. Other types of cycles inequalities were investigated by \citet{Bengtsson2007} for $n=7$ and by \citet{Cabello:2010XXX} for arbitrary $n$. The general form of the KCBS-like inequality for odd cycles of length $\geq 5$ was introduced by \citet{Cabello:2010XXX} and by \citet{Liang:2011PR}. The $n$-cycle inequalities were proven to be tight (see Sec.~\ref{polyt} for a formal definition) for any $n$-cycle contextuality scenario for $n\geq 5$ by~\citet{Araujo:2013PRA}. The case $n=4$ (CHSH) was already proven by \citet{Fine:1982PRL}. More generally, for  even $n$ with $n>4$, the NC inequalities can also be interpreted as Bell inequalities, the {\it chained} 
inequalities~\cite{BraunsteinAP1990}, but  they are not tight in the Bell scenario.

\subsubsection{State-independent contextuality}\label{sssec:state-indep}

State-independent contextuality is directly related to proofs of the KS theorem. In fact,
\citet{Badziag:2009PRL} proved that each KS set can be converted into an inequality showing SI-C, and \citet{XDYuetalNJP2015} developed  a similar method for proofs of the KS theorem based on more general algebraic conditions, as is the case for the PM square. 

\citet{Yu:2012PRL} proved a stronger statement, which was already partially discussed here. They provided 
a set of projectors admitting a $\{0,1\}$-assignment, according to the constraints (\textbf{O'}) and (\textbf{C'}), 
but that nevertheless demonstrates SI-C. From the set of vectors $\{\ket{v_i}\}_i$ listed in 
Fig.~\ref{fig:yo13int}, one constructs the observables ${A_i\equiv 2\ketbrac{v_i}-\openone}$. The compatibility 
relations among the observables $A_i$ follow from the orthogonality relations of the corresponding vectors, and 
are again summarized in the graph in Fig.~\ref{fig:yo13int}. Each pair of observables $A_iA_j$ such that $(i,j)$ is 
an edge of the graph is compatible. One can thus write the following NC inequality
\begin{equation}\label{yuoh}
\sum_i \mean{A_i} - \tfrac{1}{2}\sum_{\text{ edges }} \mean{A_iA_j}\leq 8,
\end{equation} 
 where the NCHV bound $8$ is simply computed by trying all possible $2^{13}$ noncontextual value assignments for $\{A_i\}_i$. However, using the vectors in Fig.~\ref{fig:yo13int} one can easily compute the quantum value for the operator 
\begin{equation}
L=\sum_i {A_i} - \frac{1}{2}\sum_{\text{ edges }} {A_iA_j}= \frac{25}{3}\openone,
\end{equation} 
giving
\begin{equation}
\mean{L}_\rho= \frac{25}{3} > 8, 
\end{equation}
for any quantum state $\rho$. The inequality associated with the Yu-Oh set has been further improved to maximize the gap between NCHV and the quantum values \cite{Kleinmann:2012PRL}.

Subsequently, several other SI-C sets that are not KS proofs have been proposed \cite{Bengtsson:2012PLA,Xu:2015PLA} and a systematic construction of state-independent contextuality inequality based on anti-distinguishable sets of vectors was proposed by \citet{Leifer:2020PRA}. Graph-theoretical methods allow for an exhaustive search of SI-C sets \cite{Cabello:2016JPA,RamanathanHorodecki2014,CKB2015} that is discussed in more detail in Sec.~\ref{graphth}.


\subsubsection{Logical and strong contextuality}\label{log_str_cont}


In the previous discussion, we distinguished between two types of statistical contextuality arguments: state dependent and state independent. A different classification has been proposed by \citet{Abramsky:2011NJP} was instead based on the existence of an extension of certain classical models and was independent of possible quantum violation of noncontextuality inequalities. More precisely, this classification adds to the standard notion of contextuality, two other stronger ones, namely, {\it logical contextuality} and {\it strong contextuality}. The notion of contextuality coincides with the one previously defined: the impossibility of a NCHV  [see Eqs.~\eqref{eq:nchvdef} and \eqref{eq:fineth}], namely, the impossibility of interpreting a given set of marginals in terms of a global distribution. 

To define logical and strong contextuality, we need the notion of possibilistic collapse of a probability distribution: Given a probability distribution associated with a marginal scenario $\mathcal{G}$, we replace every nonzero entry by $1$. In simple terms, a possibilistic collapse only distinguishes between possible (associated with $1$) and impossible (associated with $0$) events, without assigning a probability to them. Still, one is able to obtain contextuality for possibilistic models. A typical example of this type of contradiction was given by   \citet{HardyPRL1993}, who obtained a contradiction between quantum predictions and local hidden-variable models by simply looking at possible and impossible events. \citet{Abramsky:2011NJP} defined this type of arguments \textit{logical contextuality}, and showed that, in some sense, it is a stronger form of contextuality with respect to the standard one. In fact, the impossibility of a noncontextual description of a possibilistic assignment, in particular, implies the impossibility of a noncontextual description of the original probabilistic assignment; \cite[see][Proposition 4.4.]{Abramsky:2011NJP}. 

The possibilistic collapse of a distribution naturally introduces the notion of {\it support} of a distribution, namely, the possible events. \citet{Abramsky:2011NJP}  defined {\it strong contextuality} as the impossibility of any deterministic assignment that is {\it consistent} with the support of a distribution. To be more precise, by \textit{consistent} we mean that whenever an event is not in the support, we assign the value $0$ to it, when it is in the support, we may assign $0$ or $1$. The simplest example of strong contextuality is given by the Popescu-Rohrlich (PR) box \cite{Popescu:1994FPH}: a distribution between four variables $A_1,A_2,B_1$, and $B_2$ in a Bell scenario, i.e., with contexts $\{A_i, B_j\}$, such that all pairs are perfectly correlated except one that is perfectly anticorrelated, i.e., $\mean{A_1 B_1}=\mean{A_1B_2}=\mean{A_2B_1}=-\mean{A_2B_2}=1$. A perfect correlation (anticorrelation) means that only the events for which $a_i=b_j$ ($a_i\neq b_j$) are in the support of the distribution. It is then clear that there is no deterministic assignment satisfying these constraints. PR-box correlations are beyond what is possible in quantum mechanics; however, for two parties with three or
more settings and for three or more parties strong contextuality is possible with correlations realized in quantum mechanics. This is the case of \citet*[GHZ][]{GHZ89}-type arguments  and more generally the so-called all-versus-nothing arguments \cite{Mermin90b,Cabello:2001PRLb,Cabello:2005PRL}.

Another notion, closely related but based on the convex structure of noncontextual models, is that of {\it maximal contextuality}. Any probabilistic model $p$ satisfying the nondisturbance condition in Eq.~\eqref{eq:p_nondist} can be decomposed as
\begin{equation}
p= \alpha p_{\rm NC} + (1-\alpha) p_{\rm ND}
\end{equation}
for some $0\leq \alpha \leq 1$, where $p_{\rm NC}$ is a noncontextual distribution whereas $p_{\rm ND}$ is a generic nondisturbing one; see Eq.~\eqref{eq:p_nondist}. The maximal $\alpha$ such that this decomposition exists is called the {\it noncontextual fraction} of $p$ \cite{Abramsky:2011NJP,Amselem:2012PRL}, in analogy with the {\it local fraction} in Bell nonlocality~\cite{Elitzur:1992PLA}; see also \citet{Barrett:2006PRL,Aolita:2012PRA}. To a maximal $\alpha$, i.e., the noncontextual fraction, it corresponds a minimal $(1-\alpha)$, which is called the {\it contextual fraction}. A model is {\it maximally contextual} if $\alpha=0$. \citet{Abramsky:2011NJP} showed that a probabilistic model is strongly contextual if and only if it is maximally contextual. 

Note that a similar classification, based on Hardy-type (or ``definite prediction sets''), GHZ-type (or ``partially noncolorable sets''), and KS-type contextuality was introduced by \citet{Cabello:1996JPA} and used by \citet{Xu:2020PRL}. This hierarchy presents some analogies with the one by \citet{Abramsky:2011NJP}. All these classifications, however, can be considered in some sense incomplete as they take into account only classical models and thus fail to recognize the importance of contextuality arguments such as  the one by \citet{Yu:2012PRL}. 

We see in Sec.~\ref{polyt} how the contextual fraction is connected to noncontextuality inequalities and the polytope description of noncontextual correlations. Moreover, its connection to the resource theory of contextuality and advantages in quantum computation are discussed in Secs.~\ref{res_th} and \ref{sec:contextualityincomputation}, respectively. Finally, we remark that the sheaf-theoretical approach  of \citet{Abramsky:2011NJP} has inspired several lines of research connecting quantum contextuality to topological and algebraic topological methods \cite{Abramsky2011,Abramsky2015, Caru2016, Beer:2018PRA, Okay:2020Q, Raussendorf:2019QIC}.


\subsubsection{Other approaches to noncontextual hidden-variable models}\label{sssec:NCHVE}


The previously  discussed definitions  represent the minimal requirements for a model where outcomes have a context-independent assignment. Closer to the original formulation of the KS theorem, one can add {\it exclusivity}, arising from the condition ($\mathbf{O'}$), or {\it completeness}, arising from the condition ($\mathbf{C'}$) 
in Sec.~\ref{sec:KS-sets}. Noncontextuality inequalities derived   using both these two assumptions were usually called {\it Kochen-Specker inequalities} by~\citet{Larsson:2002EPL}, and a similar argument was presented by \citet{Simon:PRL2001}. 
More recent results do not use these extra conditions but they often appear  in parallel with a derivation of general NC inequalities as given by KCBS~\cite{Klyachko:2008PRL} and~\citet{Yu:2012PRL}, and in several subsequent works. It is thus worth mentioning such approaches to contextuality and emphasizing the difference and connection to the previously mentioned one.

A typical example is the following. Given a set of rank-1 projectors $\{ \ketbrac{v_i}\}_i$ in dimension $d$, the relations of compatibility (i.e., joint measurability) between the observables that they represent in quantum mechanics correspond to orthogonality relations; i.e., $\{\ketbrac{v_i}, \openone - \ketbrac{v_i}\}$ is compatible with $\{\ketbrac{v_j}, \openone - \ketbrac{v_j}\}$ if and only if $|\braket{v_i|v_j}|\in\set{0,1}$. If we denote $a_i=1,0$ as the classical value associated with the positive operator valued measure (POVM)  $\{\ketbrac{v_i}, \openone - \ketbrac{v_i}\}$
in the NCHV model, then the assumptions of exclusivity and completeness correspond, respectively, to $p(a_i= a_j=1|\lambda)=0$ whenever $\mean{v_i|v_j}=0$, and to $\sum_{i\in I} p(a_i=1|\lambda) = 1$ whenever $\mean{v_i|v_j}=0$ for all $i,j\in I, i\neq j$ and $\sum_{i\in I} \ketbrac{v_i}=\openone$. 

Any inequality valid for a NCHV model is also valid for an NCHV model with the previously mentioned extra assumptions. Conversely, an inequality valid for a NCHV model with such extra assumptions can be transformed into an inequality valid for NCHV models with the extra assumptions by adding extra terms, as we discuss in the following for the case of the extra assumption of exclusivity. A similar argument can be constructed for the case in which  the completeness condition ($\mathbf{C'}$) is also assumed. However, this does not necessarily mean that the bound is preserved. An example was provided by \citet{Bengtsson:2012PLA}, who presented a noncontextuality inequality with a bound of $63/5$ under the assumption of noncontextuality, but only $61/5$ if one adds the requirements of exclusivity and completeness. 

By assuming exclusivity, we obtain an inequality of the form
\begin{equation}
\sum_{i\in \mathcal{I}} \mu_i p(a_i=1) \stackrel{\rm NCHV+E}{\leq} \Omega, 
\end{equation}
with the superscript indicating that it holds when the extra exclusivity assumption is made. We can then 
add on the lhs pairwise correlation terms $-\mu_{ij}p(a_i = 1, a_j=1)$  with appropriately chosen 
weights $\mu_{ij} \geq 0$, such that the total value of the expression decreases when the 
exclusivity condition is violated. For  suitably chosen weights we have
\begin{equation}\label{eq:nchve-e}
\sum_{i\in \mathcal{I}} \mu_i p(a_i=1) -\sum_{(i,j)\in \mathcal{I'}}\mu_{ij}p(a_i = 1, a_j=1)\stackrel{\rm NCHV}{\leq} \Omega, 
\end{equation}
making Eq.~\eqref{eq:nchve-e} valid also for general NCHV models. 

It is straightforward to use this conversion in the KCBS inequality \cite{Klyachko:2008PRL}
 \begin{equation}\label{eq:kcbs-exc}
\sum_{i=0}^4 p(a_i = 1) \stackrel{\rm NCHV+E}{\leq} 2,
\end{equation} 
transforming it into an inequality valid for general NCHV models, namely
\begin{equation}\label{eq:kcbs-exc_tr}
\begin{split}
\sum_{i=0}^4 p(a_i = 1) -\sum_{i=0}^4 p(a_i=1,a_{i+1}=1) \\
= \sum_{i=0}^4 p(a_i=1, a_{i+1}=-1) \stackrel{\rm NCHV}{\leq} 2,
\end{split}
\end{equation}
where the sum in $a_{i+1}$ is modulo 5 and we simplify the expression by using the marginal condition in Eq.~\eqref{eq:fineth}. Not only does this transformation not change the classical bound, it also does not modify the quantum value obtained from the quantum observables $A_j=2\ketbrac{v_j} -\openone$ discussed in Sec.~\ref{sssec:state-dep}, since it satisfies by construction $\mean{v_i|v_{i+1}}=0$ giving $p(a_i=1,a_{i+1}=1)=0$.
This idea was exploited in several works \cite{Yu:2012PRL,CKB2015,AsadianPRL2015}, 
and a completely general treatment of the problem was presented by~\citet{Yu-Tong14,CabelloPRA2016}. 
Experimental tests of contextuality are challenging, so it is useful to  reduce the set of assumptions to an absolute minimum when designing such tests. Finally, a construction of further noncontextuality inequalities that use extra exclusivity assumptions can be obtained using the graph-theoretical approach to quantum contextuality; see Sec.~\ref{graphth}.


\subsection{Operational definitions and physical assumptions: ideal measurements}
\label{ssec:compatible}
In the previous sections,  an abstract notion of a context was enough to introduce 
the mathematical structures of NCHV models and the Kochen-Specker theorem; only some intuition on its physical meaning was provided. 
This is no longer sufficient when discussing possible experimental tests. In particular, the physical implications of a violation of a noncontextuality inequality actually depend on the physical assumptions at the basis of the chosen notions of context, the type of measurements considered, the details of their experimental implementation, etc. All these details must be verified to check their consistency
with the notion of the NCHV model one wants to test. Consider the CHSH inequality. If the measurements appearing in it are performed on distant particles, such that the input generation on one side and the outcome generation on the other side are spacelike separated events, then one can interpret its violation as a disproof of local hidden-variable models. In contrast, if the measurements are performed on the same system, the interpretation of the violation will have a different meaning, depending on the notion of context chosen. Are the measurements sharp? Are they commuting? These and similar questions must be addressed and their answers motivated by the terms of the actual experimental setup.
 
In the following, we address questions such as the following: What assumptions must be fulfilled by this implementation, in order to make an experiment a reasonable test of contextual behavior? What models can be disproved by such experiments? 
To achieve this, we proceed in two steps. First, we need an operational definition of contexts that allows us to identify them with certain experimental joint measurements. More precisely, we focus on the notion of disturbance for sequential measurements. This  is done in the framework of ideal measurements, where properties such as perfect nondisturbance are achievable. Second, we extend the notion of noncontextuality to  the case of nonideal measurements. We show in Sec.~\ref{ssec:exper_imp} how this can be achieved via an explicit quantification of the disturbance.

\subsubsection{Two perspectives: Observables and effects}\label{sssec:obs_eff}
To describe experimental realizations of contextuality tests, we  adopt as much as possible a ``black-box'' description of the measurements. Such a description does not presuppose the validity of quantum mechanics, even though the design of the operations carried out to obtain 
the measurement results (such as which laser pulses to use in ion experiments and where to put 
beam splitters and polarizers in photonic experiments) may be motivated by 
a quantum mechanical description. Each measurement apparatus is seen as a box that takes as input a physical system in a certain state and return a classical output and, possibly, the physical system in a new state. We are not interested in the details of the functioning of the apparatus, however, we still need some physical assumptions on how to combine the different experimental apparatuses to observe joint probability distributions. In fact, a prerequisite for contextuality, in the sense of the previously derived inequalities, is the possibility of performing  two or more measurements together, corresponding to nontrivial marginal scenarios. 

Note that different notions of contextuality exist in the literature.
To clarify the origin and relation between the different notions of contextuality, 
it is helpful to return to the original discussion by \citet{Kochen:1967JMM}. The starting point of their argument is that it is always possible to construct a hidden-variable model for a set of observables, if such a theory does not need to satisfy functional relations among them. At the same time, they were dissatisfied by the impossibility proof by \citet{vonNeumann:1932SPR,vonNeumann:1931AM}, which used linear relations among incompatible observables. In contrast, they chose an intermediate perspective inspired by Gleason's approach \cite{Gleason:1957JMM}, where functional relations are assumed only among compatible measurements since they can be experimentally tested in a joint measurement. As a notion of compatibility, they define the notion of {\it commea\-su\-ra\-bi\-lity}, meaning that the 
statistics of a set of observables $\{A_i\}_i$ can be recovered as a function of a single measurement $B$. In particular, for ideal measurements, this notion was shown to be equivalent to pairwise commutativity~\cite{Kochen:1967JMM}.
Using more modern terminology, we may identify this idea with the notion of {\it joint measurability} \cite{BuschBook,BuschBook2016, BLW_review}, which is valid for more general measurement. A generalized measurement is 
represented by a POVM: a collection of  effect operators $\{E_i\}_i$ 
such that $E_i\geq 0$ and $\sum_i E_i=\openone$ with the computation of probabilities via 
${\rm Prob}(i)=\tr(\rho E_i)$, and quantum instruments $\{\mathcal{I}_i\}_i$ for the computation of the 
state-update rule, i.e., $\rho \mapsto \mathcal{I}_i(\rho)$, where $\mathcal{I}_i$ is a 
completely positive map. That is ,${\rm id}_{\mathbb{C}^k}\otimes \mathcal{I}_i$ is a positive map for each $k$, where ${\rm id}_{\mathbb{C}^k}$ is the identity map on the Hilbert space $\mathbb{C}^k$, and $\sum_i \mathcal{I}_i$ is a trace-preserving map, i.e., $\tr[\sum_i \mathcal{I}_i(\rho)]=\tr(\rho)$ for all $\rho \geq 0$, also known as a quantum channel; see, e.g., \citet{Heinosaari:2012} for an introduction.
Two POVMs $A$ and $B$ with effects $A_i$
and $B_j$ are called {\it jointly measureable}, if there is a third POVM $G$ with effects $\{G_{ij}\}_{ij}$, such that $A_i = \sum_j G_{ij}$
and  $B_j = \sum_i G_{ij}$. Equivalently, one can substitute the sum over one index with a more general classical postprocessing~\cite{AliFOP2009}. 
The equivalence between joint measurability and commutativity is no longer true for nonideal measurements, this point  plays  an important role in the discussion of experimental tests. 

One may summarize here by saying that the basic elements of the KS theorem are dichotomic observables, i.e., the 
$\{P_i\}_i$ of Sec.~\ref{sec:KS-sets} with values $\{0,1\}$ or $\{$false, true$\}$, and contexts are defined as sets of commea\-su\-ra\-ble observables. On the other hand, as discussed  by \citet{Kochen:1967JMM}, the simplest way of performing a joint measurement of three observables, belonging to a context,  is given by a single trichotomic measurement, where the three orthogonal projectors $P_i,P_j$, and $P_k$ are interpreted as its effects. Thus, the KS theorem can  equivalently be analyzed from the {\it observable perspective} (OP) or from the {\it effect perspective} (EP), namely,
\begin{itemize}
\item[OP] The basic objects of contextuality are observables and their compatibility (joint-measurability) relations. A context is defined by a set of compatible observables. A noncontextual hidden-variable model is one that  assigns values to each observable regardless of which joint measurement they appear in.
\item[EP] The basic objects of contextuality are effects and their relation of being part of the same generalized measurement. A context is defined by a single measurement. A noncontextual hidden-variable model is one that assigns  values to each effect regardless of which measurement they appear in.
\end{itemize}

Note that this distinction between OP and EP has been introduced here to clarify and separate different ideas and different approaches to contextuality. In some older works on contextuality this distinction was not so sharp and the two perspectives were often interchanged. A typical example was given by~\cite{Klyachko:2008PRL}, who derived noncontextuality inequalities for both joint measurement and single effects. Moreover, we remark that OP should not be confused with the {\it Observable-based approach} discussed by \citet{AcinCMP2015}.

If one wants to pass from the scenario with idealized measurements to actual experimental tests the two perspectives present different challenges. Several questions arise, such as: What happens if the measurements are noisy? How can we operationally identify contexts in an experimental scenario? Possible ways of generalizing these two different perspectives to deal with actual experiments gave rise to different notions of contextuality. The OP is the the perspective we primarly consider in this review. The EP was the most common in the initial works on the KS theorem; see e.g., the discussion in Sec.~\ref{sec:KS-sets} in terms of value assignments to triples of orthogonal vectors. In more recent times, \cite{Spekkens:2005PRA,Spekkens:FOP2014} analyzed this approach to the Kochen-Specker theorem and contextuality, challenging the assumption of determinism and arguing against the possibility of applicability of such notions to unsharp measurements.  Such ideas have led to, among other things, a definition of 
contextuality and an approach to contextuality tests unlike from the one presented thus far.

For completeness, we want to clarify the difference between OP, 
the most common alternative EP, and the Spekkens definition of contextuality in some detail. 
We  discuss EP in Sec.~\ref{sssec:op_def_ep}. In particular, we  discuss Spekkens' 
criticism of the latter perspective. In Sec.~\ref{ssec:other_cont}, we  provide a more 
detailed discussion of Spekkens's approach to contextuality.

\subsubsection{Operational definitions of contexts: OP}\label{sssec:op_def_op}
Given the two discussed perspectives, a natural question is how to translate such notions into experimental procedures in the lab. For instance, an observable can be identified with an experimental measurement procedure (such as a sequence of laser pulses and detection for an ion-trap experiment, or optical elements and detection for a photonic one) and an effect can be connected to the probability of a certain outcome in a measurement. How can we say that we implement the ``same mathematical object'' in ``different experimental contexts''? The mathematical object in question is an observable in OP or an effect in EP.

We start with OP. In this case, it is easy to identify the basic objects, i.e., the observables, whereas a harder task is to identify contexts and compatible observables. The minimal requirement for a context is to be given by a joint measurement. In fact, a large fraction of theoretical works on contextuality simply discuss joint measurements, without entering into the details of their experimental realization; see, e.g., \citet{Abramsky:2011NJP,CSWPRL2014,AcinCMP2015}. If one goes the lab and perform a test of contextuality, however, different options arise. A possible way to perform joint measurements is simply to perform the measurements in a sequence. This is the perspective adopted in several contextuality experiments \cite{Lapkiewicz:2011NAT,Kirchmair:2009NAT,Amselem:2009PRL}, and the one we present  in the following from a theoretical point of view. An analysis of such experiments and similar ones is presented in Sec.~\ref{ssec:experiments}. Other approaches that do not use sequential measurement are possible, including that of \citet{Zhan:17}, which we review in the following. 

\begin{figure}

\includegraphics[width=0.23\textwidth]{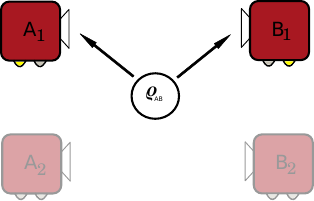} \hspace{.3cm}\includegraphics[width=.23\textwidth]{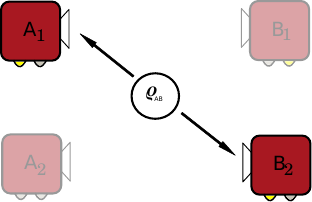}

\large{(a)}

\vspace{.5cm}
\includegraphics[width=0.35\textwidth]{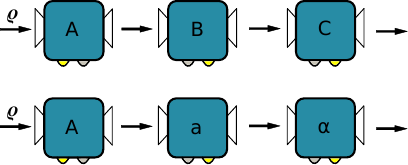}

\large{(b)}

\vspace{.5cm}

\includegraphics[width=0.35\textwidth]{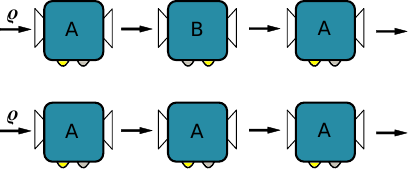}

\large{(c)}
\caption{\label{fig:scen}%
Different experimental procedures in Bell and more general noncontextuality scenarios.
(a) Two contexts in Bell scenario: The same measurement $A_1$ is performed in two different contexts, either with $B_1$ or with $B_2$. 
(b) Two contexts in the PM scenario: The measurement of $A$ is performed in two contexts, either with $B$ and $C$ or with $a$ and $\alpha$. As in the Bell scenario, the same measurement procedure, represented by the box with label $A$, is repeated in different contexts.
(c) Additional measurements to quantify experimental imperfections. The measurement of $A$ is repeated alone or together with $B$. Subsequent measurements of $A$ must confirm the same outcome, represented by the yellow light on the bottom of the device.
}
\end{figure}

In a sequential implementation of a joint measurement, it is clear how to identify the ``same observable'' in ``different contexts'', since an observable is given by a specific set of measurement procedures (laser pulses, beam splitters, etc.). One simply has to repeat the same procedures in different sequences. 
What, then, are the joint measurements in a sequential implementation? The simple act of performing one measurement after the other is not enough to consider it a joint measurement.  Intuitively, one would need some notion of nondisturbance, in the sense that the physical property revealed by a measurement $B$ is not altered if $A$ is measured first, and the same if the order is exchanged. This intuition is particularly clear when one considers a Bell scenario: the choice of measurement performed by one party (say, Alice) cannot have any influence on the measurement outcome of the other party (say, Bob) since these events are assumed to happen in spacelike separated regions. Special relativity guarantees that no ``influence'' or ``disturbance'' could propagate from one space-time region to the other. Bell nonlocality can be considered a particular form of contextuality where the assumption of context-independence is identified with the locality assumption; Alice's outcome is independent of what measurement Bob is performing. 
One may relax such a constraint by assuming that two measurements are performed on two systems few meters apart in the same lab and still do not disturb each other, or even measurements on different degrees of freedom on the same system, etc. This motivates us to further develop this idea to identify contexts as sets of measurements that are in some sense nondisturbing, and to deal with experimental imperfection from this perspective, as we discuss in  Sec.~\ref{ssec:exper_imp}. Similar ideas of classical hidden-variable models based on the notion of nondisturbance among measurements were developed by Leggett and Garg in the category of {\it macrorealist models} \cite{Leggett:1985PRL,Emary:Review2014}.

Before defining a notion of nondisturbance, we clarify which type of measurements are relevant. A wide range of measurement are possible in quantum mechanics, from sharp measurement to 
the trivial POVM $\{\openone/2,\openone/2\}$, always giving one of two outcomes with equal probability. We want the observables to represent the measurement of a property rather than a 
random coin flip. A possible condition is that of {\it self-repeatability}: namely, if we perform the same measurement twice, we obtain the same outcome. This is a prerequisite for speaking 
about nondisturbance for single runs of an experiment. In fact, if the outcome changes randomly when repeating a measurement, as in the case of the previously mentioned trivial POVM, it does not make sense 
to speak about the outcome not being disturbed. Notice that this notion of disturbance for POVMs, which was introduced by \citet{HW2010} should not be confused with the notion of disturbance at the level of the probability distributions [see \citet[][]{RamanathanPRL2012}], which does not presuppose the sequential realization of the joint measurement. To better understand this point, we consider the notion of nondisturbance for quantum sequential measurements  \cite{HW2010}:
 $A$ does not disturb $B$, if it is impossible to detect through a measurement of $B$ whether or not $A$ 
was measured (and its outcome discarded) before $B$. However, this property does not imply that $A$ has no effect on 
$B$. For instance, imagine that we perform the sequence of measurements $BAB$. If we obtain the outcome $+1$ for the first 
measurement of $B$ and $-1$ for the second, we may conclude that $A$ ``disturbed'' $B$. This notion makes sense only if $B$ is self-repeatable. A given set of observables $\{A,B,C,\ldots\}$  satisfy the {\it outcome repeatability} property if for any sequence of them, in any possible order and with each observable appearing multiple times, the outcome of their first appearance is always repeated in all subsequent measurements. Hence, in the sequential measurement implementation of joint measurements, a set of observables that satisfy outcome repeatability, and in addition the statistical nondisturbance conditions of \citet{HW2010} and their generalization to arbitrary sequences [see \citet[][]{Uola2018}] is what we call a {\it context} (in OP). 

The previous definition was inspired by the quantum mechanical notion of projective measurements. In fact, for projective measurements all  properties capturing some idea of simultaneous measurability, i.e., joint measurability, nondisturbance, outcome repeatability, etc., are equivalent to the commutativity of observables. This notion is the basis of the NCHV models presented in Sec.~\ref{sssec:NCHV}, such as Eq.~\eqref{eq:nchvdef}, since one requires the single outcomes, i.e., the corresponding deterministic response functions,  not to be disturbed.

In summary, in OP, combined with the sequential realization of joint measurements, observables are given by experimental measurement procedures, contexts are defined by sequential measurements of outcome-repeatable observables, and each single-measurement procedure is repeated in an identical way in each sequence of measurements. The different steps in the realization of a contextuality test can  be listed as follows
\begin{itemize}
\item[S.1] Define experimental measurement procedures and associate with each one a classical random variable with the same values as the possible outcomes.
\item[S.2] Identify contexts in terms of outcome-repeatable and statistical-nondisturbing measurements.
\item[S.3] Perform measurements in different sequences, according to the defined contexts. For each measurement the same procedure is repeated in different contexts.
\item[S.4] Compare the observed statistics for contexts (sequences) with the one predicted using the NCHV for the corresponding classical variables.
\end{itemize}
This is the perspective explicitly adopted by 
\citet{Kirchmair:2009NAT,Guhne:2010PRA} and \citet{Larsson:2011AQT} among others.
These steps are defined for ideal measurements. In actual experiments, the 
outcome repeatability property is never exactly satisfied due to unavoidable errors and imprecision in the 
experimental implementations. Nevertheless, having this theoretical framework in mind, one can devise practical 
methods to deal with experimental imperfections. In real experiments, then, we need an additional step
\begin{itemize}
\item[S.5] Perform additional experimental runs to quantify deviations from ideal (outcome-repeatable and nondisturbing) measurements and compare them to the classical models accordingly.
\end{itemize}
See Fig.~\ref{fig:scen} for some examples of the different measurement procedures. The problem of quantifying deviations from ideal measurements and the comparison to the classical models is discussed in Sec.~\ref{ssec:exper_imp}.

Before concluding this section, we comment on the possibility of simply 
using the notion of joint measurability, as in the original 
work by \citet{Kochen:1967JMM}, by extending it to nonideal measurements. 
\citet{Larsson:2011AQT} extensively argued in favor 
of using sequential measurements for contextuality tests. The main 
motivations can be summarized as follows. For joint measurement devices, 
a change of context corresponds to a physically entirely 
different setup even if one of the settings within the context 
remains unchanged. It is difficult to argue that the outcome 
of the unchanged setting is unchanged from physical principles, as 
already noted by \citet{Bell:1966RMP,Bell:1982FP}; see also the discussion
by \citet{Cabello:2009FPP}. In the previously outlined 
black-box scenario, one can always decide whether one performs
a measurement alone or in sequence with other measurements.  The existence of
``contextless'' devices, associated with the single-measurement setting and
then combined in the sequential measurement 
setup, is then the argument for noncontextual behavior. In a sequential 
measurement one uses the same observables for which one wants to verify 
the contextual behavior. This allows for a direct identification of the single observables 
in each context and a change of the context occurs by substituting only some observables in the sequence. 
This is in contrast to the joint measurement scenario where the entire device changes and other means of 
identifying observables have to be applied. Still, it is possible to identify observables in a joint measurement, without sequential measurements, in terms of their statistics, namely, if they provide the same distribution of outcomes for any state preparation. This approach presents different challenges, such as making sense of the expression ``for any state preparation''. Further details are provided later.

Finally, an argument against the use of joint measurability {\it alone}
to define contexts for nonideal measurements can be formulated by applying a construction by \citet{Fritz:2012RMP} to the contextuality scenario. Even though 
the original argument dealt with a different problem related to nonlocality, one can straightforwardly adapt it to contextuality.  
We present the main idea in the following, by considering the CHSH scenario as a contextuality scenario. We denote the measurements effects as $\{A_{a}^x\}$ and $\{B_{b}^y\}$, where $a$ and $b$ are the outputs and $x$ and $y$ are the measurement settings, and define contexts as set of jointly measurable observables. In the CHSH scenario, each context consists of a pair $\{A^x, B^y\}$. If we assume that contexts are defined by jointly measurable observables, this implies that there is a joint measurement $G_{ab}^{xy}$ for each pair of settings $x$ and $y$. The joint-measurability conditions amounts to
\begin{equation}
A_a^x = \sum_b G_{ab}^{xy},\quad B_b^y = \sum_a G_{ab}^{xy}, \quad \text{ for all } a,b,x,y.
\end{equation}
Notice that such operators automatically give rise to a nonsignaling distribution $p(ab|xy)$ \cite{Popescu:1994FPH}: namely, it satisfies $\sum_b p(ab|xy)=\sum_b p(ab|xy')$ for all $a,x,y, y'$, and $\sum_a p(ab|xy)=\sum_a p(ab|x'y)$ for all $b,x,x',y$. On the other hand, for any given nonsignaling distribution $\{p(ab|xy)\}_{abxy}$, we can construct such joint measurements simply as
\begin{equation}\label{eq:cont_JM}
G_{ab}^{xy}:=p(ab|xy) \openone, \quad A_a^x := \sum_b G_{ab}^{xy},\quad B_b^y := \sum_a G_{ab}^{xy}.
\end{equation}
Since all operators are a multiple of the identity, one can simply take a one-dimensional Hilbert space. At the level of correlations, the conditions of nonsignaling precisely amounts to the condition of joint-measurability for one-dimensional quantum systems.

This construction implies that by defining contexts simply in terms of jointly measurable observables all nonsignaling correlations~\cite{Popescu:1994FPH} can be obtained using one-dimensional quantum systems. The argument extends straightforwardly to any arbitrary contextuality scenario, where the counterparts of the nonsignaling correlations are the {\it nondisturbing}  \cite{RamanathanPRL2012} or the {\it nonsignaling-in-time} \cite{KoflerPRA2013} correlations. In other words, by defining contexts simply in terms of joint measurability, all maximally contextual correlations, defined as the extreme point of the {\it nondisturbing polytope}~\cite{RamanathanPRL2012}, can be reached using one-dimensional (and hence classical simulable) quantum systems.

Intuitively,  this argument shows that when one uses too broad of a notion of contexts, i.e., joint-measurability alone, contextuality becomes a trivial property. This is the necessary conclusion if one adopts the definition of NCHV given in Sec.~\ref{ssec:NC_ineq}, where Fine's theorem allows any distribution to be decomposed in terms of deterministic ones. For a different perspective on the problem of trivial POVMs that does not necessarily support the previous conclusions, see \citet{Henson:2015PRA} and  \citet{KunjwalQuantum2019}.

This does not mean that the possibility of defining contexts through joint measurability and applying this definition to experimental implementations has not been explored and 
that other approaches are not possible. For instance, the notion of joint measurability  was used as a 
definition of context for nonideal measurements by~\citet{Liang:2011PR}. Notice that \citet{Liang:2011PR} discussed both the framework of Kochen-Specker contextuality and Spekkens contextuality. They avoid the problem of trivial POVMs by introducing a  {\it sharpness parameter} and a consequent modification of 
noncontextual bounds for correlations. Observables proportional to the identity, as in the example of Eq.~\eqref{eq:cont_JM}, are then maximally unsharp, and the modified bound of the considered noncontextuality inequality then becomes the algebraic maximum~\cite{Liang:2011PR}. 
This approach was further 
investigated by \citet{KunjwalPRA2014,KunjwalPRA2015} and the corresponding inequality 
was experimentally tested by \textcite{Zhan:17}.

\subsubsection{Operational definitions of contexts: EP}\label{sssec:op_def_ep}
Different problems arise in the EP. Here, a context is easily identifiable operationally, as it 
consists simply of a single measurement. The difficult part is to identify the same effect in different 
measurements. A solution is an identification of effects based on the observed statistics. As an example, we can consider Spekkens's approach, first proposed in 2005 \cite{Spekkens:2005PRA} and further clarified in 2014 \cite{Spekkens:FOP2014}, which was based on the notion of statistical indistinguishability. In simple terms, one identifies the effect associated with the outcome $k$ of the measurement $M$ with the effect associated with the outcome $k'$ of 
the measurement $M'$ if  
\begin{equation}\label{eq:Sp_MNC}
p(k|P, M)=p(k'|P,M')\qquad \text{ for all preparations } P,
\end{equation}
where $p(k|P,M)$ denotes the probability of the outcome $k$ of the measurement $M$ given the preparation $P$. This idea of statistical identification of effects has appeared in the literature; see  \citet{Fuchs:2002}. Here, however, it is explicitly used to constrain possible hidden-variable models. If two effects have the same statistics, they must be represented by the same mathematical object in the hidden-variable model. 

In this framework, a hidden-variable model, or an ontological model according to the terminology of \citet{Spekkens:2005PRA} [see also the extensive discussion by \citet{Harrigan:2010}], describes the observed probabilities for a given preparation $P$ and measurement $M$ 
with an outcome $k$ as
\begin{equation}
p(k|P,M)=\sum_\lambda \mu_P(\lambda) \xi_{M,k}(\lambda),
\end{equation}
where $\mu_P(\lambda)$ represents the probability distribution of the space of the hidden variable $\lambda$ 
associated with the preparation $P$, satisfying $\mu_P(\lambda)\geq 0$ and $\sum_\lambda \mu_P(\lambda)=1$, and 
$\xi_{M,k}(\lambda)$ represents the response function associated with the outcome $k$ of the measurement $M$, satisfying 
$\xi_{M,k}(\lambda)\geq 0$ and $\sum_k \xi_{M,k}(\lambda)=1$ for all $\lambda$. Noncontextuality, then, amounts to the
assumption that $\xi_{M,k}=\xi_{M',k'}$ if the condition ~\eqref{eq:Sp_MNC} is satisfied. 
One can then compare these 
assumptions with the usual assumptions of Kochen-Specker theorem~\cite{Spekkens:2005PRA,LeiferPRL2013,KunjwalPRL2015}.

We saw in Sec.~\ref{sssec:NCHV} [Eq.~\eqref{eq:nchvdef}] that the nondeterministic response functions can 
always be transformed into deterministic ones by extending the space of the hidden variable. However, if one 
assumes that two effects with the same statistics are represented by the same 
object in the hidden-variable model, one can no longer transform the nondeterministic response functions 
$\xi_{M,k}(\lambda)$ into deterministic ones, if the measurement is not sharp. A proof of this fact was given by~\citet{Spekkens:FOP2014}. The intuition at the basis of the proof could be summarized as follows. For a given 
effect $E$ of an unsharp measurement, take its spectral decomposition $\sum_i \lambda_i \Pi_i$, which has all 
eigenvalues in $[0,1]$. The projectors $\{ \Pi_i\}_i$ then constitute a projective measurement that gives $E$ 
via some postprocessing. Since $E$ is statistically indistinguishable from the  previously constructed postprocessing 
of a projective measurement, the 
corresponding response functions in the ontological model must be identical. Hence, the response function 
associated with $E$ must also arise from the same postprocessing; i.e., they are not deterministic.

This observation together with the practical impossibility of obtaining perfect projective measurements in 
actual experimental implementations led to the development of a different notion of noncontextuality 
inequalities~\cite{Spekkens:2009PRL, Mazurek2016, KunjwalPRL2015,Pusey2015,XuPRA2016,KunjwalPRA2018, SchmidPRA2018,KrishnaNJP2017} based on the identification of measurement effects according to  Eq.~\eqref{eq:Sp_MNC} and an analogous notion for 
preparations called  preparation noncontextuality \cite{Spekkens:2005PRA}.  We  review the 
approach based on the latter perspective in Sec.~\ref{ssec:other_cont}.


\subsection{Modeling experimental imperfections}\label{ssec:exper_imp}
Starting from the operational definition of contexts and compatibility provided Sec.~\ref{sssec:op_def_op} for ideal 
projective measurements, we  further develop these ideas in order to deal with imperfect and noisy measurements 
typical of any  experimental implementation of a contextuality test. We emphasize that there 
is no general recipe that can be applied to all experiments. On the contrary, for every experimental realization it 
is necessary to make some assumptions on the hidden-variable model describing the type of noise present. Typically, it is necessary to perform additional 
measurements to quantify the amount of noise and check its consistency with the previously mentioned assumptions, and 
possibly to  modify the noncontextuality inequality accordingly. Several approaches have been proposed and implemented in experimental tests of contextuality. In the 
following, we discuss the theoretical work presented by 
\citet{Kirchmair:2009NAT,Guhne:2010PRA}, by ~\citet{Szangolies:2013PRA,Szangolies:2015book}, and by 
\citet{Larsson:2002EPL,Simon:PRL2001,Kujala:2015PRL,WinterJPA14}.

\subsubsection{Quantifying disturbance in sequential measurements}\label{sssec:dist_err_mod}

To deal with actual experiments, we need to relax the assumption of perfectly compatible measurements by admitting measurements that produce a (certain type of) disturbance on subsequent measurements, and then trying to quantify such a disturbance. This was the approach proposed by \citet{Kirchmair:2009NAT} and further developed by \citet{Guhne:2010PRA}; we  mostly follow the latter. Such an approach can be summarized as follows. Probabilities associated to perfectly compatible measurements are still described by NCHV models; however, we admit the possibility of incompatible measurements that introduce disturbance and change the outcomes of subsequent measurements in a sequence, giving rise to context-dependent outcomes. The amount of disturbance is then estimated experimentally, under the physical assumption that the amount of noise introduced by the measurements is cumulative; i.e., it does not cancel out by performing more measurements.

To keep the notation simple, we  consider only the case of $\pm 1$-valued observables, a generalization to arbitrary (finite) outcomes is straightforward.  Consider a hidden-variable model describing the probabilities for all possible sequences  $\mathcal{S}_{AB}=\{A,B,AA,AB,BA,BB, AAA,\ldots\}$ of two $\pm 1$-valued observables $A$ and $B$. We  denote the outcome probabilities for single measurements as $p[\pm|A], p[\pm|B]$ and, similarly, for sequences $p[\pm \pm |AA],  p[\pm \pm |AB], \ldots$ for, respectively, measurement of the sequence $AA$ and $AB$, where the temporal ordering of the sequence is from left to right. We admit the possibility of a discarded outcome such as $p[+ \bigcdot -| B A B]$, which  denotes the probability for the outcomes $+$ for the first measurement of $B$, a discarded outcome ($\bigcdot$) for the measurement of $A$ and $-$ for the final measurement of $B$. 

As anticipated, our model  departs from the NCHV model discussed in Sec.~\ref{sssec:NCHV}. If $A$ and $B$ are compatible observables, it must necessarily be that $p[+ \bigcdot -| B A B]=0$, as it follows immediately from Eq.~\eqref{eq:nchvdef}. Consider now the case where $A$ and $B$ are not compatible. Terms that are not experimentally accessible are still well defined in this model, such as  $p[(+|A) \text{\&}  (+|B)]$, namely, the probability that the first measurement gives the outcome $+1$, in the case both of a measurement of $A$ and of a measurement of $B$. However,  the outcome probability $p[+ +|  A B]$ for their sequential measurement does not correspond to the previously mentioned one, and it is, in fact, not included in the description given by the NCHV model. In this sense, the present discussion extends the NCHV framework to a new hidden-variable model that includes the description of sequences of incompatible measurements,  and in which certain outcomes are allowed to depend on the particular sequence of 
measurements, if such a sequence involves incompatible measurements. This is a central point common to all analyses of experimental imperfections.

For measurements that are assumed to be compatible, say $A$ and $B$, the hidden-variable model satisfies the usual noncontextuality conditions:
\begin{equation}\label{eq:nc_seq}
\begin{split}
v(A_1| S_1,\lambda)=v(A_2|S_2,\lambda), \text{ for all } \lambda,  \\
\text{ for sequences } S_1=\{A\},\ S_2=\{ BA \},
\end{split}
\end{equation}
where $v(A_i| S,\lambda)$ denotes the value assigned by the hidden variable model to the observable $A$ in position $i$ in the sequence $S$ for a given $\lambda$.
Notice that Eq.~\eqref{eq:nc_seq} is a condition on the hidden-variable model that is not directly experimentally testable, hence, it cannot be taken as an operational definition of compatibility. How can we quantify the disturbance introduced by a measurement in this model? The first observation is that for this model the following inequality holds
\begin{equation}\label{eq:ineflip}
p[(+|A) \text{\&} (+|B)] \leq p[++|AB] + p[(+|A) \text{\&} (\bigcdot - |AB)].
\end{equation}
Intuitively, given a specific value $\lambda$ of the hidden variable such that it contributes to the lhs, either
 the value of $B$ either stays the same, i.e., $\lambda$ contributes to $p[++|AB]$, or the value of $B$ is flipped by the measure of $A$, i.e., $\lambda$ contributes to $p[(+|A) \text{ and } (\bigcdot - |AB)]$.

The correlator $\mean{AB} = \sum_{ab=\pm 1} ab\ p[(a|A) \text{\&} (b|B)]$, representing the correlation between $A$ and $B$ assigned by the hidden-variable model, can be written as
\begin{equation}
\mean{AB} = 1 - 2p[(+|A) \text{\&} (-|B)] -2p[(-|A) \text{\&} (+|B)].
\end{equation}
In contrast, the correlation between $A$ and $B$ observed in an experiment where we measure the sequence $S=\{AB\}$ denoted by  $\mean{A_1B_2}$, is given by $\mean{A_1B_2}= \sum_{ab=\pm 1} ab\ p[ab|AB]$. In general, $\mean{A_1B_2}\neq \mean{AB}$ if $A$ and $B$ are incompatible.

By definining $p^{\rm flip}[AB]$ as the probability that the outcome of $B$ is flipped by the measurement of $A$, namely, 
\begin{equation}
p^{\rm flip}[AB] :=  p[(+|B)\text{\&}(\bigcdot - |AB)] + p[(-|B) \text{\&} (\bigcdot + |AB)],
\end{equation}
and using Eq.~\eqref{eq:ineflip}, we can bound $\mean{AB}$ with the experimentally observable correlator $\mean{A_1B_2}$ as follows:
\begin{equation}\label{eq:flipbound}
\mean{A_1B_2} - 2p^{\rm flip}[AB] \leq \mean{AB} \leq \mean{A_1B_2} + 2p^{\rm flip}[AB].
\end{equation}
From Eq.~\eqref{eq:nc_seq}, $p^{\rm flip}[AB]=0$ for compatible measurements.

Applying this reasoning to the CHSH inequality \cite{Clauser:1969PRL},
\begin{equation}
\mean{AB} + \mean{BC}+\mean{CD} - \mean{AD} \leq 2.
\end{equation}
We have the corresponding expression for the case of sequential measurements~\cite{Guhne:2010PRA}
\begin{equation}\label{eq:chsh_flipbound}
\begin{split}
\mean{\chi_{\rm CHSH}^{\rm seq}}:= \mean{A_1B_2}+\mean{C_1B_2}+\mean{C_1D_2}-\mean{A_1D_2} \\
\leq 2 \left(1 + p^{\rm flip}[AB] + p^{\rm flip}[CB] + p^{\rm flip}[CD] + p^{\rm flip}[AD]\right).
\end{split}
\end{equation} 

What is left is to bound the unobservable quantity $p^{\rm flip}$. We introduce
\begin{equation}\label{eq:p_err_def}
p^{\rm err}[BAB]:= p[+ \bigcdot - |BAB] + p[-\bigcdot +|BAB],
\end{equation}
corresponding to the probability of flipping the value of $B$ in a sequence of three measurements with an intermediate measurement of $A$. In contrast to $p^{\rm flip}[AB]$,  $p^{\rm err}[BAB]$ is experimentally measurable.

At this point, one needs an assumption on the hidden-variable (HV) model describing the experimental noise, namely
\begin{itemize}
\item[({\bf CN})] Cumulative noise assumption:
\begin{eqnarray}
p[(\pm |B) \text{ and } (\bigcdot \mp |AB)] 
&\leq &p[(\pm |B) \text{ and } (\pm \bigcdot \mp |BAB)]
\nonumber
\\
& =& p[\pm \bigcdot \mp |BAB].
\label{eq:err_flip_bound}
\end{eqnarray}
\end{itemize}
Notice that Eq.~\eqref{eq:err_flip_bound} is not directly testable as it contains some inaccessible correlations that are defined only at the level of the HV model. Nevertheless, one can have a physical intuition of what this means. In fact, this assumption corresponds to the idea of a \textit{cumulative noise}, i.e., the noise always increases with additional measurements. It is more likely to flip the outcome of $B$ if we perform a measurement of both $B$ and $A$ than it is if we perform only a measurement of $A$. This seems to be a reasonable assumption if we want to model experimental imperfections, where the measurements are not supposed to ``collude'' to cancel out the noise when arranged in specific sequences. Similar ideas were explored by Wilde and Mizel in their discussion of the so-called {\it clumsiness loophole} for Leggett-Garg inequalities \cite{Wilde2012}.

Equation \eqref{eq:err_flip_bound} then directly implies $p^{\rm flip}[AB]\leq p^{\rm err}[BAB]$, and allows us to rewrite Eq.~\eqref{eq:chsh_flipbound} as
\begin{equation}\label{eq:chsh_err_bound}
\begin{split}
\mean{\chi_{\rm CHSH}^{\rm seq}}- 2\left(  p^{\rm err}[BAB] + p^{\rm err}[BCB] \right.\\
+ \left.    p^{\rm err}[DCD]  + p^{\rm err}[DAD]\right)  \leq 2 ,
\end{split}
\end{equation}
which involves only experimentally testable quantities.  \citet{Kirchmair:2009NAT} reported an experimental violation of Eq.~\eqref{eq:chsh_err_bound} for a sequential measurement of the CHSH expression with the value 
\begin{equation}
\begin{split}
\mean{\chi_{\rm CHSH}^{\rm seq}} -2(p^{\rm err}[BAB] + p^{\rm err}[BCB] + p^{\rm err}[DCD]\\
 + p^{\rm err}[DAD])=2.23(5) \ .
\end{split}
\end{equation}
A similar analysis was performed in the experiment by \citet{JergerNC2016}.

\subsubsection{Context-independent time evolution}\label{sssec:time_evol}

An implicit assumption hidden in the definition of the previous model is that the hidden variable $\lambda$ is static, i.e., not evolving during the time passing between one measurement and the subsequent one. 
\citet{Szangolies:2013PRA} and \citet{Szangolies:2015book} proposed a relaxation of such conditions admitting a hidden variable changing in time, but with an evolution that is still context-independent, in the sense that it does not depend on the specific measurements performed.  This investigation led to modified noncontextuality inequalities satisfied by such an extended set of noncontextual correlations, thus, able to demonstrate quantum contextuality in a broader framework.

The notion of {\it noncontextual evolution}  has been formalized as follows \cite{Szangolies:2015book}:
The system evolves according to a sequence of hidden variable states $\lambda_i \rightarrow \lambda_j \rightarrow \lambda_k \rightarrow \cdots $ that is independent of the measurements performed. 
To understand this notion, it is helpful to consider a simple example of a noncontextuality inequality that can be maximally violated by such models.  Consider the CHSH inequality, but now evaluated according to the following sequential measurement scheme:
\begin{equation}\label{eq:chshjoc}
\mean{A_1B_2} + \mean{B_1C_2} + \mean{C_1D_2} - \mean{D_1A_2} \leq 2,
\end{equation}
where, as in Sec.~\ref{sssec:dist_err_mod}, we denote via the symbol $\mean{A_1 B_2}$ the fact that we  perform the measurement $A$ first, then the measurement $B$, and take the expectation value of the product of their outcome. \citet{Szangolies:2013PRA} constructed the following noncontextual model with evolution. The hidden variable $\lambda$ evolves from an initial state $\lambda_1$ to $\lambda_2$, regardless of the measurement performed at the initial time. The measurements are chosen such that $A,B,C$ and $D$  always give the outcome $+1$ on the state $\lambda_1$ and likewise for $\lambda_2$, with the only exception being $A$, which gives the value $-1$ on $\lambda_2$. It is then straightforward 
to verify that Eq.~\eqref{eq:chshjoc} can be violated up to the algebraic maximum $4$. 
On the other hand, forcing the measurements to always appear in the same order, namely 
\begin{equation}\label{eq:chshjocord}
\mean{A_1B_2} + \mean{B_1C_2} + \mean{C_1D_2} - \mean{A_1D_2} \leq 2,
\end{equation}
restores the classical bound $2$. Intuitively, since observables are forced to appear always in the same position in the sequence, they are  always drawn from the same distribution for $\lambda=\lambda_1$ or $\lambda_2$, hence, they are not affected by the evolution of $\lambda$. This behavior is analogous to what happens in Leggett-Garg tests \cite{Leggett:1985PRL,Emary:Review2014}, where the hidden variable $\lambda$ is allowed to evolve freely in time.

Analogous reasoning can be applied to different noncontextuality inequalities~\cite{Szangolies:2013PRA}, such as the PM inequality seen in Eq.~\eqref{s2:PMineq}
\begin{equation}
\begin{split}
\mean{A_1B_2C_3} + \mean{c_1a_2b_3} + \mean{\beta_1 \gamma_2 \alpha_3} + \mean{A_1 a_2 \alpha_3} \\
+ \mean{\beta_1 B_2b_3} - \mean{c_1\gamma_2 C_3} \leq 4,
\end{split}
\end{equation}
where the observables are always forced to appear  in the same position in each sequence of measurements: $A$ always appears first, $a$ always appears second, $\alpha$ always appears third, etc. The same inequality has also been investigated from the perspective of dimension witnesses based on quantum contextuality~\cite{Guhne:2013PRA} and for a comparision
between the spatial and temporal scenario~\cite{XuPRA2017}.

Moreover, similar ideas can even be applied  to noncontextuality inequalities where observables cannot be forced to  always be in the same position, such as  KCBS inequality \cite[cf. Eq.~\eqref{eq:kcbs}]{Szangolies:2013PRA}
\begin{equation}
\begin{split}
\mean{A_1 B_2} + \mean{B_1 C_2} + \mean{C_1 D_2} + \mean{D_1 E_2} \\
+ \mean{E_1 A_2} - \mean{A_1 A_2} \geq -4,
\end{split}
\end{equation}
where the extra term $\mean{A_1 A_2}$ is designed to give a ``penalty'' whenever the value $A$ changes with a change in $\lambda$. 
In summary, this approach provides a simple method of dealing with context-independent time evolution of the hidden variable, which can be easily combined with the one presented in Sec.~\ref{sssec:dist_err_mod}

\subsubsection{First proposals of experimentally testable inequalities}\label{ssec:KS_ineq}

The first attempts to derive experimentally testable noncontextuality inequalities were made in the early 2000's by \citet{Larsson:2002EPL} and \citet{Simon:PRL2001}. They relaxed Kochen-Specker assumptions by requiring that the noncontextuality assumption and the completeness condition ($\mathbf{C'}$) of Sec.~\ref{sec:KS-sets} are only approximately satisfied. More precisely, the model proposed by \citet{Simon:PRL2001}  considered a relaxation of ($\mathbf{C'}$), whereas 
\citet{Larsson:2002EPL} considered a relaxation of both ($\mathbf{C'}$) and the noncontexual value assignment. In the following, we  discuss the approach of \citet{Larsson:2002EPL}.

For any pair of intersecting contexts appearing in a KS set, i.e., triples of vectors $(a,b,c)$ and $(a,b',c')$, we make the assumption that the value assignment is approximately context-independent, i.e.,
\begin{equation}
    p\left[v_{(a,b,c)}(a)\neq v_{(a,b',c')} (a)\right]\le \varepsilon,
\end{equation}
where $v_{(a,b,c)}(a)$ denotes the value assigned to $a$ in the context $(a,b,c)$.

Models that obey the previous assumption were called $\varepsilon$-{\it ontologically faithful noncontextual} ($\varepsilon$-ONC) models by~\citet{WinterJPA14}.
A formal statement of the second assumption is that the value assignments on any given context approximately satisfy the  condition ($\mathbf{C'}$), i.e.,
\begin{equation}
    p\bigg(\sum_{i=a,b,c} v_{(a,b,c)}(i)\neq 1\bigg)\le \varepsilon'.
\end{equation}
We now now write $M$ for the number of interconnections between contexts [such as $M=1$ for 
the single shared vector between $(a,b,c)$ and $(a,b',c')$], and $N$ for the number of contexts.
A Kochen-Specker proof consists of a set of vectors for which it is impossible to assign a noncontextual 
value satisfying all logical relations. In the previous language, this implies the impossibility of an 
assignment with $\varepsilon=\varepsilon'=0$. This is also expected to hold for small 
disturbances. This intuition can be made quantitative, with the following inequality derived  by \citet{Larsson:2002EPL} from the previous assumptions:
\begin{equation}
  M\varepsilon+N\varepsilon'\ge 1,
  \label{eq:KSinequality}
\end{equation}
implying that if the errors in the logical relations ($\varepsilon'$) are small, noncontextuality must often fail  ($\varepsilon$ must be large). 
The presence of the logical relations associated with the KS-theorem makes it a KS-inequality such as those discussed in Sec.~\ref{sssec:NCHVE}.

\begin{table*}[t]
  \centering
  \setlength\tabcolsep{1ex}
  \begin{tabular}{l r c r r c c}
    & $d$ & $n$ & $N$ & $M$ & 
    $\varepsilon'$ ($\varepsilon=0$) 
    & $\varepsilon'$ ($\varepsilon=0.01$) 
    \\
    \hline
    \citet{Peres:1993}                        & 3 & 57 (33) & 40 & 96 & 0.0250 & 0.0010\\
    Kochen and Conway \cite{Peres:1993}             & 3 & 51 (31) & 37 & 91 & 0.0270 & 0.0024\\
    Sch{\"u}tte \cite{Bub:1997}                    & 3 & 49 (33) & 36 & 87 & 0.0278 & 0.0036\\
    \citet{Kernaghan:1995PLA}    & 8 & 36      & 11 & 72 & 0.0909 & 0.0255\\
    \citet{Kernaghan:1994JPA}             & 4 & 20      & 11 & 30 & 0.0909 & 0.0636\\
    \citet{Cabello:1996PLA}    & 4 & 18      &  9 & 18 & 0.1111 & 0.0911\\
    \citet{LisonekPRA2014} & 6 & 21      &  7 & 21 & 0.1429 & 0.1129\\
  \end{tabular}
  \caption{Various Kochen-Specker proofs: the dimension $d$, the
    number of projectors $n$ (the number inside parantheses is the number used in the contradiction, while number outside them counts all vectors when completing the bases), 
    the number of contexts $N$, and the
    number of context changes $M$. The final two columns are lower bounds for the probability of failure of the logic relation $\varepsilon'$ given a noncontextual model ($\varepsilon=0$), and when there is a small probability of failure of noncontextuality from experimental causes (such as $\varepsilon=0.01$). }
  \label{tab:41-1}
\end{table*}

Using experimental estimates of the probability of failure of the logical relations, one can draw conclusions on how far from  noncontextual the data are. 
For the Kochen-Specker proofs available at the time, the numbers $M$ (interconnections) and $N$ (contexts) were high, so   even small values of $\varepsilon$ and $\varepsilon'$ would give a value on the left-hand side of Eq.~\eqref{eq:KSinequality} that is larger than 1; see the numbers listed in Table \ref{tab:41-1}. An experiment to violate one of these inequalities would be   challenging, given that errors in the directions would translate to an increased probability of failure of noncontextuality, and would therefore lower the bound on $\varepsilon'$. However, there is no direct connection between directional accuracy and $\varepsilon$, the probability of failure of noncontextuality, and no immediate way to estimate this failure probability from experimentally measurable quantities. 



\subsubsection{Approximate quantum models}\label{ssec:appr_QM}

An attempt to use the quantum description of the system and measurements for the estimate of the failure probability of Sec.~\ref{ssec:KS_ineq} was presented by \citet{WinterJPA14}. Winter considered quantum effects $Q_i$ that are $\varepsilon$ close in the operator norm to the ideal projectors $P_i$ associated with each direction $i$, such that
\begin{equation}\label{eq:eps_precise_qm}
||Q_i-P_i||\le\varepsilon.
\end{equation}
\citet{WinterJPA14} called this $\varepsilon$-precise quantum ($\varepsilon$-PQ) model. Note that in principle $\varepsilon$ can be estimated through a tomographic characterization of the measurement. 
It is now tempting to equate the $\varepsilon$ from the $\varepsilon$-ONC model (as previously defined) with the $\varepsilon$ from the $\varepsilon$-precise quantum model, the reason being that they each constitute a distance measure in their respective realms. In the following, we review how this identification can play a role in contextuality tests according to \citet{WinterJPA14}. 

We consider an $\varepsilon$-ONC model consisting of a set of $\{0,1\}$-valued classical random variables $X_i^C$,  each associated with a rank-1 projector $P_i$ and a context $C$, such that 
\begin{equation}
\begin{split}
&\sum_{i\in C} X_i \leq 1, \text{ for all contexts } C, \\
&{\rm Prob}(X_i^C \neq X_i^{C'}) \leq \varepsilon, \text{ for all }  i, C, C',\\ &\text{ such that } i\in C\cap C'.
\end{split}
\end{equation}
In other words, the model is explicitly contextual; i.e., for each $i$ and each context $C$, we have a different random variable $X_i^C$, but $X_i^C$ and $X_i^{C'}$ take different values at most with probability $\varepsilon$. The case $\varepsilon=0$ clearly coincides with the usual NCHV model.

From the previously mentioned explicitly contextual model one can define context-independent variables $Y_i$ as
\begin{equation}\label{eq:Y_i}
Y_i := \prod_C X_i^C.
\end{equation}
From Eq.~\eqref{eq:Y_i} it is clear that the probability that $Y_i$ is different from $X_i^C$ for some $C$ is equal to the probability that  $X_i^C$ are not all equal for different $C$; hence, it is smaller than $(k_i-1)\varepsilon$, where $k_i$ is the number of contexts in which $i$ appears. We thus have
\begin{equation}
\mean{X_i^C}\leq \mean{Y_i} + (k_i-1)\varepsilon,
\end{equation}
which corresponds to the modified bound for $\varepsilon$-ONC models
\begin{equation}
\sum_i \lambda_i \mean{P_i} \leq \beta_{{\rm \varepsilon-ONC}} := \beta_{0{\rm -ONC}} + \varepsilon \sum_i \lambda_i (k_i -1),
\end{equation} 
where $\beta_{{\rm 0-ONC}}$ denotes the usual NCHV bound and $\beta_{{\rm \varepsilon-ONC}}$ represents the modified bound for a $\varepsilon$-ONC model.

For a given violation $\beta_q$ of the NCHV bound $\beta_{{\rm 0-ONC}}$ we can have a contradiction up to the precision  $\varepsilon_0 = (\beta_q - \beta_{{\rm 0-ONC}})/[\sum_i \lambda_i (k_i-1)]$, i.e., the violation cannot be explained using a model with imprecision $\varepsilon \leq \varepsilon_0$. 
\citet{WinterJPA14} also provided some estimates of the $\varepsilon_0$ associated with maximal quantum violation of different noncontextuality inequalities, such as the KCBS inequality with $\varepsilon_0^{\rm KCBS} < 0.047$ and the PM-square inequality with $\varepsilon_0^{\rm PM} < 0.0138$. 

By equating the $\varepsilon$ of the $\varepsilon$-ONC model with the $\varepsilon$ of the $\varepsilon$-PQ model, 
e.g., with the latter extracted via a tomography of the quantum effects, one could use the previous argument to 
disprove noncontextuality for imprecise measurements. However, there is no direct connection between the two as 
both definitions (of an approximate quantum model and of an ontologically faithful noncontextual classical 
model) are independent. In fact, the measures arise in different domains and measure distances between 
different conceptual object types: in one case the operator-norm distance between an ideal measurement ($P_i$) 
and the realized effect ($Q_i$), and in the other case the statistical distance between an outcome assignment in 
one context [$v_{(a,b,c)}(a)$ in the previous example] and the corresponding outcome assignment in another 
context [$v_{(a,b',c')}(a)$].

\subsubsection{Maximally noncontextual models}\label{ssec:max_NC}

Another approach to the quantification of measurement disturbance in contextuality tests was proposed by \citet{Kujala:2015PRL}, via the notion of \textit{maximal noncontextuality}.
They  relaxed the definition of a NCHV model by allowing for some disturbance in the measurements, not necessarily seen as sequential measurements, leading to an apparent context-dependence, namely, the observation of different marginal distributions for the measurement of the same observable in different contexts. 
In this way, they obtained an explicitly contextual classical model, in which observables in different contexts were represented by different classical random variables. 

It is instructive to show a simple example. Consider the KCBS scenario discussed in Sec.~\ref{sssec:state-dep}: five $\{+1,-1\}$-valued observables $A_0,\ldots, A_{4}$ with compatible pairs $A_i,A_{i+1}$, with the sum modulo $5$. Consider the following version of the KCBS inequality
\begin{equation}\label{eq:KCBS_err_mod}
\sum_{i=0}^3 \mean{A_i A_{i+1}}-\mean{A_0 A_4}\leq 3,
\end{equation} 
where addition in the indices is interpreted modulo $5$. The NCHV model was extended by  \citet{Kujala:2015PRL} by taking a copy of each classical variable for each context. In this case, there are five contexts $\{A_i, A_{i+1}\}$ for $i=0,\ldots,4$ and each $A_i$ appears twice, i.e.,  in the $i$-th and the $(i+1)$-th contexts. They constructed an explicitly contextual model by taking context-dependent copies, i.e., $A_i^{(i)}$ and $A_i^{(i+1)}$, where the superscript indicates the context. This construction of a contextual model is similar to that discussed in Sec.~\ref{sssec:dist_err_mod}, where the value assignment of an observable explicitly depends on the sequence it appears in, see Eq.~\eqref{eq:nc_seq}. However, to facilitate an easier comparison with the original paper, we follow the notation by \citet{Kujala:2015PRL}.

As in the previous cases, in order to interpret experimental results one needs to make an assumption regarding the type of disturbance present. The authors choose to introduce the notion of {\it maximally noncontextual} model. A rigorous definition was provided by \citet{Kujala:2015PRL}. Here, we may reformulate it in simple terms as follows: variables representing observables in different contexts are equal to each other with the maximum probability allowed by the observed marginals.
Intuitively, this notion states that there is no disturbance other than that observed in the marginals. In a manner similar to that discussed in Sec.~\ref{sssec:dist_err_mod}, it is reasonable to apply this model if the noise is supposed to arise from some clumsiness of the measurements; i.e., the measurement apparatuses are not colluding to cancel out the noise when combined in a certain way. 

\citet{Kujala:2015PRL} derived a class of inequalities valid for maximally noncontextual models, for what they called {\it cyclic systems}, also referred to as the $n$-cycle scenario \cite{Araujo:2013PRA}, namely, a collection of $\{+1,-1\}$-valued observables $A_0,\ldots, A_{n-1}$ with compatible pairs $A_i,A_{i+1}$, with the sum modulo $n$. This scenario includes the Leggett-Garg inequality \cite{Leggett:1985PRL}, the CHSH inequality~\cite{Clauser:1969PRL}, and the KCBS inequality~\cite{Klyachko:2008PRL}, corresponding to, respectively, $n=3,4,5$. The KCBS inequality
in Eq.~\eqref{eq:KCBS_err_mod} becomes
\begin{equation}\label{eq:KCBS_max_NC}
\sum_{i=0}^3 \mean{A_i^{(i)} A_{i+1}^{(i)}}-\mean{A_0^{(4)} A_4^{(4)}} - 
\sum_{i=0}^4 | \mean{A_i^{(i)}} - \mean{A_i^{(i-1)}}|\leq 3.
\end{equation}
Such inequalities can be derived using methods similar to those associated with standard noncontextuality inequalities (see Sec.~\ref{polyt}): under the assumption of a joint probability distribution over all variables $\{A_i^{(c)}\}_{i,c}$, one computes the projection of the corresponding probability simplex onto the space of observable marginals [in this case, the correlators $\{\mean{A_i^{(i)} A_{i+1}^{(i)}}\}_i$ and expectation values (marginals) 
$\{\mean{A_i^{(c)}}\}_{i,c=i,i+1}$]. A rigorous derivation of Eq.~\eqref{eq:KCBS_max_NC} can be found in  \citet{Kujala:2015PRL}.

In contrast to the proposal in Sec.~\ref{sssec:dist_err_mod}, the present method does not require one to perform additional measurements, as the experimental data obtained for the usual test of the KCBS inequality already contains all the information necessary to evaluate the lhs~of Eq.~\eqref{eq:KCBS_max_NC}.  In fact,  \citet{Kujala:2015PRL} compared this expression with the experimental results for the test of the KCBS inequality given by \citet{Lapkiewicz:2011NAT}  by computing a $99.99999999\%$ confidence interval for the lhs of Eq.~\eqref{eq:KCBS_max_NC} and obtaining the interval $[3.127,4.062]$, thereby confirming a violation of the inequality.
This result can then be interpreted as a disproof of maximally noncontextual models. \citet{Amaral:2018JMP,AmaralPRA2019} investigated general methods to derive inequalities such as Eq.~\eqref{eq:KCBS_max_NC} for arbitrary scenarios.


\subsection{Experimental realizations}\label{ssec:experiments}
In this section, we discuss some  experimental tests of quantum contextuality. 
We cannot give a detailed description 
of the experimental techniques. Instead, we explain some typical
experiments and their underlying assumptions.

\subsubsection{Early experiments}

We first mention some of the early experiments aiming at a test of 
quantum contextuality \cite{Michler:2000PRL, Huang:2003PRL, Hasegawa:2006PRL, 
Bartosik:2009PRL}. These early experiments were characterized by the 
fact that they did not measure one of the contextuality inequalities from
Sec. \ref{ssec:NC_ineq}. Instead, some predictions of quantum mechanics
are assumed to be correct in order to interpret the observations as a 
refutation of noncontextuality.

As an example, we discuss the experiment by \citet{Bartosik:2009PRL}
in some detail. We consider the following six observables on a two-qubit system, 
\begin{eqnarray}
A &=& \sigma_x \otimes \openone,
\quad
B = \openone \otimes \sigma_x,
\quad
a =  \sigma_y \otimes \openone,
\nonumber \\
b &=& \openone \otimes \sigma_y,
\quad
G= \sigma_x \otimes \sigma_y,
\quad
g= \sigma_y \otimes \sigma_x
\end{eqnarray}
Thus, for any noncontextual model the inequality
\begin{equation}
\mean{\Theta} = - \mean{AB}- \mean{ab} + \mean{G A b} +  \mean{g a B} - \mean{G g} \leq 3 
\end{equation}
holds. This can be directly checked by considering all the $\pm 1$ assignments
to the measurements.  For a two-qubit Bell state in an appropriate basis, one
can reach the value $\mean{\Theta}=5$, as the Bell state can be a common 
eigenstate of $G$ and $g$ with eigenvalue $-1$.

As our observables are defined locally, one can assume the relation 
$\mean{G A b}=\mean{g a B}=1$. Note, however, that this assumption 
is justified only for the given definition of observables, it does not 
necessarily hold if the measurements are considered to be black boxes. 
But in that case the inequality simplifies to
\begin{equation}
\mean{\theta} = - \mean{AB}- \mean{ab} - \mean{G g} \leq 1.
\label{eq-simplifiedksinequality}
\end{equation}
For testing this inequality, \citet{Bartosik:2009PRL}
used neutron interferometry. Here the two qubits are represented by the spin
and the path of a neutron in an interferometer. These $2$ degrees of freedom
are independently accessible, and the terms $\mean{AB}$ and $\mean{ab}$ can 
 be measured directly. 

Under the assumption of the validity of quantum mechanics, the term $\mean{G g}$
can be measured by performing a Bell measurement (i.e., a measurement in the basis 
of all the four Bell states) on the two qubits. This also allows one  to reconstruct 
the values of $G$ and $g$ separately. Typically, such a Bell measurement is nonlocal
and therefore difficult, but since the two qubits are encoded on a single neutron, 
this is feasible here. Finally, a value of $\mean{\theta}= 2.291 \pm 0.008$ was found, 
resulting in a clear violation of Eq.~(\ref{eq-simplifiedksinequality}).

\subsubsection{A test of the Peres-Mermin inequality with trapped ions}

One of the first experiments testing contextuality inequalities 
was performed with ion traps \cite{Kirchmair:2009NAT} and it can
be considered the prototypical example from which the general description
in Sec.~\ref{ssec:compatible} has been developed, as well as the basis
for many other subsequent experiments.
This experiment  aimed at an implementation of the inequality coming 
from the PM square, see Eq.~\eqref{s2:PMineq}.

\begin{figure}[t]
\begin{center}
\includegraphics[width=0.5\textwidth]{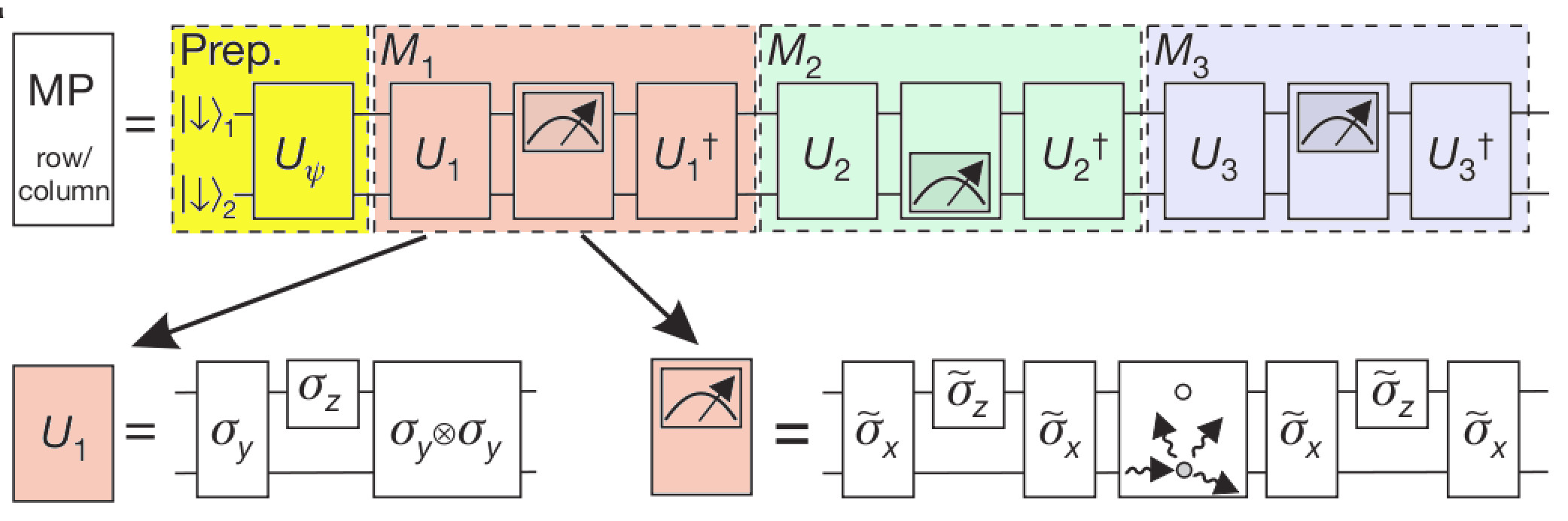}
\hspace{0.5cm}
\includegraphics[width=0.35\textwidth]{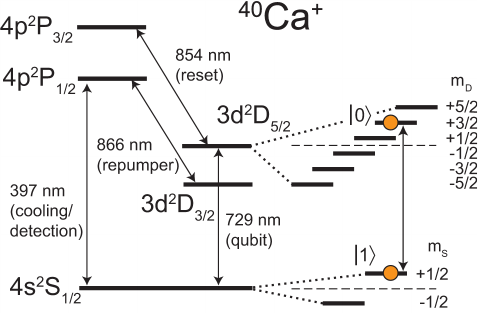}
\end{center}
\caption{Upper panel: level scheme of a single $^{40}$Ca$^+$ ion, 
highlighting the electronic levels used for the qubit.  \\
Lower panel: Implementation of 
the measurement sequence of one column or row.  To retrieve
only 1 bit of information, a measurement is implemented by first
performing a nonlocal unitary transformation, then only one qubit is
read out, and finally the unitary transformation is reversed.  
From  \citet{Guhne:2010PRA}.
}
\label{fig-exp-1}
\end{figure}

We start by describing the experimental setup: A pair of $^{40}$Ca$^+$
ions in a Paul trap was used to model the four-dimensional Hilbert space. 
For each ion, the qubit is represented by the states 
$\ket{1}=\ket{S_{1/2},m_S=1/2}$ and  $\ket{0}=\ket{D_{5/2},m_D=3/2}$, see 
also Fig.~\ref{fig-exp-1}. Manipulating and reading out this single 
system can be done using laser pulses with high fidelity. For performing
nonlocal gates, a M{\o}lmer-S{\o}rensen gate was used where both ions
were illuminated with the same laser. This allows one to perform nonlocal gates
even if the ion-crystal is not in the thermal ground state. This is important, 
as the later measurements require state detection of one ion, which can excite
the motional quantum number. 

A crucial point of the experiment is the appropriate implementation of the 
measurements of the PM square. Here, it is important that these are
global measurements on a four-dimensional system with only two outcomes $(\pm 1)$.
This means that a measurement like $C= \sigma_z \otimes \sigma_z$ cannot 
be implemented by measuring $\sigma_z$ on both particles separately, as this
would give four different results and destroy the coherence in the subspaces 
corresponding to the eigenvalues of  $\sigma_z \otimes \sigma_z$, as discussed
in our first presentation of the PM square in Sec.~\ref{sec:nutshell}. 
To circumvent this, one can write $C$ and any other observable in the 
PM square as
\begin{equation}
C= \sigma_z \otimes \sigma_z = 
U_{C}^\dagger [\sigma_z \otimes \openone] U_{C},
\end{equation}
where $U_{C}$ is some nonlocal unitary gate. Physically, this allows one to 
implement $C$ by first applying the $U_{C}$ to the state, then
reading out only the first ion, and finally undoing the transformation
$U_{C}$ again. In the experiment, the internal state of the 
second ion was in addition transferred to a different level during the 
readout of the first ion, in order to protect it from fluorescence light
during the detection process. 

In this way, all the nonlocal measurements on the PM square can be 
implemented, but note that the measurement of the third row
and the third column requires the implementation of six nonlocal gates within the
sequence. Consequently, the fidelity of a single nonlocal gate must be 
high (in the experiment it was around $98\%$) in order to observe the desired
results. For the interpretation of the experiment, the details of the
decomposition of $C= U_{C}^\dagger [\sigma_z \otimes \openone] U_{C}$ are not
relevant: A measurement like $C$ is seen as a black box, where a state is 
subjected to certain measurement procedures and a result $\pm 1$ is obtained. 
The details of the decomposition are inside the black box (see also 
Fig.~\ref{fig-exp-1}). One has to determine in the experiment whether
these black boxes represent repeatable and nondisturbing measurements; this has
also been done as later discussed.

\begin{figure}[t]
\begin{center}
\includegraphics[width=0.35\textwidth]{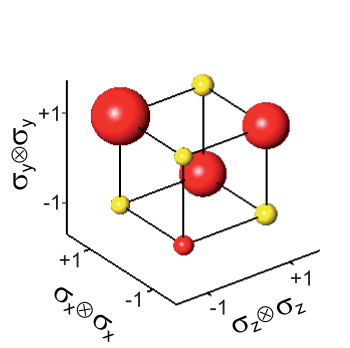}
\hspace{0.5cm}
\includegraphics[width=0.5\textwidth]{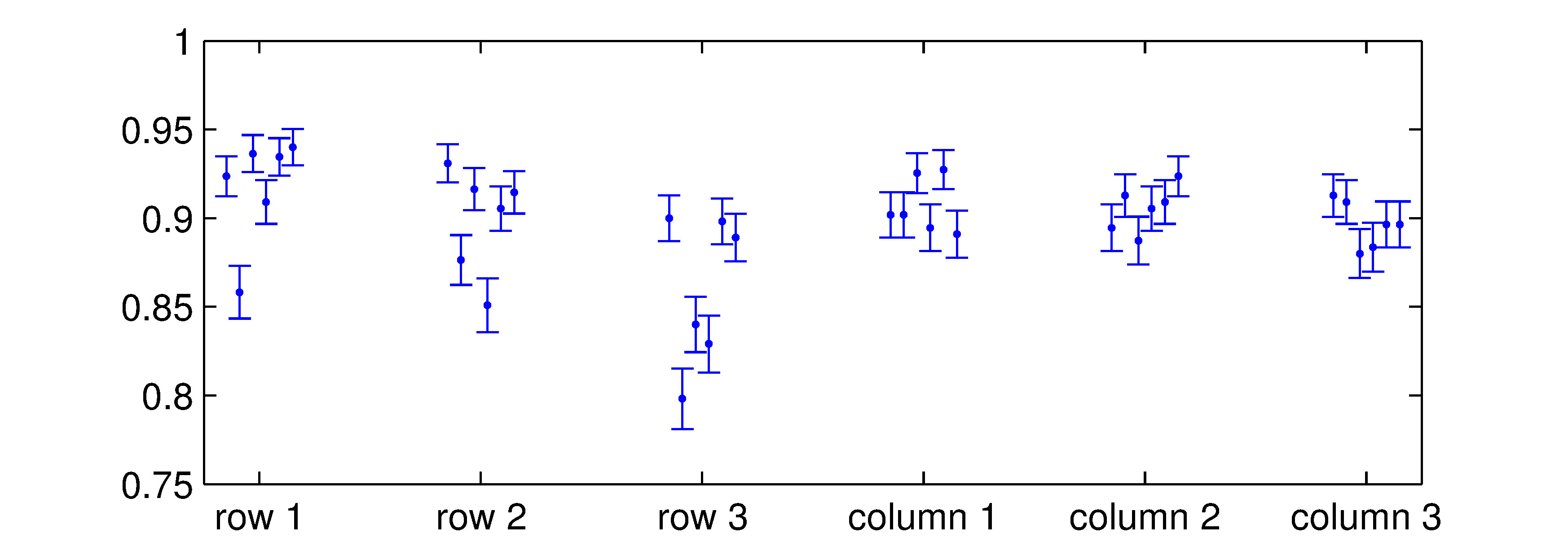}
\end{center}
\caption{Upper panel: measurement correlations for a sequence
of measurements for a partially entangled input state. The color choice
indicate whether the product of the three results gives $+1$ (yellow)
or $-1$ (red). The volume of a sphere is proportional to
the likelihood of finding the corresponding measurement outcome.\\
Lower panel: permutations
of the observables within the rows and columns can serve as a test of the
compatibility of the measurements. The measured absolute 
values of the products of observables for all the six possible permutations are shown. 
For each permutation, 1100 copies of the singlet state were used.
From~\citet{Kirchmair:2009NAT}.
}
\label{fig-exp-2}
\end{figure}

With these measurements, one can start to test the noncontextuality inequality. 
Performing a sequence of measurements, one obtains eight possible results, 
since any of the measurements results in a $\pm 1$ outcome; see also 
Fig.~\ref{fig-exp-2}. Multiplying the results gives the total value, 
which is then used for computing the total expectation value of an inequality. 
As a first result, if one takes a two-qubit singlet state as an input state, 
a violation of 
\begin{equation}
\mean{\mathsf{PM}} = 5.46\pm 0.04 > 4
\end{equation}
has been found, displaying a clear violation. A further central prediction 
of quantum mechanics is the state independence of the violation. For that, 
ten different states have been tested, including mixed states and separable 
states. In all cases, a violation has been found with values of 
$\mean{\mathsf{PM}}$ ranging from $5.23 \pm 0.05$ to $5.46 \pm 0.04$.

For it to be a complete contextuality test, some more issues have 
to be discussed. One first has to test and quantify the degree to which 
the implemented measurements are indeed compatible. Closely related to 
that is the question as to why the observed violation was not the one expected
from quantum mechanics. All this allows one to finally exclude 
hidden-variable
models, albeit with additional assumptions besides noncontextuality.

Concerning the compatibility of the measurements within one row or column, 
the experiment by \citet{Kirchmair:2009NAT} made several tests.  First, 
for compatible measurements the order of the measurements within one row or 
column should not matter. This was tested (see Fig.~\ref{fig-exp-2}) and 
confirmed. Second, for compatible measurements the values within a sequence
of measurements should not change. For that, one can consider the measurement
$A= \sigma_z \otimes \openone$ and a sequence  of measurements 
$A_1 A_2 A_3 \cdots$ of it, where we again  use the notation of 
Sec.~\ref{sssec:dist_err_mod}, with $A_i$ denoting the measurement of $A$
in the sequence position $i$ (to avoid confusion,  
this notation is not used in the following sections). 
The question is whether the results of 
$A_i$ and $A_j$ are the same, this can be quantified by the mean value 
$\mean{A_i A_j}$. Here, values from $\mean{A_1 A_2}=0.97\pm 0.01$ to 
$\mean{A_1 A_5}=0.95\pm 0.01$ have been reported \cite{Guhne:2010PRA}. 
For a nonlocal measurement such as $c=\sigma_x \otimes \sigma_x$ the 
imperfections are larger and one finds, for instance 
$\mean{c_1 c_3} = 0.88\pm 0.01$. In addition, measurement sequences 
like $c_1 C_2 c_3$ can be tested in order to test the compatibility of
$C$ and $c$. Here, values of $\mean{c_1 c_3}=0.83\pm 0.02$ have been found. 

The observations confirm that the implemented measurements are to a certain
extent repeatable and compatible. The question remains as to whether this is 
sufficient to rule out hidden-variable models with some extra assumptions. 
For that, \citet{Kirchmair:2009NAT} used a model where certain error
probabilities for short sequences are assumed to to be bounded by the error
probabilities of longer sequences; see Sec.~\ref{sssec:dist_err_mod} and  \citet{Guhne:2010PRA} 
for a detailed discussion. With that, the inequality 
\begin{equation}\label{eq:PM_exp_err}
\begin{split}
\mean{\chi} = 
\mean{BC}+\mean{bc}+\mean{Bb}-\mean{Cc}
-2 p^{\rm err}[CBC]\\
-2 p^{\rm err}[cbc]
-2 p^{\rm err}[bBb]
-2 p^{\rm err}[cCc]
\leq 2
\end{split}
\end{equation}
can be derived. In Eq.~\eqref{eq:PM_exp_err} $p^{\rm err}[CBC]$ denotes the probability that
the value of $C$ is flipped, if the sequence $C_1 B_2 C_3$ is measured. 
Experimentally, a value of $\mean{\chi}= 2.23 \pm 0.05$ was found, ruling
out this type of hidden-variable model.


\begin{figure*}[t]
\begin{center}
\includegraphics[width=0.8\textwidth]{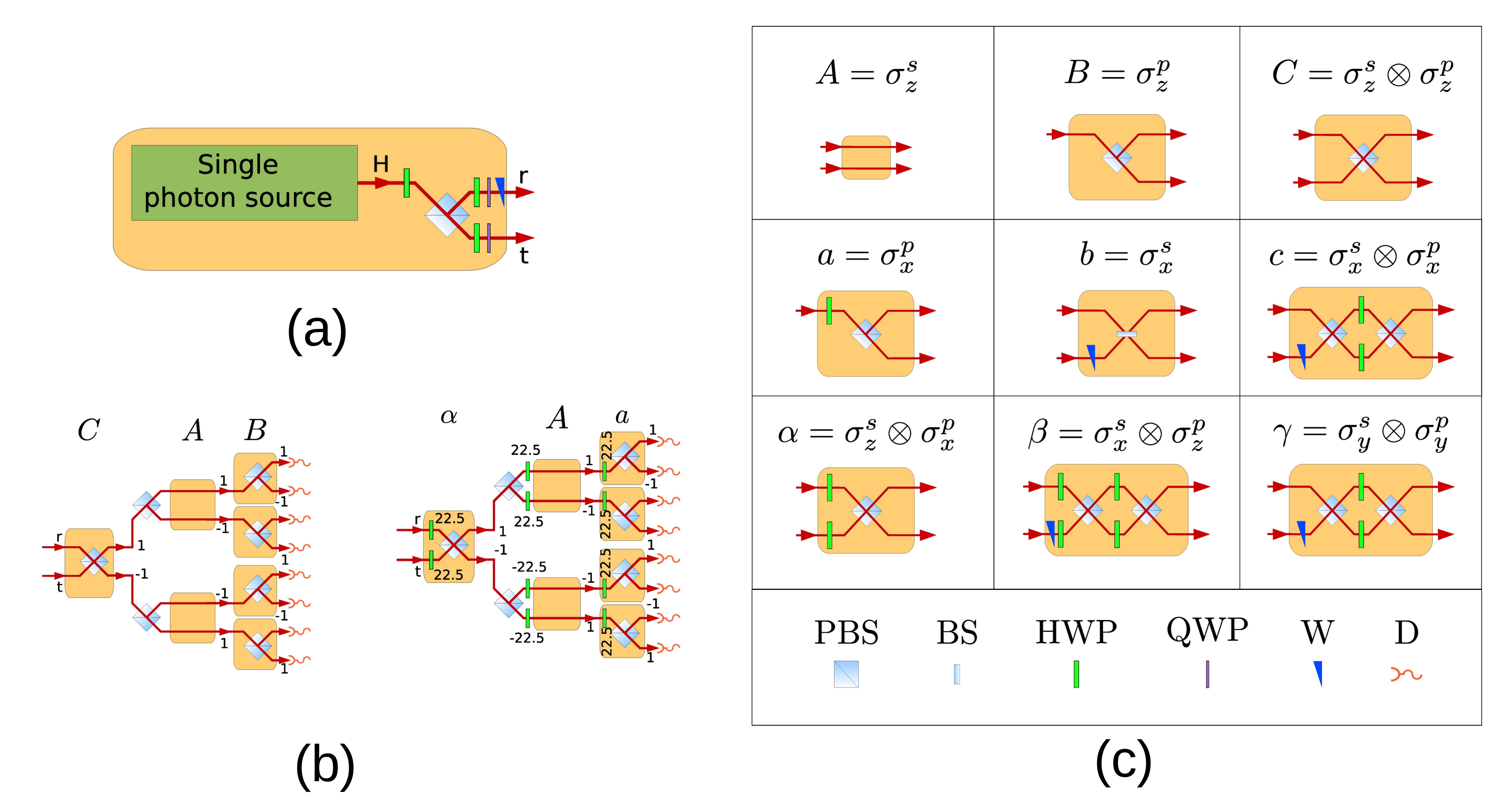}
\end{center}
\caption{(a) Encoding of two qubits in one photon. A tunable polarizing 
beam splitter distributes the photon over two spatial modes, resulting
in the spatial qubit. The polarization qubit is adjusted by half- and 
quarter-wave plates. 
(b) A row or column of the PM square is measured by a sequence of
interferometers. After the sequence, the photon can be in one of eight different 
outputs, representing the eight outcomes of the sequence of measurements. 
(c) Detailed interferometric setups for all  nine measurements in the PM square. From \citet{Amselem:2009PRL}. 
}
\label{fig-exp-3}
\end{figure*}


\subsubsection{A test of the Peres-Mermin inequality with photons}

A further test of the PM square was performed with photons 
\cite{Amselem:2009PRL}. A single photon was used there to carry two qubits: 
One qubit was encoded in the polarization, and a second one was encoded in the path
of the photon; see also Fig.~\ref{fig-exp-3}.

Given this two-qubit system, one has to implement the nine measurements
$A, \dots , \gamma$ from the PM square. Note that a standard
measurement of the polarization or path with photon detectors is not 
suitable, as then the photon is absorbed and no further sequence of 
measurements can be carried out. \citet{Amselem:2009PRL} did this 
by constructing an interferometric setup for each measurement, where the 
result of the measurement was encoded in the output port of the interferometer; 
see Fig.~\ref{fig-exp-3} (c). For instance, $A= \sigma_z^s$ on the spatial qubit
is effectively an empty interferometer: the photon leaves the setup in direct
correspondence to the input. The polarization measurement $B= \sigma_z^p$ 
can be implemented with a polarizing beam splitter, which makes the output 
port dependent on the input polarization. These are only simple examples; certain
other measurements essentially require an entangled Bell measurement for
their implementation. 

For measuring a sequence like $CAB$ one has to concatenate these interferometers,
see Fig.~\ref{fig-exp-3} (b). Here one needs to build 2 times the setup 
for measuring $A$, one for each possible output port of $C$. Moreover, one 
needs 4 times the setup for $B$, as $CA$ can have four possible results, 
i.e., the photon may be in four different paths. Finally, the result of the
entire sequence is measured by a click in one of eight detectors. 
The inequality was checked for 20 different input states. On
average, a value of 
\begin{equation}
 \mean{\mathsf{PM}} = 5.4550 \pm 0.0006 > 4
\end{equation}
was found. 


\subsubsection{A test of the KCBS inequality with photons}
\begin{figure*}[t]
\begin{center}
\includegraphics[width=0.8\textwidth]{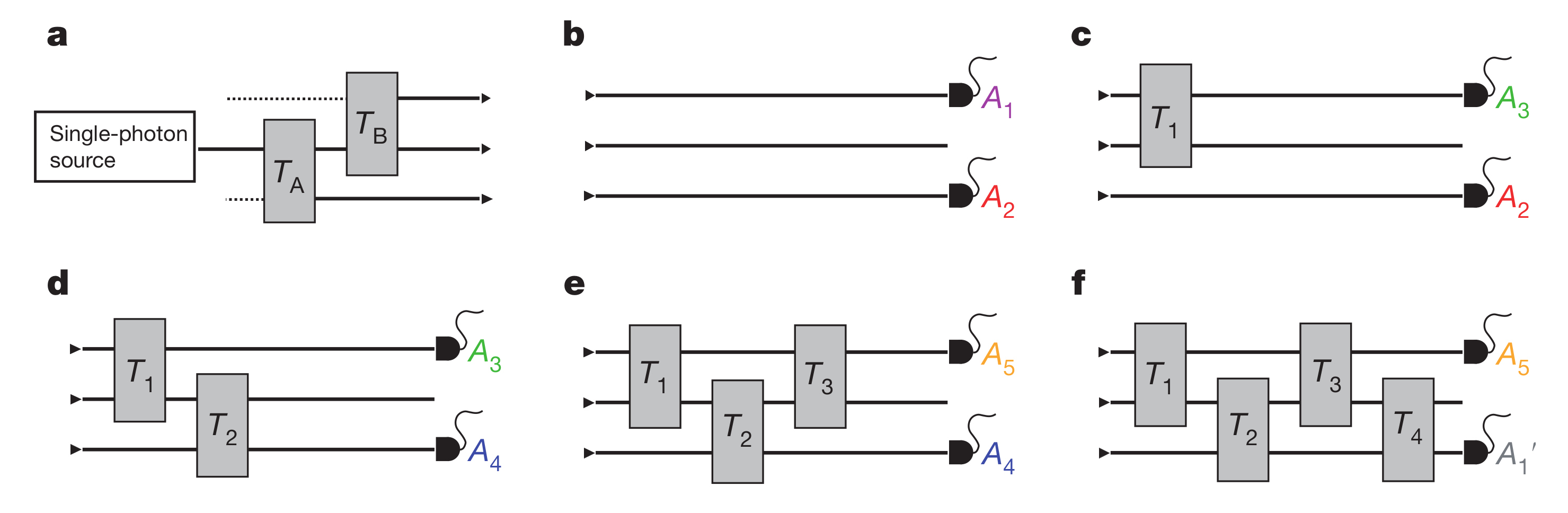}
\hspace{0.3cm}
\includegraphics[width=0.8\textwidth]{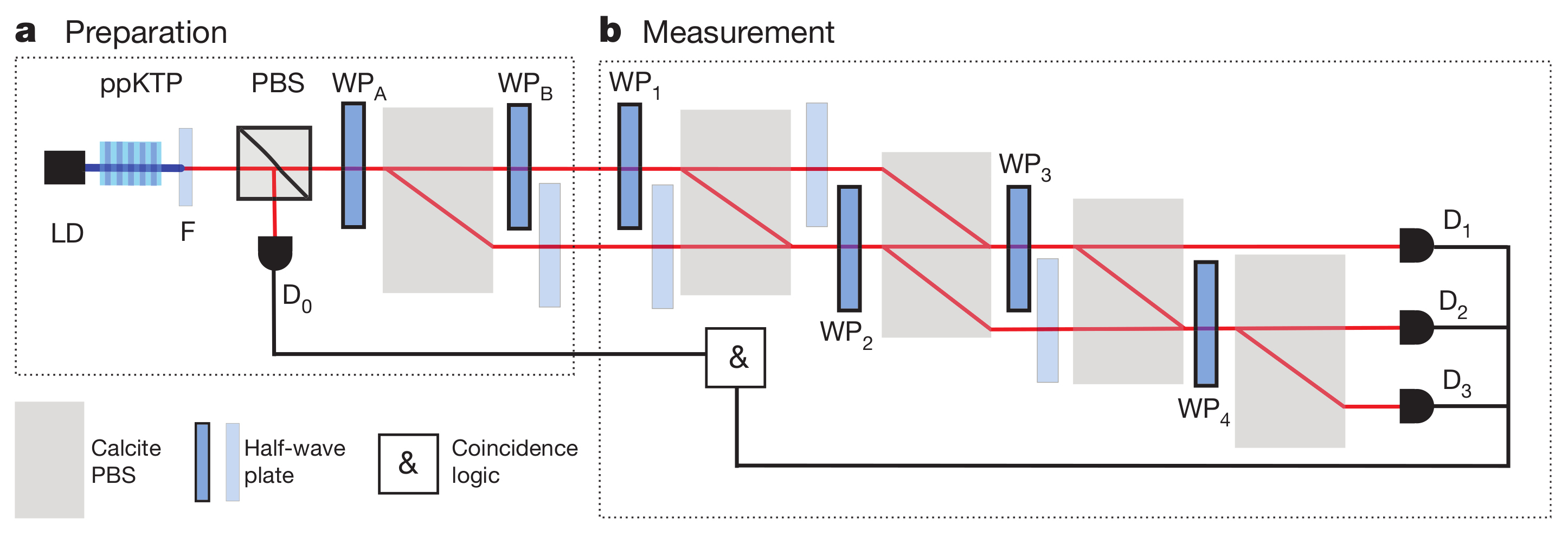}
\end{center}
\caption{Left panels: etup for the measurement of the correlations $\mean{A_i A_j}$
in the KCBS inequality in Eq.~(\ref{eq-exp-kcbs}). The measurement $A_i$ has 
the result $-1$ if the corresponding detector clicks; otherwise, the value $+1$
is assigned. Consequently, the product $A_i A_j$ has the value $+1$ if both 
or no detectors register a photon, otherwise, the value is $-1$.
Right panels: concrete experimental setup. 
The transformations $T_i$ are implemented via insertion of the half-wave plates WP$_i$, 
which in combination with polarizing beam splitters distribute the photons across 
the modes. From \citet{Lapkiewicz:2011NAT}.
}
\label{fig-exp-4}
\end{figure*}

The KCBS inequality from Sec.~\ref{ssec:NC_ineq} was first 
tested in an experiment using photons \cite{Lapkiewicz:2011NAT}.
Here, the three basis states of the Hilbert space were given by 
three possible paths of a photon in an interferometer; see also 
Fig.~\ref{fig-exp-4}. 

A single photon is first coherently distributed over the three 
paths via beam splitters, thus preparing the initial state. A 
measurement is done by detection in a possible path (result $-1$),  
and if no photon is detected the measured value is $+1$. A pair of 
observables from the KCBS inequality is measured simultaneously 
by marking two paths. If the photon is in such a path, the product 
is assigned the value $-1$; no detection corresponds to the value $+1$. 
For example, $A_1 A_2$ can be directly be measured [Fig.~\ref{fig-exp-4}(b)], 
but for the next term in the inequality, optical elements are used to manipulate 
the two paths which are not needed for the measurement of $A_2$. The two
paths are then used for $A_3$ and the next term in the inequality 
(Fig.~\ref{fig-exp-4}(c)). 

The setup has to ensure, for instance, that the observable $A_2$ that is measured 
both in the term $A_1 A_2$ and in the term $A_2 A_3$ corresponds exactly to the same 
experimental setup. In the experiment this problem was solved by a careful design
of a measurement sequence; see the right-hand side of Fig.~\ref{fig-exp-4}. However, 
in the last correlation one does not measure $A_5 A_1$, but instead $A_5 {A}_1'$ where 
${A}_1'$ has a different structure than the measurement $A_1$ in the correlation 
$A_1 A_2$. One can, however, compare the properties of $A_1$ and  ${A}_1'$ and 
then argue that it is effectively the same measurement. In the experiment there was 
a small deviation between $A_1$ and  ${A}_1'$, but it was suggested to take 
this into account with a correction term in the KCBS inequality, leading to
a modified classical bound of $-3.081\pm 0.002$. For the KCBS correlation, there is a
value of 
\begin{equation}
\begin{split}
 \mean{A_1 A_2} + 
 \mean{A_2 A_3} + 
 \mean{A_3 A_4} + 
 \mean{A_4 A_5} \\
 + 
 \mean{A_5 A_1'}  
=
-3.893 \pm 0.006,
\end{split}
\label{eq-exp-kcbs}
\end{equation}
which violates the contextuality inequality by
120 standard deviations.

\subsubsection{Final considerations on KS contextuality experiments}

To conclude this section, we first mention some other experimental 
tests of contextuality inequalities. The contextuality inequality from the PM 
square has also been tested using nuclear magnetic resonance 
\cite{Moussa:2010PRL}, with photons in its entropic version~\cite{QuPRA2020},
a similar inequality has been tested with photons \cite{LiuPRA2009}, and the Mermin star inequality
has been tested with nitrogen-vacancy centers in diamond~\cite{vanDam:2019PRL}.
The KCBS inequality and its generalizations have been tested with superconducting qubits \cite{JergerNC2016}, photons \cite{AriasPRA2015,BorgesPRA2014}, and ions  \cite{Malinowski:2018PRA},
and \citet{UmSR2013,UmPRA2020} explored the connection to randomness generation.

The inequality of Yu and Oh was first tested with photons by \citet{ZuPRL2012};
see also the discussion given by \citet[][]{Amselem:2013PRL,Reply_to_comm2013}. 
Further tests have been implemented with a single trapped ion \cite{Zhang:2013PRL} 
and nitrogen-vacancy centers in diamond \cite{Kong2012,Kong2016}. 
In a more
recent experiment with a trapped ion, the compatibility relations of the observables were
studied in detail \cite{Leupold:2018PRL}. 
Finally, there are experimental works 
that aim at an observation of contextuality effects in classical systems \cite{FrustagliaPRL2016,ZhangPRL2019,LiSR2017}; see also the discussion given by \citet{Markiewicznpj2019}.

In all the previously listed experiments, several different approaches have been proposed to test contextuality. It is important to separate them into three different main categories. In Secs.~\ref{ssec:compatible} and \ref{ssec:exper_imp}, we argued in favor of the use of sequential measurements, such as  the experiments by \citet{Kirchmair:2009NAT} discussed in this section. The alternative is that of joint measurements. Some experiments adopted this approach, such as the one by \citet{Lapkiewicz:2011NAT} described in this section. Even in the joint measurement approach, effort has been put into the identification of which part of the device corresponds to each single measurement, as well as the quantification of experimental imperfections and their consequences on the noncontextual bound, in a similar spirit to the approaches discussed in Sec.~\ref{ssec:exper_imp}. 
Finally, there are other experiments where the single measurements in each context are not as well characterized and are sometimes even implemented in different ways in different contexts, making the experimental procedure itself context dependent. This can lead to discussions about the interpretation of
the experiment \cite{ZuPRL2012,Amselem:2013PRL,Reply_to_comm2013}.

The experiments chosen as examples are simply representatives of a broad range of different experiments with analogies and differences. The variability of the different approaches is arguably due to the lack of a clear and comprehensive theoretical description of a quantum contextuality experiment. This review aims to fill the gap.

Finally, there are two types of experiments we have not mentioned thus far, namely, experiments of Spekkens's contextuality, which are covered in Sec.~\ref{ssec:other_cont}, and the Liang-Spekkens-Wiseman approach \cite{Liang:2011PR}, which was tested by ~\citet{Zhan:17} and which we do not cover in this review.


\subsection{A different notion of contextuality: Spekkens's approach}\label{ssec:other_cont}
\subsubsection{Spekkens's definition of noncontextuality}
A different notion of contextuality was introduced by \citet{Spekkens:2005PRA} and further explored and developed in subsequent papers \cite{Spekkens:FOP2014,KunjwalPRL2015,Pusey2015,Mazurek2016,XuPRA2016,KunjwalPRA2018,KunjwalQuantum2020, SchmidPRA2018, Schmid:2021PRX, Schmid2020}. The starting point is an operational interpretation of a physical theory, namely, a construction where the primitive elements are preparation, transformation, and measurement procedures. Such procedures are intended as a list of instructions for ``operations'' that can be performed in a laboratory. For the case of a prepare-and-measure scenario (i.e., ignoring for the moment transformation procedures), the basic elements are preparations and measurement effects together with rules for calculating probabilities [i.e., $p(k|P,M)$], representing the probability of obtaining the outcome $k$ for the measurement $M$ given the preparation procedure $P$. 

The notion of a noncontextual operational theory is based on the idea of the statistical indistinguishability of procedures. In simple terms, one may call {\it operationally equivalent} those procedures that give rise to the same statistics and may require that they should be represented by the same elements of the theory. Consequently, Spekkens defines equivalence classes of preparations and measurements as follows: 
\begin{align}
\label{equivp}
P & \sim P' \Longleftrightarrow p(k|P,M) = p(k|P',M) \ \\
&\nonumber \text{ for all measurements and outcomes } k,M,\ \\
\label{equivm}
(M,k) & \sim (M',k') \Longleftrightarrow p(k|P,M) = p(k'|P,M') \ \ \\
&\nonumber  \text{ for all preparations } P, 
\end{align}
where the symbol $\sim$ denotes operational equivalence.
If one applies the equivalence in Eq.~\eqref{equivp} to experimental procedures described according to quantum mechanics, one obtains that each equivalence class $[P]$ is associated with a quantum state $\rho$ since there is no way to distinguish via a quantum measurement two preparations that give rise to the same state $\rho$. A typical example is that of a spin-$1/2$ particle: it is not possible to distinguish the preparation of an equal mixture of states polarized along $z$, corresponding to the quantum state $\rho=(\ketbra{0}{0} + \ketbra{1}{1})/2$, from the preparation of an equal mixture of states polarized along $x$,  corresponding to the quantum state $\rho=(\ketbra{+}{+} + \ketbra{-}{-})/2$. Similarly for Eq.~\eqref{equivm} each measurement $M$ is associated with a POVM $\{ E_k\}_k$ and each outcome $k$ is associated with a single element $E_k$, with $E_k\geq 0$ and $\sum_k E_k = \openone$, regardless of the particulars of the experimental implementations of the measurement.

A hidden-variable description of preparations and measurement procedures is given by an {\it ontological model} that plays a role similar to that of a NCHV in Sec.~\ref{sssec:NCHV}. A crucial difference, however, is that such procedures involve only a preparation and a measurement. In fact, we restrict ourselves to the prepare-and-measure scenario to keep the presentation simple and concise. Notice, however, that the original formulation of Spekkens \cite{Spekkens:2005PRA} also considered transformations and that recent developments of Spekkens contextuality included also the case of, e.g., sequential measurements, or more complex compositions of operations \cite{Schmid2020a}.  An ontological model for the probability $p(k|P,M)$ is then given by 
\begin{equation}\label{spont}
p(k|P,M)=\int d\lambda \mu_P(\lambda) \xi_{M,k}(\lambda),
\end{equation}
where $\mu_P: \Omega \rightarrow [0,1]$ is the probability density associated with the preparation procedure $P$, i.e., $\int d\lambda \mu_P(\lambda)=1$, and $\xi_{M,k}:\Omega \rightarrow [0,1]$ represent the indicator function associated with the outcome $k$ of $M$, satisfying ${\sum_k \xi_{M,k}(\lambda) =1}$, for all $\lambda\in \Omega$. 

Equation~\eqref{spont} is reminiscent of the expression for joint measurements or local hidden state models arising in the context of quantum steering~\cite{Uola:2020RMP}. In fact, some formal equivalence between Spekkens's preparation contextuality and these two phenomena has been shown~\cite{Tavakoli:2020PRR}. This is no longer true if the set of states under consideration does not include all quantum states or if one also considers  measurement noncontextuality, in addition to preparation noncontextuality; see \citet{Selby2021} for further details. Finally, we recall that another notion of steering, based on the original argument by \citet{Schrodinger35}, was shown to hold for a Spekkens-noncontextual toy theory by \citet{Spekkens:2007PRA}. This notion of steering, however, does not coincide with the one introduced by \citet{Wiseman:2007PRL} and used by \cite{Tavakoli:2020PRR}.

The condition of noncontextuality, then, amounts to the requirement that the same description in the ontological model corresponds to each equivalence class in the operational model. In other words, if $P\sim P'$ then $\mu_P=\mu_{P'}$, (condition of preparation noncontextuality) and if $(M,k)\sim (M',k')$, then $\xi_{M,k}=\xi_{M',k'}$  (condition of {\it measurement noncontextuality}). It is then possible to obtain a contradiction between the previously stated assumptions and the predictions of quantum mechanics, hence, showing the impossibility of a noncontextual ontological model.

\begin{figure}[t]
\includegraphics[width=0.4\textwidth]{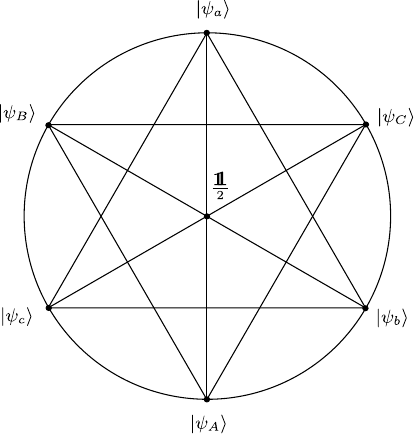}
\caption{\label{fig:spek}Different decomposition of the maximally mixed state $\tfrac{1}{2}\openone$, represented in the $(x,z)$ plane of the Bloch sphere, in terms of the pairs along opposite lines, (such as $\ket{\psi_a},\ket{\psi_A}$), or triples in the same triangle (such as $\ket{\psi_a},\ket{\psi_b},\ket{\psi_c}$).}
\end{figure}

In the simplest example of the impossibility of a preparation-noncontextual ontological model, \citet{Spekkens:2005PRA} wrote the maximally mixed state of a qubit, i.e., $\openone/2$, as a convex combination of different rank-1 projectors, namely,
\begin{align}
\tfrac{\openone}{2} & =\tfrac{1}{2}(\ketbrac{\psi_a}+ \ketbrac{\psi_A}) \nonumber \\
& =\tfrac{1}{2}(\ketbrac{\psi_b}+ \ketbrac{\psi_B}) = \frac{1}{2}(\ketbrac{\psi_c}+ \ketbrac{\psi_C})
\nonumber\\
 & =\tfrac{1}{3}(\ketbrac{\psi_a}+ \ketbrac{\psi_b} + \ketbrac{\psi_c})
 \nonumber\\
&= \frac{1}{3}(\ketbrac{\psi_A}+ \ketbrac{\psi_B} + \ketbrac{\psi_C}),
\label{eq-spekkens-constraint1}
\end{align}
where the vectors (depicted in Fig.~\ref{fig:spek} in the Bloch representation) are defined as 
\begin{align}
\ket{\psi_a}  & =(1,0),\ \ket{\psi_A}=(0,1), 
\nonumber \\
\ket{\psi_b} & =\tfrac{1}{2}\left({1},{\sqrt{3}}\right),\ 
\ket{\psi_B}=\tfrac{1}{2}\left({\sqrt{3}},-{1}\right),
\nonumber \\
\ket{\psi_c}&=\tfrac{1}{2}\left(1, -{\sqrt{3}}\right), 
\ket{\psi_C}=\tfrac{1}{2}\left({\sqrt{3},1}\right).
\end{align}
Under the assumption of preparation noncontextuality, the corresponding probability measures in the ontological models must coincide.
\begin{equation}\label{eq:pnc_conv}
\begin{split}
\frac{\mu_a(\lambda) + \mu_A(\lambda)}{2}= \frac{\mu_b(\lambda) + \mu_B(\lambda)}{2} = \frac{\mu_c(\lambda) + \mu_C(\lambda)}{2}\\
=\frac{\mu_a(\lambda) + \mu_b(\lambda)+\mu_c(\lambda)}{3} = \frac{\mu_A(\lambda) + \mu_B(\lambda)+\mu_C(\lambda)}{3}.
\end{split}
\end{equation}
Moreover, owing to the orthogonality conditions, i.e., $\mean{\psi_a|\psi_A}=\mean{\psi_b|\psi_B}=\mean{\psi_c|\psi_C}=0$, the corresponding distributions in the ontological model should not overlap, namely,
\begin{equation}\label{eq:pnc_null}
\mu_a(\lambda)\mu_A(\lambda)=\mu_b(\lambda)\mu_B(\lambda)=\mu_c(\lambda)\mu_C(\lambda)=0 \text{ for all } \lambda,
\end{equation}
since they can be distinguished with certainty with a single-shot measurement.
By checking all possible assignments of null values according to the previously mentioned product conditions, one immediately realizes that the only possible solution to Eqs.~\eqref{eq:pnc_conv} and \eqref{eq:pnc_null}  is $\mu_a(\lambda)=\mu_A(\lambda)=\mu_b(\lambda)=\mu_B(\lambda)=\mu_c(\lambda)=\mu_C(\lambda)=0 $.
The same argument can be extended to show preparation contextuality of any mixed state~\cite{Banik:2014FOP}. The above ideas have been further elaborated on to devise experimental tests of Spekkens's contextuality, as we discuss in Sec.~\ref{sssec:Spekkens_ine}.

\subsubsection{Inequalities for Spekkens' noncontextuality}\label{sssec:Spekkens_ine}
Spekkens' notion of noncontextuality was tested experimentally by \citet{Spekkens:2009PRL, Mazurek2016,Hameedi:2017PRL,Anwer:2019XXX}. To do so, it is first necessary  to derive noncontextuality inequalities that are testable against the observed statistics. This was done in several works~\cite{Spekkens:2009PRL, Mazurek2016, KunjwalPRL2015,Pusey2015,XuPRA2016,KunjwalPRA2018,SchmidPRA2018,KrishnaNJP2017}. In the following, we present the results by \citet{Mazurek2016}, on both  the theoretical and experimental sides, as they managed to overcome some difficulties present in the first experiment \cite{Spekkens:2009PRL},  particularly, in relation to the operational equivalence of preparations and measurements. The noncontextuality inequality by \citet{Mazurek2016}, based on both preparation and measurement noncontextuality, is presented in the following. One first needs to introduce six preparations $P_{t,b}$, for $t=1,2,3$ and $b=0,1$ , such that 
\begin{equation}\label{eq:Maz_P_cond}
P_* := \frac{1}{2}(P_{t,0} + P_{t,1})=\frac{1}{2}(P_{t',0} + P_{t',1})   \text{ for all } t,t'=1,2,3,
\end{equation}
and three dichotomic measurements $M_t$ for $t=1,2,3$, such that the average measurement is the ``fair coin flip'' measurement, i.e.,
\begin{equation}\label{eq:Maz_M_cond}
M_*:=\frac{1}{3}\sum_t M_t, \text{ with }  p(b|M_* , P)=\frac{1}{2}\ \forall P \text{ and } b=0,1,
\end{equation}
where, as usual, $p(b|M , P)$ denotes the probability of an output $b$ given the preparation $P$ and the measurement $M$.
The noncontextuality inequality reads
\begin{equation}\label{eq:Maz_NC_ineq}
A =\frac{1}{6}\sum_{t=1,2,3} \sum_{b=0,1} p(b|M_t , P_{t,b})\leq \frac{5}{6}.
\end{equation}
This upper bound can be computed in terms of the ontological model as follows
\begin{align}
\frac{1}{6} & \sum_{t=1,2,3} \sum_{b=0,1}\sum_\lambda \xi(b|M_t, \lambda)\mu(\lambda|P_{t,b})\nonumber \\
& \leq \frac{1}{3}\sum_{t=1,2,3} \sum_\lambda \eta(M_t, \lambda)\Big( \frac{1}{2} \sum_{b=0,1}\mu(\lambda|P_{t,b})\Big)
\nonumber \\
& =\sum_\lambda \mu(\lambda|P_*)\Big(\frac{1}{3} \sum_{t=1,2,3} \eta(M_t, \lambda)\Big)
\nonumber \\ 
&\leq \max_\lambda \Big( \frac{1}{3} \sum_{t=1,2,3} \eta(M_t, \lambda) \Big), 
\end{align}
where $\eta(M_t, \lambda):=\max_{b=0,1} \xi(b|M_t, \lambda)$. The assumption that $M_*$ is the  fair coin 
flip, implies that $1/3\sum_t \xi(b|M_t, \lambda) = \xi(b|M_*, \lambda) = 1/2$ for $b=0,1$, 
which constrains the three-dimensional vector $\left(\xi(0|M_t, \lambda)\right)_{t=1,2,3}$ on a two-dimensional polytope 
inside the $[0,1]^3$ cube, whose extremal points are given by $(1,1/2,0)$ and the coordinate permutations. 
Taking the outcome maximization defining $\eta$, i.e., flipping one or more outcomes, one obtains at most the
assignments $(1,1/2, 1)$, and permutations, which give the upper bound $5/6$. Notice that the 
derivation of Eq.~\eqref{eq:Maz_NC_ineq} uses both preparation noncontextuality 
[$\sum_{b}\mu(\lambda|P_{t,b})/2 = \mu(\lambda|P_*)$] and measurement noncontextuality 
[$\sum_t \xi(b|M_t, \lambda)/3 = \xi(b|M_*, \lambda)$]. In fact, the inequality can be violated by models that 
violate at least one of the constraints, as shown by \citet{Mazurek2016}. Moreover, quantum 
theory can violate it up to the algebraic maximum $1$. This is done  using as 
$P_{t,b}$ the six preparations in 
Fig.~\ref{fig:spek}, with the pairs $b=0,1$, for fixed $t$, associated with antipodal points (i.e., 
$\ket{\psi_a},\ket{\psi_A}$ etc.), and as measurements 
the three $M_t$ projective measurements with two outcomes, rotated by 
$2\pi/3$ on the Bloch sphere: where the three effects,  each for the 
$0$ outcome of $M_t$ for $t=1,2,3$, form the triangle $\ket{\psi_a},\ket{\psi_b},\ket{\psi_c}$. 
General methods for computing maximal violations of such noncontextuality inequalities have been developed~\cite{Chaturvedi:2021Quantum,Tavakoli:2021PRX}.

\subsubsection{Experimental tests of Spekkens' contextuality}
In the derivation of Eq.~\eqref{eq:Maz_NC_ineq},  the assumptions of both preparation 
and measurement noncontextuality enter. Notice that, without further assumptions, to 
infer operationally indistinguishability one needs to perform all possible measurements 
as in Eq.~\eqref{equivm}. To avoid this problem, a minimal set of measurements is 
assumed to be necessary to infer that two preparation procedures are operationally 
indistinguishable. This minimal set is said to be {\it tomographically complete}. 
Similarly, a tomographically complete set of preparations is assumed to exist, in 
order to infer that two measurements are operationally indistinguishable. 
To characterize this with minimal assumptions, \citet{Mazurek2016} first 
analyzed the dimension needed to describe the state preparations and 
measurements of a single polarized photon. In the experiment, four measurements 
on eight input states were performed, but then it was found that the experimental 
data can be described by assuming three independent measurements and a state space 
given by a four-dimensional hyperplane, in the sense that assuming more independent 
parameters does not allow for a more accurate description of the experimental data. Given 
this observation, one can avoid testing the operational equivalence over infinitely many 
preparations and measurements. The observed numbers of parameters are the same that
quantum mechanics uses to describe qubit systems.

\begin{figure}[t]
\includegraphics[width=0.5\textwidth]{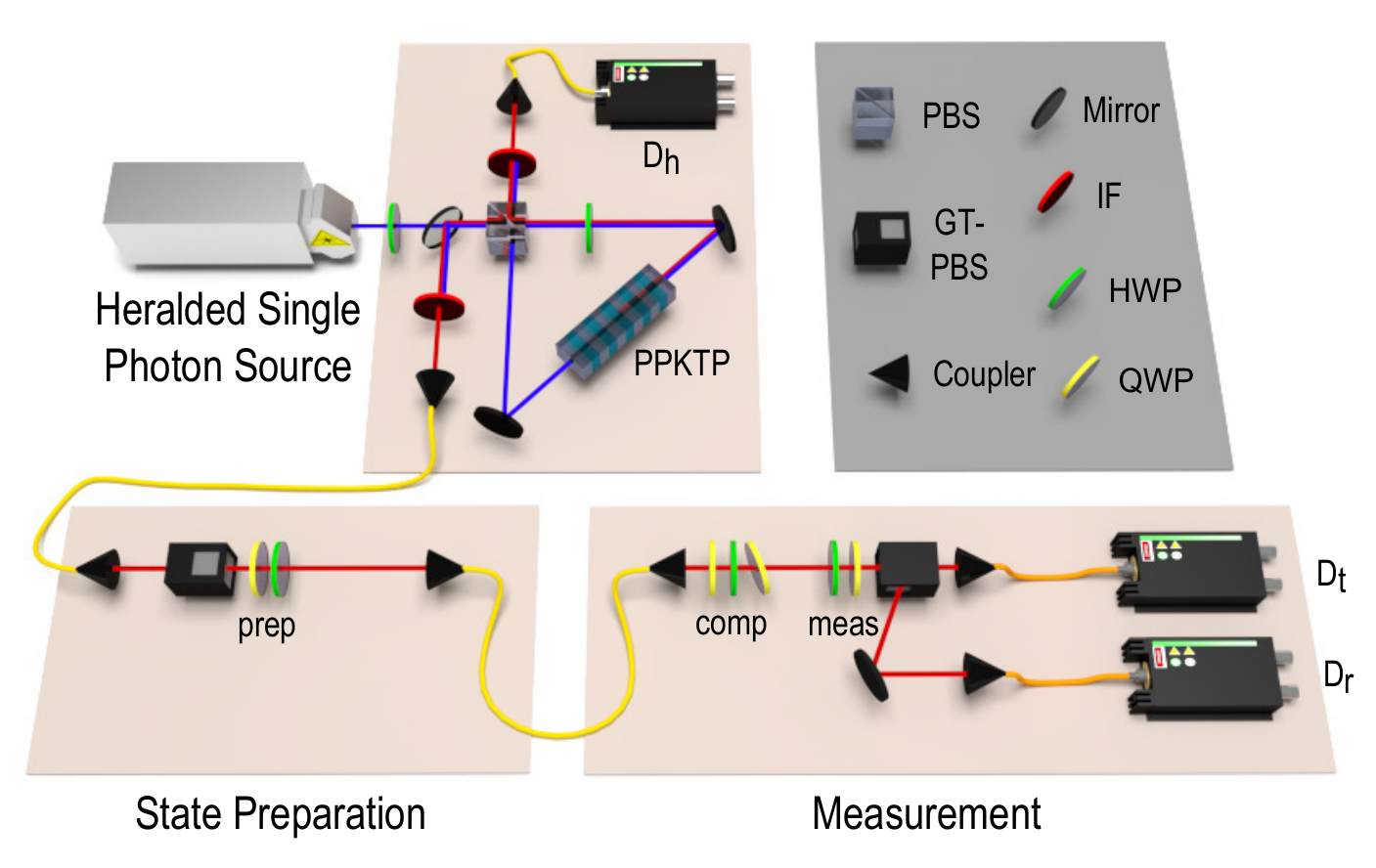}
\caption{Experimental setup used by \citet{Mazurek2016}. The quantum system consists of a single photon in which a specific polarization is prepared via a polarizer and two waveplates (preparation $P_{t,b}$) and then measured via two waveplates, a polarized beamsplitter and two detectors (measurement $M_t$).
From~\citet{Mazurek2016}.
}
\label{fig-exp-Maz}
\end{figure}

A second problem arises, namely that due to experimental imperfections, it is not possible to find pairs of 
preparations $\{P_{t,b}\}_b$ with the same average preparation $P_*$ and a triple of measurements with the 
same average measurement $M_*$ corresponding to a fair coin flip. This problem was addressed by
\citet{Mazurek2016} by 
constructing the so-called {\it secondary} preparations and measurements, as convex mixtures of the primary 
ones (eight state preparations and four measurements), which were those directly implemented in the experiment. 
In other words, from the primary preparations and measurements performed, one can infer what would have been 
the value of their convex combinations. Among the secondary operations, one can select those that are the 
closest to the primary ones and at the same time satisfy exactly the conditions in 
Eqs.~\eqref{eq:Maz_P_cond} and \eqref{eq:Maz_M_cond}. To avoid any reference to quantum theory, secondary 
preparations and measurements are described in terms of a generalized probability theory, with the above 
assumption of tomographical completeness for three measurements and four preparations. The experiment was 
performed by preparing and measuring the polarization degree of freedom of a single photon, as depicted in 
Fig.~\ref{fig-exp-Maz}. An experimental value of $A=0.99709 \pm 0.00007$ for the parameter $A$ in 
Eq.~\eqref{eq:Maz_NC_ineq} is then observed, based on the inferred values of the secondary preparations and 
measurements, violating the noncontextual bound $5/6$.\\

\subsubsection{Relation with different notions of hidden-variable models}
In Spekkens' notion of contextuality, a fundamental role is played by two 
properties, namely, the existence of a nonunique decomposition of quantum 
mechanical mixed states into pure states, and the requirement that the 
indistinguishability present at the operational level, identified here with 
quantum mechanical predictions, is satisfied at the level of the ontological 
model. In contrast, if one assumes a classical model (specifically a 
measurement contextual hidden-variable model), then each mixed state is a 
probability distribution and hence has a unique convex decomposition into 
determinisitic assignments, which take here the role of pure states.
It follows that hidden-variable models for two-level quantum systems, such as the 
models by \citet{Bell:1966RMP} and by \citet{Kochen:1967JMM}, turn out to be 
preparation contextual \cite{LeiferPRL2013}. In these models, in fact, if we 
were able to directly measure the hidden variable $\lambda$, we could 
distinguish between an equal mixture of $\ket{0}$ and $\ket{1}$ and an equal 
mixture of $\ket{+}$ and $\ket{-}$, even though both mixtures give rise to same 
quantum state $\rho=\openone/2$. This is an explicit example of preparation 
contextuality. According to the hidden-variable model such preparations are 
indeed different, even though the theory does not contain measurements that are 
able to distinguish them.

\cite{Spekkens:2005PRA} defined a noncontextual ontological model of an operational 
theory as one where any two procedures that are operationally 
equivalent (in the theory) have identical representations in the ontological 
model. This definition is defended by appealing to a methodological principle, 
referred to as Leibniz’s principle of the ontological identity of empirical 
indiscernibles \cite{Spekkens2019}, which can be stated as follows: If an 
operational theory predicts that two procedures are indistinguishable but, in 
the theory, they have distinct representations, then the theory should be 
discarded and replaced by a new theory relative to which the two procedures 
have identical representation.

For a defense of this methodological principle for constructing physical 
theories, the reader is referred to \cite{Spekkens2019}. For a discussion on 
how this methodological principle is used to motivate the notion of 
noncontextual ontological model, see
\cite{Spekkens:2005PRA}. The program of hidden-variable theories for quantum 
mechanics  \cite{Belinfante1973} provides examples of theories that do not 
adhere to this methodological principle. Note that it is 
still under debate as to whether some proposed hidden-variable theories predict 
deviations from quantum mechanics and, if they do, whether these 
deviations are observable. The former is an open problem in Bohmian 
mechanics, in relation to Bell experiments \cite{Correggi2002, Kiukas:2010JMP} 
and tunneling times \cite{Hauge:1989RPM,Landauer:1994RMP,Stomphorst:2002PLA}, 
and in some classes of hidden-variable models analyzed from a thermodynamical 
perspective \cite{Cabello_Th_PRA2016}, whereas the latter is an open problem 
in  collapse models \cite{Bassi:2013RMP}. For a criticism of the notion 
of preparation contextuality, see
\citet{Ballentine:2014XXX}.


\section{Advanced topics and methods}\label{sec:Advanced}


In this section, we discuss more advanced topics and methods associated with quantum contextuality. We address questions such as how to compute noncontextuality inequalities for a given scenario, what the corresponding maximal quantum violation is, which scenarios give rise to contextuality and to state-independent contextuality, how contextuality is related to nonlocality, etc.

Here, since we discuss mostly theoretical results where no direct connection with experimental procedures is made, no additional assumptions on the measurements, such as nondisturbance and repeatability, are necessary. In most cases, it is enough to think about collections of joint measurements, regardless of the way they may be implemented in the lab. Similarly, the distinction between the OP and the EP is mostly irrelevant here, and we often shift between the two points of view, sometimes privileging observables and sometimes privileging effects. Finally, in some sections, such as Sec.~\ref{ssec:graph_CSW}, \ref{sssec:chrome},   \ref{KS_Bell} and \ref{nullif}, we explicitly consider projective measurements.

The section is organized as follows. In Sec.~\ref{polyt}, we introduce the noncontextuality polytope, which describes noncontextual correlations associated with a given scenario, i.e., a fixed set of measurements and contexts, and allows one to compute noncontextuality inequalities. 
In Sec.~\ref{graphth}, we discuss the connection between graph theory and noncontextuality: since contexts can be represented as graphs, or more generally hypergraphs, several properties of contextuality scenarios can be investigated in terms of graph theoretical properties.
In Sec.~\ref{KS_Bell}, we discuss the connection between Bell nonlocality and contextuality.  In Sec.~\ref{class_sim}, we discuss different approaches to the classical simulation of contextual correlations.
In Sec.~\ref{nullif}, we present the debate on the nullification of the KS theorem. 


\subsection{The noncontextuality polytope}\label{polyt}

The set of correlations that can be achieved within a noncontextual 
 theory forms a polytope in the space of probability assignments. 
 Analogously to the case of Bell nonlocality,
 the study of these polytopes plays a fundamental role in the investigation of noncontextuality inequalities.
We introduce the basic notions and discuss some results that have been achieved using 
 this approach. Correlation polytopes were introduced in the study of Bell inequalities by  \citet{Froissart:1981}, \citet{GM:1984}, and \citet{Pitowsky:1986JMP}; see also the book \citet{Pitowsky:1989}, which has been the most widely used reference. \citet{Horn:1948TAMS} had already provided a solution to the marginal problem many years earlier, which turned out to be equivalent to the correlation polytope approach; see \citet[][]{DeSimone:2015LAA}.
  In the context of noncontextuality inequalities, the first researchers to systematically use these notions were \citet{Klyachko:2008PRL}, 
and \citet{Kleinmann:2012PRL}.

\subsubsection{The simplest example}\label{sssec:polytope}
In this section, we introduce, in basic terms and by means of the simplest example, the notion of the correlation polytope. In Sec.~\ref{sssec:LP}, we  discuss their basic mathematical properties. 

In the case of a finite number of measurement settings and outcomes, a probability distribution is described by some positive numbers $p_i\geq 0$, $i=1,\ldots,n$ such that $\sum_i p_i=1$. We can interpret them geometrically as the set ${\mathcal{S}_n=\{ \pp \in \mathbb{R}^{n} | p_i\geq 0, \sum_i p_i=1\}}$, also known as a {\it simplex}: a $n-1$-dimensional polyhedron with $n$ facets and $n$ extremal points, i.e., the generalization of the triangle, tetrahedron, etc. 
Equivalently, it can be seen as the set of convex combinations of the elements of the canonical basis of $\mathbb{R}^n$, $\{ \ee_i\}_{i=1}^n$, namely $\mathcal{S}_n=\{\sum_i p_i \ee_i | p_i\geq 0, \sum_i p_i=1\} =: {\rm conv}(\{ \ee_i\}_i )$. Each extremal point $\ee_i$ can be interpreted as a probability assignment of $1$ to the $i$-th {\it event} and $0$ to the others. 

\begin{figure}[t]%
\centerline{\includegraphics[width=0.7\linewidth]{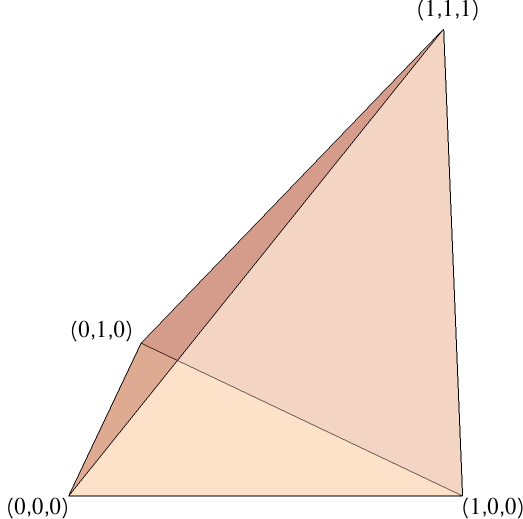}}
\caption{\label{Fig_poly2} Polytope associated with two measurements $A_1$ and $A_2$. The four vertices are the  deterministic assignments, with corresponding coordinate labeling. Equations (\ref{inepol2}) are associated with the four faces of the tetrahedron, for instance $p_{12}=0$ is the plane tangent to vertices $(0,0,0),(1,0,0)$, $(0,1,0)$, etc.}
\end{figure}

Ultimately, we want to represent probabilities of outcomes for a certain set of measurements, hence, each single 
event $i$ is associated with a specific sequence of outcomes. For instance, we may have the case of two 
measurements $A_1$ and $A_2$ with values $0$, or $1$. We then define events as $\{00, 01, 10, 11\}$ and their 
probabilities as $p_{00}:={\rm Prob}(A_1=0,A_2=0)$, 
 $p_{01} := {\rm Prob}(A_1=0,A_2=1)$, $p_{10}:={\rm Prob}(A_1=1,A_2=0)$  and $p_{11} := {\rm Prob}(A_1=1,A_2=1)$.  
It is straightforward to verify that the corresponding polytope has dimension 3 since there is an equality 
constraint (normalization of probability). We can then perform a linear transformation 
$(p_{00},p_{01},p_{10},p_{00}) \mapsto (p_1,p_2,p_{12})$, by computing the marginals, i.e., as 
$p_i = {\rm Prob}(A_i=1)$ and $p_{12}={\rm Prob}(A_1=A_2=1)$. The four vertices of the polytope are then the 
four vectors $p=(\varepsilon_1,\varepsilon_2,\varepsilon_1 \varepsilon_2)$ for 
$\varepsilon_1,\varepsilon_2 =0,1$, corresponding to the deterministic assignments of values to $A_1$ and $A_2$.
These vectors form the tetrahedron plotted in Fig.~\ref{Fig_poly2}.  It is straightforward to verify that the 
faces of the tetrahedron  are given by the following inequalities
\begin{equation}\label{inepol2}
\begin{split}
{p}_{12}\geq 0,\\
 {p}_1-{p}_{12}\geq 0, \\
 {p}_2-{p}_{12}\geq 0,\\ 
  1-{p}_1-{p}_2+{p}_{12}\geq 0,
 \end{split}
\end{equation}
which simply represent the constraints of positivity of the four probabilities ${\rm Prob}(A_1=x,A_2=y)$ for 
$x,y=0,1$ rewritten in terms of the marginals  $p_1$, $p_2$ and $p_{12}$.

\subsubsection{Basics of convex polytopes, affine geometry, and linear programming}\label{sssec:LP}

The general case of an arbitrary number of events is not  far from the previous simple one. Before proceeding, we first recall some basic notions about convex polytopes; for a more detailed exposition, see 
\citet{Grunbaum:2003}.
A convex polytope can be defined in two different ways.
In the vertex representation, one specifies the extremal points of the polytope, i.e.,
the vertices. If the set of vertices is finite, then the convex hull of those points is a 
convex polytope. The other way to specify a polytope is as the intersection of a finite 
family of closed half-spaces 
$H_i = \set{ \xx |\mm_i\cdot \xx\le  b_i}$. If the resulting set is bounded, it is also a convex polytope, 
otherwise, it is simply called a  polyhedron, or a polyhedral set.
An intersection of a polytope with an affine subspace (a section) and the image  
of a polytope under an affine map (such as a projection) both again yield a polytope.

A family of vectors $( \xx_1, \xx_2, \dotsc )$ is affinely independent if 
there is only a trivial solution to the equations 
$\sum_k \lambda_k \xx_k = \mathbf{0}$ and $\sum_k \lambda_k= 1$. Accordingly, the affine dimension of 
a family of vectors is $d= n-1$ if $n$ is  the maximal number of affinely independent vectors from the family.
A facet $F$ of a $d$-dimensional polytope is the intersection of an affine  
$(d-1)$-dimensional hyperplane with the polytope, so one of the open 
half-spaces defined by the hyperplane does not contain any part of the 
polytope. If the polytope is specified by a minimal set of closed half-spaces $\{H_i\}_i$, then these 
hyperplanes are $H_i\cap -H_i$. A facet is a $(d-1)$ polytope and the extremal points of the facet are exactly 
those extremal points of the polytope that belong to the facet.
Pitowsky's construction \cite{Pitowsky:1989} makes use of these facts:
The intersection of a half-space $H$ with a $d$ polytope $P$ is a facet of that 
polytope if and only if the extremal points of $P$ within $H$ span a 
$(d-1)$-dimensional affine subspace.

For the theory of Bell inequalities or noncontextuality inequalities 
the theory of linear optimization is central.
A {\it linear program} (LP) is an optimization problem of the type ``minimize 
$\cc\cdot \xx$ over $\xx\in K$''  where $\cc$ is a constant vector and $K$ 
is a polyhedral set, i.e., a finite intersection of closed half-spaces.
The set of optimal solutions again forms a polyhedral set or, if the set $K$ is a 
polytope,  is again a polytope.
If $K$ is specified by the vertices, then solving the program is simple, since 
the optimum is attained at one of the vertices.
The most important insight about linear programs is that, even if $K$ is 
specified as an intersection of half-spaces, the optimization can be solved by 
numerical means efficiently and with a certificate of optimality \cite{Boyd2004convex}.

\subsubsection{Noncontextuality inequalities}

In the following, we present an explicit construction of the correlation polytope based on the work of \citet{Pitowsky:1989}. Different, but ultimately equivalent, constructions are possible; see e.g., \citet{Abramsky2011} and \citet{AcinCMP2015}.  Given a set of observables $\set{A_i}_{i=1}^N$, we denote their possible value assignments as $\mathcal V= \mathcal V_1\times \mathcal 
 V_2\times \dotsm \times \mathcal V_N$, where $\mathcal V_k$ is the set of 
 possible values that the observable $A_k$ can assume; for instance, $\mathcal V_1= 
 \set{0,1}$ when $A_1$ is a dichotomic observable. The corresponding probability simplex is the convex hull of all assignments on the set $\mathcal{V}$, i.e.,  $\{ \pp \in \mathbb{R}^{|\mathcal{V}|}\ |\ p_{v}\geq 0, \sum_{v} p_v =1\}$, where $v=(v_1,\ldots,v_n)\in \mathcal{V}$ and $p_v={\rm Prob}(A_1=v_1,\ldots,A_n=v_n)$. 

The set of all possible contexts of $\set{A_i}_{i=1}^N$ defines the marginal scenario $\mathcal{M}$, i.e., the set of marginals that can be experimentally accessible, such as the pairs $\{A_i,A_{i+1}\}$ in the KCBS scenario~\cite{Klyachko:2008PRL}.  The correlation polytope, therefore ,is the projection of the probability simplex onto the coordinates corresponding to observable probabilities. To do so, we first need a change of coordinates. The new coordinates are obtained by considering the marginals $P(A_1=v_1), \ldots, P(A_i=v_i,A_j=v_j),\ldots,$.  Alternatively, one can choose any affine transformation (such as representations in terms of expectation values, correlators, etc.). An invertible affine transformation guarantees that the obtained conditions are still necessary and sufficient for a probability vector to have a noncontextual hidden-variable model. If the transformation is not invertible, one obtains only necessary conditions. Notice that not all coordinates of $\pp$ are independent, due to normalization ($\sum_{v} p_v =1$) and nondisturbing  conditions \cite{RamanathanPRL2012}; hence, invertibility must be checked with respect to this subspace. Moreover, such linear constraints together with the positivity of probability $p_v\geq 0$ for all $v$, define a new polytope, called the {\it nondisturbing polytope} \cite{RamanathanPRL2012}, which contains the noncontextuality polytope.

For instance, in the case of $V_i=\{0,1\}$ for all $i$, the correlation polytope associated with a marginal scenario $\mathcal{M}$ is the convex hull of vectors 
\begin{equation}\label{eq:ueps}
 \uu_\varepsilon = (\varepsilon_1,\ldots,\varepsilon_N,\ldots,\varepsilon_i\varepsilon_j,\ldots, 
 \varepsilon_{i_1}\varepsilon_{i_2}\cdots\varepsilon_{i_m},\ldots),
\end{equation}
where $\mathbf{\varepsilon}=(\varepsilon_1,\ldots,\varepsilon_n)\in\{0,1\}^N$ and $\varepsilon_i$ represents a $\{0,1\}$-valued assignment to $p(A_i):={\rm Prob}(A_i=1)$, and the marginals $p(A_i,A_j),\ldots$ are those appearing in the marginal scenario $\mathcal{M}$. These extreme points are precisely the projection of the extreme points of the simplex onto the subspace of observable probabilities.

Once these extreme points are defined, the corresponding noncontextuality inequalities can be obtained by computing the half-space representation of the polytope. There are several algorithms for performing this transformation  and several implementations of them, such as those given by \citet{cdd,lrs,porta,panda}. 

A noncontextuality inequality is of the form ${{\pmb \lambda} \cdot \pp\le 
 \eta}$, where the inequality holds true for any $\pp$ in the 
 noncontextuality polytope.
That is, a noncontextuality inequality is a half-space containing the 
 noncontextuality polytope. 
This inequality is useful only, if it can be violated by a 
 quantum system.
We write $\pmb{\Pi}$ for the vector of projectors $\pmb{\Pi}:=(P_1,\ldots,P_N, \ldots, P_i P_j,\ldots, P_{i_1} P_{i_2}\cdots P_{i_m}, \ldots)$, in analogy with Eq.~\eqref{eq:ueps}, such that 
the $i$th entry of the probability vector $\pp$ can be computed from the $i$th entry of $\pmb{\Pi}$ as $p_i=\tr(\varrho \Pi_i)$. In particular, this analogy requires that each projector (or product of projectors) is associated with an observable (or set of compatible observables), such that the structure of compatibility relations defined by the marginal scenario is reproduced by the projectors $P_1,\ldots,P_N$. For a nontrival noncontextuality inequality we have $\pmb{\lambda} \cdot \tr(\varrho {\pmb\Pi})>\eta$. 

The violation of an inequality for a fixed $\pmb\Pi$ is defined 
 as
\begin{equation}
 \Gamma(\pmb\lambda)= \frac{
  \max\set{\pmb\lambda\cdot \tr(\varrho\pmb\Pi) | \varrho \text{ quantum 
 state}}
 }{
  \max\set{\pmb\lambda\cdot \pp | \pp \text{ in  NC
 polytope}}
 }-1.
\end{equation}
Hence $\Gamma= 0$ corresponds to the situation where the inequality does not 
 have a violation for the projectors $\pmb\Pi$.
Note, that both maximizations may be restricted to the extremal points, i.e., 
the maximization for the quantum value can be performed over the pure states, 
while for the noncontextuality polytope, it is sufficient to consider all 
extremal points of the polytope. 
As a consequence, the validity of a noncontextuality inequality $\mm \cdot \xx \leq b$ can be checked by verifying that it is not violated by any vertex of the polytope, for the vertex representation ${\rm conv}(v_1,\ldots,v_k)$, and by linear programming if the description of the polytope is given in terms of half-spaces $\{ \xx | A\xx\leq \bb\}$, i.e., as $\max_\xx \mm \cdot \xx$ subject to $A\xx\leq \bb$.

Similarly, given a probability vector $\pp$, one can check to see if it belongs to a given noncontextuality polytope via a LP. If not, this LP provides, via its dual formulation, a noncontextuality inequality violated by $\pp$. One example of this is given by the {\it contextual fraction} ($\mathsf{CF}$) LP \cite{Abramsky:2011NJP,Abramsky:2017PRL,Amselem:2012PRL}, which we encountered in Sec.~\ref{log_str_cont} in connection with the notion of {\it strong contextuality}.  It is instructive to repeat its definition here, in order to directly connect it to the noncontextual polytope. In simple terms, the noncontextual fraction ($\mathsf{NCF}=1-\mathsf{CF}$) is the maximum  $\alpha\in [0,1]$ such that $\pp$ can be decomposed as $\pp = \alpha \pp_{\rm NC} + (1-\alpha) \pp_{\rm C}$, where $\pp_{\rm NC}$ is a vector belonging to the noncontextuality polytope and $\pp_{\rm C}$ a vector belonging to the nondisturbing polytope. Since both $\pp_{\rm NC}$ and $\pp_{\rm C}$ can be characterized in terms of LP,  that the noncontextual fraction can be computed as a LP. The contextual fraction can also be interpreted as a geometric quantification of contextuality. In Sec.~\ref{res_th}, we discuss its role in the resource theory of contextuality. 
Several related questions are addressed in the following sections, such as the identification of the interesting contextuality scenarios, i.e., giving rise to some $\Gamma > 0$, or even SI-C scenarios, the computation of quantum bounds, etc.

An analogous approach, based on ideas on convex optimization, polyhedral sets and linear programming, can be 
developed for the analysis of entropy, rather than probability. Following the idea initially developed by
 \citet{Braunstein:1988PRL},
 \citet{Chaves:2012PRA, FritzIEEE13,RaesiPRL2015,Kurzynski:2012PRL, Durucan:2020XXX, Chaves:2013PRA} investigated entropic noncontextuality inequalities. In 
particular, \citet{Chaves:2012PRA}  and \citet{FritzIEEE13} developed a systematic method to derive noncontextuality 
inequalities for an arbitrary marginal scenario, which can be  described as follows. In the entropic approach, one can derive 
the entropic inequalities by projecting the entropic cone, describing the joint entropies over all variables,
onto the variables corresponding to the observed marginals, in analogy with the projection of the previously described probability simplex. A complete characterization of the entropy cone is 
not known for more than three variables. However, an outer approximation in terms of the 
so-called Shannon inequalities is known; see \citet{Yeung2008} for an introduction. In 
contrast to the probability case, entropic inequalities provide only a necessary condition for noncontextuality, 
except in some special cases~\cite{Chaves:2013PRA}. 

Finally, a case not covered in the correlation polytope approach is the continuous-variable (CV) case. The first proposal of a CV contextuality test was presented by \citet{Plastino:2010PRA} for a CV version of the PM square based on modular variables. The argument was further improved by \cite{AsadianPRL2015}, removing one assumption from the NCHV model (classical complex variables of modulo $1$). The same scenario was further explored by \cite{Laversanne_Finot:2017JPA}, who considered more general observables. More recently, \citet{Barbosa:2019XXX} presented a general framework for the investigation of CV contextuality.


\subsection{Graph theory and contextuality}\label{graphth}

Since the original paper of \citet{Kochen:1967JMM}, graphs have played a central role in contextuality arguments. In the following, we discuss the connection between contextuality and graph 
theory, with particular emphasis on two types of graphs, namely, {\it compatibility graphs} and {\it exclusivity graphs}. We review several problems that can be formulated in terms of graph properties and graph-theoretic results. This comprises
the following questions: Which compatibility structures always admit a noncontextual hidden-variable model? Or, equivalently, which structures are interesting for contextuality? How can we derive noncontextuality inequalities and compute the corresponding quantum bound efficiently? Which scenarios give rise to state-independent contextuality? 


\subsubsection{Basic notions}\label{sssec:graph_basic}
We start by introducing basic notions and definitions in graph theory. Extensive discussions of this topic were given by in 
\citet{Diestel2018,Bretto2013,Beeri1983, Lauritzen1996}. A {\it graph} is a pair $G=(V,E)$ where $V$ is the set of vertices, or nodes, and $E$ is the set of edges, i.e., unordered pairs $(i,j)$ for some $i,j\in V$. Two vertices $i,j\in V$ of a graph are {\it adjacent}, or {\it connected}, if $(i,j)\in E$. A set of mutually connected vertices is called a {\it clique} of the graph. 
A set of vertices such that no two of them are connected is called an {\it independent set}. A {\it path} is a sequence of distinct vertices $v_0,\ldots, v_n$ such that $v_i$ is connected to $v_{i+1}$, for $i=0,\ldots,n-1$. A {\it cycle} is defined in the same way, but with $v_0=v_n$. A graph is an {\it acyclic}, or a {\it tree}, graph, if it contains no cycle. A graph is {\it triangulated}, or {\it chordal}, if every cycle of length $n\geq 4$ contains a {\it chord}, i.e., an edge connecting $(v_i,v_{i+2})$. The {\it complement} of a graph $G=(V,E)$ is a graph $\bar{G}=(V,\bar{E})$ where $\bar{E}=\{(i,j)|i,j \in V \} \backslash E$, i.e., every pair of connected vertices in $G$ is disconnected in $\bar{G}$ and vice versa.


\begin{figure}[t]
\begin{center}
 \includegraphics[width=0.23\textwidth]{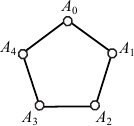}
\end{center}
\caption{\label{fig:p_com} Compatibility graph associated with the observables of the KCBS scenario corresponding to the marginal scenario $\{(A_i,A_{i+1}) \}_{i=0}^4$. This graph can also be used to illustrate the basic notions of path, cycles, and independent sets. A path is given by any sequence of sequentially connected vertices, such as $(A_i, A_{i+1}, A_{i+2})$, for any $i$ and with sum modulo $5$. A cycle is a closed path such as $(A_0,A_1,\ldots,A_4,A_0)$. An independent set is a set of disconnected vertices such as $(A_i, A_{i+2})$. The pentagon contains independent sets of at most size 2. One can easily show that the complement of a pentagon is again a pentagon with edges $(A_i, A_{i+2})$, for $i=0,\ldots,4$ and sum modulo $5$.}
\end{figure}


A {\it hypergraph} is a generalization of the above idea obtained by allowing edges to connect more than two vertices, namely, a pair $H=(V,E)$, where $V$ is the set vertices and $E$ the set of hyperedges, i.e., $E\subset 2^{V}$, with $2^{V}$ the power set of $V$. Hypergraphs can also arise from graphs; for instance, the clique hypergraph $H$ of a graph $G$ is defined by the same set of vertices and has as hyperedges the cliques of $G$. If a hypergraph contains only maximal hyperedges, i.e., for each hyperedge $E$ there is no hyperedge $E'$ such that $E'\subset E$, the graph is said to be {\it reduced}. Given a hypergraph $H$, we say that $H'$ is the {\it reduced hypergraph} of $H$ if it is obtained from $H$ by removing all nonmaximal hyperedges.

As opposed to the case of graphs, different notions of acyclicity are possible for hypergraphs. The relevant one for us is given by the following two equivalent definitions. First, a hypergraph is acyclic if it has the {\it running intersection property}, i.e., 
if there exists an ordering of the hyperedges, $E_1,\ldots,E_n$, such that
\begin{equation}
E_i \cap (E_1\cup \cdots \cup E_{i-1})\subset E_j, \text{ with } j<i, \text{ for all } i.
\end{equation}
Namely, there exists an ordering such that the intersection with any new hyperedge is completely contained in one of the previous hyperedges. Second, an equivalent definition is that a hypergraph is acyclic if it is the clique hypergraph of a triangulated graph.

Their equivalence is not obvious, see ~\citet{Beeri1983, Lauritzen1996}; however, one can easily verify that these definitions coincide in the case of hyperedges of cardinality 2 with that of trees for graphs. We see here that the running intersection property plays a central role in the construction of NCHV models.
This notion is usually called $\alpha$ {\it acyclicity} in the literature \cite{Beeri1983, Lauritzen1996}. In the following, we  refer to it simply as acyclicity.

\subsubsection{Graphs, hypergraphs, and marginal scenarios}
\label{sssec:graph_hyp}

In the abstract formulation of NCHV in Sec.~\ref{sssec:NCHV}, we defined a marginal scenario $\mathcal{M}$ as the set of of all contexts for a given set of measurements $A_1,\ldots,A_n$. A natural representation of a marginal scenario is given by a hypergraph $H$: vertices represent measurements, whereas hyperedges represent contexts; see also \citet{Amaral2018Book,AcinCMP2015}. Here we consider the most general structure possible without entering into the details of the specific way of realizing such contexts in practice, as discussed in Sec.~\ref{ssec:compatible}. Given its relevance, we often discuss the specific case of sharp measurements. For sharp measurements in quantum mechanics, \textit{Specker's principle} applies~\cite{Specker:1960D, Kochen:1967JMM,Cabello:2012XXX}, namely that 
pairwise compatibility is equivalent to global compatibility. For this reason for sharp measurements it is
enough to represent the marginal scenario as a graph, interpreting edges as pairwise compatibility relations and cliques as contexts. For the case of sharp measurements, we call such graphs {\it compatibility graphs}. 
It is interesting to notice that any graph can be interpreted as such a compatibility graph for sharp measurements, 
in the sense that these compatibility relations can be realized by a set of sharp observables on a Hilbert 
space \cite{HeunenPRA2014}. Similarly, if one considers contexts simply as sets of jointly measurable 
observables, then 
every hypergraph can be interpreted as a set of joint-measurability relations for a given set of 
POVMs~\cite{KunjwalPRA2014b}. We recall that we discussed in Sec.~\ref{ssec:compatible} the problems associated with possible definitions of contexts requiring joint-measurability alone.

Notice that the previous notion of compatibility hypergraph should not be confused with the hypergraph approach of \citet{AcinCMP2015}, who instead represented effects as nodes and some results, such as the identification of cliques with contexts in the case of sharp measurements, do not hold.
An example of a compatibility graph is given in Fig.~\ref{fig:p_com} for the 
KCBS scenario. Each vertex represents a measurement setting $A_0,\ldots,A_4$, and edges connect vertices 
corresponding to two joint measurement $\mean{A_i A_{i+1}}$ appearing in the KCBS inequality~\cite{Klyachko:2008PRL}. 

\citet{BudroniMorchio10, Kurzynski:2012PRL, RamanathanPRL2012} investigated graph-theoretical properties of the marginal scenario hypergraph (or the compatibility graph for sharp measurements) that directly imply the existence of a NCHV regardless of the value of the observed correlations. These represent special cases of a general result for marginal scenario hypergraphs that follows from a theorem by Vorob'ev (or Vorob'yev, depending on the transliteration from the Cyrillic alphabet used), which can be stated in our terminology as follows:
\begin{theorem*}[\citealp{Vorobev1962}]
Any marginal scenario represented by an acyclic hypergraph admits a joint probability distribution. 
\end{theorem*}
The theorem was originally stated by \citet{Vorobev1959} [translate into English as~\citet{Vorobev1967coal}] and later proven by \citet{Vorobev1962}; see also \citet{Vorobev1963markov}. The same result was also independently proven by  \citet{ Kellerer1964a, Kellerer1964b, Malvestuto88}.

The Vorob'ev theorem says that, given a set of probabilities associated with a marginal scenario and coinciding on their intersection, if their structure is represented by an acyclic hypergraph, then there is always a probability distribution for which they are the marginals. Intuitively, Vorob'ev's result can be understood as the construction of a global probability by ``gluing together'' probability distributions on their intersection, a notion referred to as ``adhesivity''~\cite{Matus_DM_2007}. The acyclicity property of hypergraphs, particularly the running intersection property, guarantees that such a construction can always be made in a consistent way. It is instructive to illustrate this idea with the simplest example, which follows. Consider three variables $A,B$ and $C$ and two distribution $p_1(a,b)$ and $p_2(b,c)$, such that $\sum_a p_1(a,b)=\sum_c p_2(b,c)=:p(b)$. This corresponds to a marginal scenario described by a line graph $A-B-C$, which is acyclic. One can explicitly construct a joint distribution on $A,B$ and $C$ 
by ``gluing'' the distributions on their intersection, namely
\begin{equation}
p(a,b,c):= \frac{p_1(a,b) p_2(b,c)}{p(b)}=p_1(a|b) p_2(c|b) p(b),
\end{equation} 
with the convention that $p(a,b,c):=0$ if $p(b)=0$. This is precisely the construction used by \citet{Fine:1982PRL} to prove that CHSH inequalities are necessary and sufficient conditions for the existence of local hidden-variable models in the case of two inputs and two outputs.

Vorob'ev's result has been  discussed in relation to contextuality~\cite{Barbosa2014,BarbosaPhD,Xu:2019PRA} and causal discovery methods~\cite{BudroniPRA2016}. This result has implication for the computation of correlation polytopes and entropic cones associated with noncontextuality scenarios \cite{BudroniJPA2012,Araujo:2013PRA,Kujala:2015PRL} and more general causal structures \cite{Chaves2014, BudroniPRA2016}. Notice that such a result is also at the basis of the derivation of non-Shannon inequalities in classical information theory \cite{Zhang2003,Matus2007}.

For the case of compatibility graphs, it is sufficient to verify that the graph is triangulated, since the corresponding hypergraph of contexts, the clique hypergraph, is acyclic according to the previous definition; see the discussion given by~\citet{Xu:2019PRA} for additional details. 
The previous result has allowed for the identification of the simplest noncontextuality scenarios. The argument presented by~\citet{Kurzynski:2012PRL} can be  summarized as follows. The simplest compatibility graph giving rise to contextual correlations must contain a cycle with a length larger than 3; i.e., it is a square corresponding to the CHSH scenario \cite{Clauser:1969PRL}. For sharp measurements, such a graph can be obtained only in dimension $d=4$. For $d=3$, one needs at least a pentagon, which precisely corresponds to the KCBS scenario~\cite{Klyachko:2008PRL}. This can be
seen as follows: A nontrivial sharp measurement in $d=3$ is represented by the POVM $\{\ketbrac{v}, \openone - \ketbrac{v}\}$ since it cannot be the identity and it cannot be nondegenerate, otherwise compatibility becomes a transitive relation \cite{Correggi2002}, giving rise only to a collection of fully connected graphs, hence, always admitting a NCHV. The compatibility of two measurements, with associated rank-1 projectors $\ketbrac{v}$ and $\ketbrac{w}$, corresponds to $\mean{v|w}=0$. Since two nonorthogonal vectors in $d=3$ have a unique orthogonal subspace, it is impossible to get a square 
compatibility graph with four different measurements.

\subsubsection{Exclusivity graphs and their independence, Lov\'asz, and fractional packing numbers}\label{ssec:graph_CSW}

In this section, we introduce the notion of an exclusivity graph (namely a graph where connected vertices represent mutually exclusive events), and discuss the significance of associated graph-theoretical quantities such as the {\it independence number}, {\it Lov\'asz number}, and {\it fractional packing number}, following the discussion presented by \citet*[CSW,][]{CSWPRL2014}.

Given a measurement context, denoted by compatible settings $(s_1,\ldots,s_n)$, an {\it event} corresponds to a given set of joint outcomes $(o_1,\ldots,o_n|s_1,\ldots,s_n)$. Two events $(o_1,\ldots,o_n|s_1,\ldots,s_n)$ and $(o'_1,\ldots,o'_n|s'_1,\ldots,s'_n)$ are said to be {\it exclusive} if there are  $i$ and $j$ such that $s_i=s'_j$ but $o_i\neq o'_j$. In other words, two events are exclusive if at least one pair of measurement settings coincides, but they have different outcomes.

\begin{figure}[t]
\begin{center}
\includegraphics[width=0.35\textwidth]{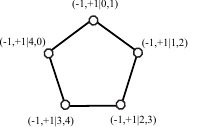}
\end{center}
\caption{\label{fig:p_ex} Exclusivity graph associated with the five events appearing in the inequality \eqref{eq:k_ev}. The notation $(-1,+1|0,1)$ refers to the event of outcome $-1$ for the measurement of $A_0$ and outcome $+1$ for the measurement of $A_1$, etc.}
\end{figure}

It is helpful to consider in detail a simple example given by the graph in Fig~\ref{fig:p_ex}: each vertex represents two possible outcomes for two settings, and two vertices are connected by an edge if the corresponding events are mutually exclusive. Such a graph represents a version of the KCBS noncontextuality, namely Eq.~\eqref{eq:kcbs-exc_tr} (discussed in Sec.~\ref{sssec:NCHVE}),
\begin{equation}\label{eq:k_ev}
S_{_{\rm KCBS}}=\sum_{i=0}^4 p(-1,+1|i,i+1)\stackrel{\rm NCHV}{\leq} 2,
\end{equation}
where $p(-1,+1|i,i+1)\equiv {\rm Prob}(A_i=-1,A_{i+1}=1)$ and the sum is taken modulo $5$.

For the specific choice of quantum observables in the KCBS scenario discussed in Sec.~\ref{sssec:state-dep}, such events are represented by projectors, such as $(-1,+1|i,i+1) \mapsto Q_i$ and 
$p(-1,+1|i,i+1)=\tr(\rho Q_i)$, where $Q_i:=\Pi_i^+ \Pi_{i+1}^-$ and $\Pi_i^\pm$ is the projector associated with the outcome $\pm 1$ of $A_i$. Mutually exclusive 
events correspond to orthogonal projectors, such as $Q_i Q_{i+1}=(\Pi_i^+ \Pi_{i+1}^-)(\Pi_{i+1}^+ \Pi_{i+2}^-)=0$. 

\citet{CSWPRL2014} noticed the similarity between Eq.~\eqref{eq:k_ev}
and the definition of the 
Lov\'asz number of a graph. The Lov\'asz number was  introduced by \citet{Lovasz79} as an upper bound on the Shannon capacity of a graph \cite{Shannon1956}. It is a well-studied object in graph theory, and it can be efficiently computed via semidefinite programming~\cite{Lovasz2009}. For a graph $G=(V,E)$ its Lov\'asz number $\vartheta$ is given by
\begin{equation}\label{eq:lov}
\vartheta(G)=\max_{v_i,\psi} \sum_{i\in V} |\mean{\psi|v_i}|^2,
\end{equation}
where the maximum is take over all vectors $\ket{\psi}$ and over all vectors $\ket{v_i}$ such that 
$\mean{v_i|v_j}=0$ whenever $i,j\in V$ are adjacent vertices. This set of vectors is also called 
an {\it orthogonal representation} (OR) of $\overline{G}$, the complement of $G$. Notice that the OR of a graph is defined by the fact that nonadjacent nodes are associated with vectors that are orthogonal~\cite{Lovasz2009}, which is why we consider the complement graph $\overline{G}$ to define the OR entering Eq.~\eqref{eq:lov}. This seemingly counterintuitive convention might be due to the original definition of Shannon capacity of a confusability graph \cite{Shannon1956}, see Sec.~\ref{sec:zero_err} for more details, where nodes represent symbols of an alphabet and edges their ``confusability'', whereas in our case edges represent exclusivity. For this reason, some researcher prefer instead to work directly with the complement graph, i.e., the nonorthogonality graph~\cite{AcinCMP2015}.

The maximum of the expression $S_{_{\rm KCBS}}$ in QM can in fact be written as
\begin{equation}\label{eq:kcbslova}
\max_{\rho,Q_i} \sum_i \tr{(\rho Q_i)}=\max_{v_i,\psi} \sum_{i\in V} |\mean{\psi|v_i}|^2=\vartheta(G),
\end{equation}
where each vertex of the graph $G=(V,E)$ corresponds to a projector appearing on the lhs~of 
(\ref{eq:kcbslova}) and two vertices are adjacent if the corresponding projectors are orthogonal. Notice that with the previous definition of the Lov\'asz number, the quantum maximum of $S_{_{\rm KCBS}}$ is given by the Lov\'asz number of the exclusivity graph. Moreover, the use of a pure state $\ket{\psi}$ instead of $\rho$ is no restriction since, by a convexity argument, the maximum of $S_{_{\rm KCBS}}$ is always achieved by pure states. Similarly, the use of one-dimensional projectors $\ketbrac{v_i}$ is no restriction, since for an arbitrary projector $Q_i$, we have $\mean{\psi |Q_i|\psi}= |\mean{\psi|v_i}|^2$ where $\ket{v_i}:=Q_i \ket{\psi}/\sqrt{\mean{\psi|Q_i|\psi}}$.

In addition to the Lov\'asz number, two other graph-theoretical quantities are central to the discussion on correlation bounds in different theories: the independence number $\alpha$ and the fractional packing number $\alpha^*$. The former is defined as the cardinality of the maximal independent set of a graph, which can be interpreted as the maximum number of $1$'s, i.e., as logically true, that can be assigned to a set of vertices without violating the exclusivity condition, namely
\begin{equation}
\begin{split}\label{eq:inde_num}
&\alpha(G) =\max_{c_i} \sum_{i\in V} c_i, \\
&\text{ such that } c_i=0,1,\ c_ic_j=0 \text{ if } (i,j)\in E.
\end{split}
\end{equation}
In terms of the NCHV models with additional exclusivity constraints (NCHV+E) discussed in Sec.~\ref{sssec:NCHVE}, $\alpha$ can be interpreted as the maximum of deterministic assignments that respects the exclusivity condition, namely, that two adjacent vertices cannot both be assigned  the value $1$.

The fractional packing number is a linear program relaxation of the independence number, namely, the maximum sum of weights such that in every clique the sum of weights is $1$,
\begin{equation}
\begin{split}\label{eq:frac_pack}
\alpha^*(G) = \max_{p_i} \sum_{i\in V} p_i,\text{ such that } p_i\geq 0,\\
 \text{ and } \sum_{i\in C} p_i \leq 1 \text{ for all cliques } C.
\end{split}
\end{equation}
The interpretation is as follows. Probabilities for single events are identified regardless of the context and the sum of probabilities of exclusive events within each context is less than or equal to $1$. This can be interpreted as a bound for {\it generalized probability theories} (GPTs) that still respect some notion of exclusivity within each context, i.e., sum of probabilities below one. We return to this notion of exclusivity later.

In summary, the different graph-theoretical quantities, i.e., $\alpha$, $\vartheta$, and $\alpha^*$, provide information on the bounds on correlations for different theories, classical, quantum, and generalized probability theories, respectively. For the expression $S_{_{\rm KCBS}}$ in Eq.~\eqref{eq:k_ev}, we know that the Lov\'asz number provides a tight bound, i.e., it can be achieved in quantum mechanics. More precisely, in the KCBS case there are sharp measurements $A_0,\ldots,A_4$, with $A_i= \{\Pi_i^+,\Pi_i^-\}$ such that the events $(+1,-1|i, i+1)$ are exclusive and the rank-1 projectors $\ketbrac{v_i}$ maximizing Eq.~\eqref{eq:lov} are given by $\ketbrac{v_i}= \Pi_i^+ \Pi_{i+1}^-$, as discussed in Sec.~\ref{sssec:state-dep}. 

Depending on the specific assumptions on the measurement scenario, however, the bounds obtained by the Lov\'asz number may not be tight. A typical example is the pentagon \cite{Sadiq:2013PRA}, which is interpreted as the exclusivity graph of a subset of events in the CHSH scenario, i.e., of the form $(a,b|x,y)$, with $x$ the setting of Alice and $y$ of Bob. The reason why this bound is not tight is that in order to interpret these events in the CHSH scenario, we need additional compatibility constraints on the measurements, in order for them to be distributed between two parties. In other words, Alice's observables are compatible with Bob's, a condition that is not encoded in the exclusivity graph. A possible extension of the exclusivity graph approach to nonlocality scenarios via multigraphs, encoding the separation into different parties, was proposed by \citet{RabeloJPA2014}.

This situation, however, is not specific to Bell scenarios, but happens also for contextuality scenarios if additional assumptions on the compatibility relations among measurements are made. In other words, the situation occurs if one wants to reconstruct not only the effect operators but also the original observables and their compatibility relations. This is similar to what happens in the Navascu\'{e}s-Pironio-Ac\'{i}n (NPA) characterization of multipartite quantum correlations~\cite{NPA_PRL,NPA_NJP}, and this is the reason why one needs to define a hierarchy of SDP conditions rather than a single one. In fact, even if the single operators $\ketbrac{v_i}$ can be reconstructed by the Lov\'asz number SDP, it is not clear that one can reconstruct the observables, such as $\{A_{a|x}\}_{a,x}$ and $\{B_{b|y}\}_{b,y}$ associated with the events $(a,b|x,y)$ in a Bell scenario, of which they are assumed to be effects, with the correct compatibility (in this case, commutativity) relations among them. 

An alternative approach involves taking the notion of observables and 
contexts as our starting point and developing from there the exclusivity 
relations: given the observables $\{A_{o|s}\}_{o,s}$, one constructs all possible events, i.e., for each context $C$ all the events $p(o_1,\ldots, o_{|C|}| s_1,\ldots, s_{|C|})$, constituting the nodes of 
the hypergraph, whereas the hyperedges are defined using the previously mentioned exclusivity relations
(at least two identical settings with different associated outcomes). 
The hypergraph-theoretic approach to contextuality introduced and extensively 
investigated by \citet*[AFLS,][]{AcinCMP2015} describes sets of exclusive events precisely keeping track of this structure. More precisely, in the AFLS approach,  the hypergraph of effects (nodes) and exclusivity relations (hyperedges) keeps information on which collection of effects correspond to the measurements performed in the specific physical situation considered. In contrast, in the CSW approach, one starts from the effects (nodes) and their exclusivity relations (edges) and tries to construct a general noncontextuality inequality, as we later explain.

We now discuss how one can find noncontextuality inequalities 
in the graph approach. Here, one starts from an exclusivity graph, such as one for which it is known that $\alpha(G)<\vartheta(G)$, and interprets it as a compatibility graph, i.e. promotes each single event to a measurement. An associated noncontextuality inequality can be constructed such that the classical and quantum bounds correspond to the independence and Lov\'asz number, respectively. A general method was presented in the original paper \cite{CSWPRL2014}, however, here we discuss a slightly different (and arguably simpler) approach since we already encountered it in the KCBS example in Sec.~\ref{sssec:NCHVE}. This approach is based on a general method to transform KS inequalities into NC inequalities; see \citet{Yu-Tong14,CabelloPRA2016} for additional details. 

We assume that we have a graph $G$ such that 
$\alpha(G)<\vartheta(G)$ and want to construct a noncontextuality inequality and  provide a state and sharp quantum observables, with the correct compatibility relations, able to show a violation of the inequality. To construct the NC model, with each node of the graph $G=(V,E)$ we associate a classical variable $P_i$ with values in ${0,1}$. We write ${\rm Prob}(P_i=1)=:\mean{P_i}$ and the joint probability ${\rm Prob}(P_i=1, P_j=1)=: \mean{P_i P_j}$. From the independence number we can derive the following bound for NCHV models with the additional exclusivity assumption among connected events (see Sec.~\ref{sssec:NCHVE}), namely,
\begin{equation}\label{eq:alpha_e}
\sum_{i\in V} \mean{ P_i } \stackrel{\rm NCHV+E}{\leq} \alpha(G).
\end{equation}
The meaning of Eq.~\eqref{eq:alpha_e} is that the bound of $\alpha(G)$ is valid only in NCHV models where events satisfy additional exclusivity relations, corresponding to those encoded in the graph $G$, namely, connected nodes cannot  both be assigned the value $1$. Following the discussion in Sec.~\ref{sssec:NCHVE}, we transform the Eq.~\eqref{eq:alpha_e} into a general noncontextuality inequality as follows
\begin{equation}\label{eq:nc_alpha}
\sum_{i\in V} \mean{ P_i } - \sum_{(i,j)\in E} \mean{P_i P_j} \stackrel{\rm NCHV}{\leq} \alpha(G).
\end{equation}
Intuitively, whenever the noncontextual assignments do not respect the exclusivity condition the lhs gets a penalty that keeps the noncontextual bound the same. We denote by $\mathcal{A}\subset\{P_i\}_i$ the subset of variable to which $1$ is assigned. It can be divided into an assignment to a maximal independent set $\mathcal{I}$ plus some extra variables $\mathcal{E}$, i.e., $\mathcal{A}=\mathcal{I}\cup\mathcal{E}$. Each variable $P_i\in\mathcal{E}$, however, must violate at least one exclusivity constraint involving an element of $\mathcal{I}$ since $\mathcal{I}$ is a maximal independent set by definition, thus  giving a factor $+1$ for the first term and a factor $\leq -1$ for the second term on the lhs of Eq.~\eqref{eq:nc_alpha}. The quantum model can be constructed from the OR of 
$\overline{G}$ as in Eq.~\eqref{eq:lov}, namely, vectors $\{ \ket{v_i}\}_i$ and a state $\ket{\psi}$ such that $\mean{v_i | v_j}=0$ if $(i,j)\in E$ and 
\begin{equation}
\sum_i |\mean{\psi |v_i}|^2 = \vartheta(G).
\end{equation}
By constructing the POVMs $\tilde P_i =\{\ketbrac{v_i}, \openone- \ketbrac{v_i}\}$ and considering the initial state $\rho=\ketbrac{\psi}$, one obtains $\mean{\tilde P_i}_\rho= |\mean{\psi |v_i}|^2$ and 
$\mean{\tilde P_i \tilde P_j}_\rho =\bra{\psi} \tilde P_i \tilde P_j \ket{\psi}=0$, whenever $(i,j)\in E$. As a consequence, one obtains
\begin{equation}
\sum_{i\in V} \mean{ \tilde P_i } - \sum_{(i,j)\in E} \mean{\tilde P_i \tilde P_j}_\rho = \sum_i |\mean{\psi |v_i}|^2 =\vartheta(G) > \alpha(G),
\end{equation}
giving a violation of the noncontextuality inequality in Eq.~\eqref{eq:nc_alpha}. 

In addition to classical, quantum and GPT bounds for a given expression, the exclusivity graph approach also allows for the definition of the set of their correlations through the notions of  stable set polytope STAB$(G)$,   theta body  TH$(G)$, and  clique constrained stable set polytope QSTAB$(G)$ of a given exclusivity graph $G$; see \citet{CSWPRL2014,Amaral2018Book} for a detailed discussion. These sets are closely related to the previously defined quantities $\alpha, \vartheta$ and $\alpha^*$: 
\begin{align}
&{\rm STAB}(G) = {\rm conv}\left\lbrace x\in\{0,1\}^{|V|}\ \Big| \ x_ix_j = 0 \text{ if }(i,j)\in E\right\rbrace,
\nonumber \\
&{\rm TH}(G) =\left\lbrace p\in \mathbb{R}_+^{|V|}\ \Big|\ p_i=|\mean{\psi|v_i}|^2, \{ \ket{v_i}\}_i \text{ OR of } \overline{G} \right\rbrace,
\nonumber \\
&{\rm QSTAB}(G) = \Big\lbrace p\in \mathbb{R}_+^{|V|}\ \Big|\ \sum_{i\in C} p_i \leq 1, \ \forall \text{ cliques } C\Big\rbrace.
\end{align}
In other words, STAB$(G)$ is given by the probability vectors in the convex hull of the deterministic assignments respecting exclusivity, i.e., of $1$ to all the elements of an independent (or stable) set and $0$ to the other elements, as in Eq.~\eqref{eq:inde_num}; TH$(G)$ is given by the assignment coming from a vector $\ket{\psi}$ and the vectors of an orthogonal representation of $\overline{G}$, as in Eq.~\eqref{eq:lov}; and QSTAB$(G)$ is the set of probability assignments such that the sum of probability on each clique is bounded by $1$ as in Eq.~\eqref{eq:frac_pack}. These sets can also be characterized in terms of the quantities $\alpha,\vartheta$ and $\alpha^*$, but in   reverse order with respect to what we have seen  \cite{AcinCMP2015}, arising from a dual approach in their description~\cite{Grotschel1993,Knuth1994}. 

The notion of the stable set polytope STAB$(G)$
and the theta body TH$(G)$ also allowed the minimal
Greenberger-Horne-Zeilinger-like proof of contextuality to be found. This 
was then shown to imply that the 18 vectors found by \citet{Cabello:1996PLA}, see also Fig.~(\ref{fig:cab18}), is the minimal Kochen-Specker set \cite{Xu:2020PRL}.

The set QSTAB$(G)$ encodes the condition that the probability of mutually exclusive events (represented by a clique in $G$) is bounded by $1$, a condition that was introduced under the name consistent exclusivity principle for contextuality \cite{Cabello:2012XXX} [see also \cite[][]{Henson:2012XXX,AcinCMP2015}],  or {\it E}-principle \cite{Cabello:2013PRL}), and {\it local orthogonality} for Bell nonlocality~\cite{Fritz2013}. This condition has been extensively investigated as a possible principle that bounds contextual and nonlocal correlations in QM \cite{AcinCMP2015,Yan:2013PRL,Fritz2013,AmaralPRA2014,Cabello:2013PRL, Cabello:2015PRL,Henson:2015PRL}.

In the hypergraph approach, \citet{AcinCMP2015} showed that consistent exclusivity cannot bound the set of quantum correlations, even in the limit of an infinite number of copies of the original hypergraph, with the following argument. They showed that the set of probability vectors, i.e., with the normalization condition $\sum_i p_i =1$, obtained in this limit is the one characterized to the Shannon capacity of a graph \cite{Shannon1956}, which includes the set ${\rm TH}(G)\cap\{ \sum_i p_i =1\}$, i.e., the theta body with extra normalization constraints. As discussed above for the CHSH case, when the scenario constraints are imposed, i.e. each event is associated to a collection of outcomes for a joint measurement, the Lov\'asz number provides only an upper bound to quantum correlations. Moreover, \citet{AcinCMP2015} show that the set ${\rm TH}(G)\cap\{ \sum_i p_i =1\}$ corresponds to the first level of a NPA-type hierarchy associated with the hypergraph; see also \citet{Navascues:2015NC}. 

By not fixing the measurement scenario and simply discussing  events and their exclusivity relations, the graph approach provides a different perspective on the derivation of quantum bounds on correlations. The results of this research direction are summarized in Sec.~\ref{sssec:quest_principle}.


\subsubsection{The graph approach and the quest for a principle for quantum correlations}\label{sssec:quest_principle}


In the context of the program initiated by \citet{Cirelson:1980LMP}  [or Tsirelson depending on the transliteration; see also \citet{Tsirelson:1993HJS}] for finding simple characterizations of the sets of quantum correlations for Bell scenarios, \citet{Popescu:1994FPH} asked the following question: Why are correlations in nature not more nonlocal? 
Principles such as nontrivial communication complexity \cite{vanDam:1999}, information causality \cite{Pawlowski:2009NAT}, macroscopic locality \cite{Navascues:2010PRSA}, and local orthogonality \cite{Fritz2013} managed to exclude some nonquantum nonlocal correlations. However, none of them managed to single out even the set of quantum correlations for the simplest Bell scenario \cite{Navascues:2015NC}. 

A different approach to the problem of finding a principle for quantum correlations is the observation that quantum theory, understood as the abstract probability theory behind quantum mechanics by \citet{Hardy:2001XXX,Chiribella:2010PRA,Masanes:2011NJP}, can be seen as a probability theory for events produced by ideal measurements. 
This follows from two observations.

On the one hand, not only is a self-adjoint operator a tool to compute the probabilities of an observable,
it also represents an ideal measurement of the observable, i.e., a measurement that does not disturb any
compatible observable and yields the same result when
repeated \cite{Kleinmann:2014JPA,Chiribella:2016IC,Chiribella:2014ARX}.

On the other hand, Naimark's (or Neumark's, depending on
 the transliteration) dilation theorem \cite{Neumark:1940IANa,Neumark:1940IANb,Neumark:1943CRA}, shows that any POVM can be obtained
from a projective measurement on a larger Hilbert space. 
This implies that in a Bell scenario nonideal measurements
cannot produce correlations that cannot be attained with ideal measurements.

The first observation points out the special role of ideal measurements in quantum theory. The second observation suggests that, to find a principle for quantum correlations (in Bell scenarios and, in the process, in KS scenarios with ideal measurements), an interesting question is as follows: Why are correlations between ideal measurements in nature not more contextual \cite{Cabello:2010XXX}?

The graph-theoretical approach introduced in \citet{Cabello:2010XXX,CSWPRL2014}  substantially departs from previous approaches to principles for quantum correlations. While the standard approach investigates principles explaining correlations once the measurement scenario is fixed, the graph-theoretic approach addresses the question of principles able to explain correlations once the graph of exclusivity relations is fixed.

Given $n$ events $\{e_j\}_{j=1}^n$ produced by a set of measurements $\{M_i\}$ (that also defines a measurement scenario) and an initial state $\rho$, one can represent the relations of mutual exclusivity between these events by a an $n$-vertex graph in which each event is represented by a vertex (node) and mutually exclusive events are connected by an edge. Recall that two events are mutually exclusive if there is a measurement that produces both of them, with each of them associated with a different outcome.

Given an $n$-vertex graph $G$, there are infinitely many measurement scenarios producing events whose graph of exclusivity is $G$. We consider all pairs $(\rho, \{M_i\})$, where $\rho$ is an initial state and $\{M_i\}$ is a set of {\em ideal} measurements, that produce $n$ events $\{e_j\}_{j=1}^n$ whose graph of exclusivity is $G$. For each pair, there is a set of probabilities $\{p(e_j)\}_{j=1}^n$. We denote by ${\cal P}(G)$ the set of all sets $\{p(e_j)\}_{j=1}^n$. 

\citet{CSWPRL2014} showed that, for any $G$, in quantum mechanics ${\cal P}(G)=$TH$(G)$. The fact that this physical set has a simple mathematical characterization suggests the following question: Why in quantum theory does ${\cal P}(G)=$TH$(G)$ for any $G$?

We define ideal measurements as those that (i) yield the same result when repeated, (ii) do not disturb any compatible observable, and (iii) can be implemented with all its coarse grainings satisfying (i) and (ii). The events produced by ideal measurements then satisfy the exclusivity principle. Given a set of events such that every pair of them is mutually exclusive, the sum of the probabilities of all of them is bounded by~$1$ \cite{Chiribella:2014ARX,Cabello:2019PRA,Chiribella:2020PRR}.

For theories allowing for statistically independent copies of any set $\{p(e_j)\}_{j=1}^n$ and events satisfying the exclusivity principle (as those originated from ideal measurements), the largest possible ${\cal P}(G)$ is TH$(G)$ for any $G$ \cite{Cabello:2019PRA}.

Given a Bell or KS scenario with ideal measurements with $G$ as the graph of exclusivity, the set $\mathcal{P}(G) = {\rm TH}(G)$ is not the set of quantum correlations for this scenario. However, the subset of ${\rm TH}(G)$ obtained after applying the constraints associated with
that scenario is the quantum set of correlations for that scenario \cite{Cabello:2019PRA}. These constraints are: normalization, nondisturbance, and the requirement that the probability of each event must only be a function of the state and measurement outcomes that define it.
\citet{Cabello:2019PTR} argued that this suggests a principle for quantum correlations: the totalitarian principle stating that anything not forbidden is compulsory, which is related to the principle of plenitude \cite{Lovejoy1936}, according to which the Universe should contain all possible forms of existence.


\subsubsection{Chromatic and fractional chromatic numbers}\label{sssec:chrome}

\begin{figure}[t]
\begin{center}
\includegraphics[width=0.45\textwidth]{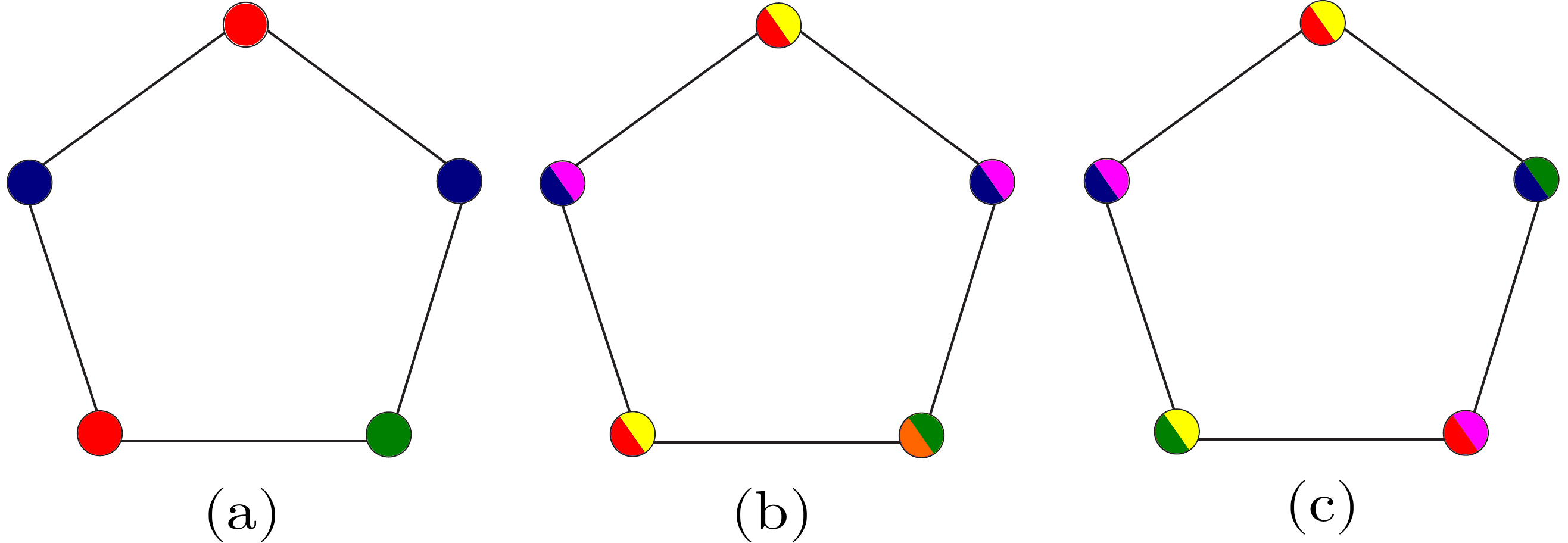}
\end{center}
\caption{\label{fig:p_frac} Different $a:b$ coloring of the pentagon, i.e., $b$ colors associated with each vertex out of $a$ total colors. (a) $3:1$ coloring of the pentagon, i.e., three colors, one for each vertex, giving a chromatic number $\chi=3$. (b) $6:2$ coloring of the pentagon obtained by doubling the colors for each vertex. (c) One color from the $6:2$ coloring can be removed, giving a $5:2$ coloring corresponding to a fractional chromatic number $\chi_f=5/2$ for the pentagon. }
\end{figure}

The chromatic and fractional chromatic numbers of a graph are also graph-theoretical quantities that play an important role in quantum contextuality, more precisely, in SI-C. In the following, we recall their definition and discuss their relation with contextuality, following the work of \citet{Cabello:2011XXX},
 \citet{RamanathanHorodecki2014} and \citet{CKB2015}. A $k$ coloring of a graph $G$ is an assignment of one out of $k$ colors to each vertex of a graph such that adjacent vertices are assigned different colors. The minimal number $k$ such that this coloring is possible is called the {\it chromatic number} of the graph and is denoted as $\chi(G)$. Equivalently, the chromatic number can be understood as the minimal number of partitions of the graph into independent sets. Similarly, the {\it fractional chromatic number} $\chi_f(G)$ is the minimum of $a/b$ such
that vertices have $b$ associated colors, out of $a$ colors, where again vertices connected by an edge have associated disjoint sets of colors. As a consequence, we have $\chi_f(G)\leq \chi(G)$. A simple example of chromatic and fractional chromatic number for the pentagon is given in Fig.~\ref{fig:p_frac}.

The chromatic number of a graph is  in general a difficult quantity to compute. It is nondeterministic polynomial-time-complete to decide whether a graph admits at $k$ coloring, except for $k=0,1,2$ and it is NP hard to compute the chromatic number~\cite{Garey2002}. The fractional chromatic number can be defined as a LP relaxation of the chromatic number; hence, it may seem easier to compute. However, computing the fractional chromatic number of a graph is NP hard~\cite{Lund1994}. Intuitively, this comes from the fact that the LP definition of the fractional chromatic number involves the knowledge of all independent sets of a graph, i.e., all sets of mutually disconnected vertices.

To discuss the connection between SI-C and the chromatic and fractional chromatic numbers, we first need to recall some basic definitions. We call a state-independent noncontextuality (SI-NC) inequality an inequality that, for a fixed set of measurements, is violated by any initial state. 
A set of elementary tests that can be used to violate such inequality, is called a SI-C set. A typical example is the Yu-Oh inequality in Eq.~\eqref{yuoh} and the Yu-Oh set in Fig.~\ref{fig:yo13int}. A related notion is that of SI-C-graph introduced by \citet{RamanathanHorodecki2014}. A SI-C-graph is a graph that for any fixed quantum state has a realization in terms of orthogonal projectors; i.e.,  a projector is associated to each vertex and two projectors are orthogonal if the corresponding vertices in the graph are connected, such that the given state violates a NC inequality. 
A SI-C set gives rise to a SI-C graph, but the converse is not always true. A typical example \cite{CKB2015} is obtained from the Yu-Oh set by increasing the dimension by $1$, i.e., $v_i \mapsto (v_i,0)$, and adding an extra vector orthogonally to all the others, i.e., $v_{E}=(0,0,0,1)$. The set is no longer a SI-C set, since by preparing the initial state $\ket{v_E}$, one would obtain a noncontextual value assignment to all variables, namely, all zero except $\ketbrac{v_E}$. On the other hand, for any pure initial state $\ket{\psi}$, one can find a realization of the graph such that the Yu-Oh NC inequality of Eq.~\eqref{yuoh} is violated. It is sufficient to choose a realization for which $\mean{\psi |v_E}=0$. According to ~\citet{RamanathanHorodecki2014}, a realization can also be found  for any mixed state.

The connection between graph coloring and SI-C has been discussed in the specific case of rank-1 projectors $\{ \Pi_i\}$ with corresponding dichotomic measurements given by $\{\Pi_i, \openone - \Pi_i\}$. The compatibility graph and the exclusivity graph then coincide, i.e., the projectors are compatible if and only if they are orthogonal, ignoring the trivial case of identical projectors. One can call the corresponding graph the orthogonality graph of $\{ \Pi_i\}$. We then have the following results proven by (i)~\citet{RamanathanHorodecki2014} and (ii)~\citet{Cabello:2011XXX}:

\begin{theorem*}[\citealp{Cabello:2011XXX,RamanathanHorodecki2014,CKB2015}]
For a set of rank-1 projectors $\{ \Pi_i\}$ in dimension $d$, the conditions (i) $\chi_f(G)>d$ and (ii) $\chi(G)>d$, for the orthogonality graph $G$, are necessary for SI-C.
\end{theorem*}

Notice that since $\chi_f(G)\leq \chi(G)$, condition (ii) is actually weaker than condition (i). However, condition (ii) has the advantage of being solvable exactly by simple integer arithmetic, while condition (i) is the solution to a linear program.

The condition $\chi(G)>d$ can be intuitively understood as necessary, since any coloring of the graph with $d$ 
different colors assigns different values to each set of $d$ orthogonal rank-1 projectors (forming a basis in 
dimension $d$), in particular, it is a consistent assignment of $0$ and $1$. The appearance of the fractional 
chromatic number is more puzzling, but it can be more or less straightforwardly derived by transforming the SDP a 
defining SI-C set $S=\{ \Pi_i\}_i$ (for rank-1 projectors) into a LP by fixing the quantum state to be the 
maximally mixed one \cite{CKB2015}. From this LP, one can extract the weights $w$ and construct the following SI-NC inequality
\begin{equation}\label{eq:nc_SIC_r1}
 \sum_i w_i \mean{\Pi_i}_\rho - \sum_{i} w_i \sum_{j\in \mathcal N(i)} \mean{\Pi_i\Pi_j}_\rho \stackrel{\rm NCHV}{\leq} 1, 
\end{equation}
which has the property that the maximal NCHV assignment is one respecting exclusivity relations and it is 
violated by the maximally mixed state with a value $\chi_f(G)/d > 1$. 

Notwithstanding the computational complexity of such problems, explicit calculations are still possible for small enough graphs. Using this result, it has been proven that Yu-Oh set is the minimal SI-C set in $d=3$ \cite{CKB2015}, namely, that there are no other SI-C-graphs with fewer than $13$ vertices in dimension $3$. 
This result was further extended by proving that any SI-C set must
contain at least 13 projectors, regardless of the dimension~\cite{Cabello:2016JPA}. The previous results are valid under the assumption of rank-1
projectors, however, they have been extended to the case of uniform
(i.e., all projectors of the same rank) rank 2 and rank 3 by \citet{XuArXiv2020}, who were also able to exclude the case of eight arbitrary projectors or fewer.


\subsection{Connections between the Kochen-Specker and Bell's theorems}\label{KS_Bell}

The connection between the proofs of the KS theorem and Bell nonlocality arguments 
has been extensively investigated since the 1970s \cite{Stairs:1978, Stairs:1983PS, Krips:1987, Redhead:1987, Brown:1990FPH, Elby:1990FPH, Elby:1990FPL, Elby:1992PLA, Clifton:1993AJP, Mermin:1990PRL, Kernaghan:1995PLA}. On the one hand, any Bell inequality can be interpreted as a noncontextuality 
inequality and there are methods to convert some noncontextuality inequalities 
into Bell inequalities violated by quantum theory, see, e.g., 
 \citet{Aolita:2012PRA, Cabello:2012PRA}.

Historically, the first results related to this question are those on the so-called KS with locality theorem \cite{Kochen70, Stairs:1983PS, HR83, Redhead:1987, Brown:1990FPH},  which later gave rise to the so-called free will theorem \cite{CK06,CK09}. Common to all these results is that the KS proof for a single spin-1 particle is expanded into related algebraic proof involving the KS set and a maximally entangled state of two spin-1 particles. 

The second wave of results connecting the KS and Bell's proofs were motivated by the GHZ proof of Bell's theorem \cite*{GHZ89}. First, it is Mermin's observation that GHZ can be converted into a tripartite Bell inequality \cite{Mermin90b} and a state-independent proof of the KS theorem \cite{Mermin:1990PRL,Mermin:1993RMP}. Second, the observation that Hardy's proof of Bell's theorem \cite{HardyPRL1992,HardyPRL1993} can be seen as state-dependent version of a KS proof \cite{Cabello:1996PLA}. Finally, it is the GHZ-like proof for two parties sharing qubits \cite{Cabello:2001PRLa}, which can be seen as originating from the PM KS proof \cite{Peres:1990PLA,Mermin:1990PRL} and which can be converted into a bipartite Bell inequality \cite{Cabello:2001PRLb} as explained later. Around all these tools, there is an extensive literature adopting different perspectives and names: ``all-versus-nothing'' proofs \cite{Cabello:2001PRLb}, ``nonlocal games'' \cite{CHTW04}, and ``quantum pseudotelepathy'' \cite{GBT05,Renner2004}.

More recently other methods have been introduced to transform inequalities associated with SI-C scenarios to Bell inequalities \cite{Aolita:2012PRA, Cabello:2012PRA,Cabello2020arxiv}. The simplest approach to the problem is arguably to map single measurements and two-time sequential measurements on a single system into bipartite measurements on a maximally entangled state. Here we  approximately follow the discussion given by \citet{Cabello2020arxiv} but with a different class of NC inequalities, namely those discussed in Sec.~\ref{sssec:chrome}. To understand this method, we start with the basic observation that, for the state $\ket{\Psi}=1/\sqrt{d}\sum_k \ket{kk}$,
\begin{equation}\label{eq:max_ent_trace}
\mean{\Psi| A\otimes B^t |\Psi}=\tr(A B)/d,
\end{equation}
where the superscript $t$ represents the transposition with respect to the basis $\{\ket{k}\}_k$. In other words, expectation values of bipartite operators on the maximally entangled state are equal (up to a transposition) to the expectation value of their product on the maximally mixed (one party) state $\openone/d$. Using the fact that in any SI-C scenario the noncontextuality inequality is violated even by the maximally mixed state, we can transform the SI-C scenario into a bipartite Bell inequality.
This idea is  general and can be applied to a wide variety of scenarios and inequalities. To make a concrete example, we  discuss the specific case of noncontextuality inequalities arising from the fractional chromatic number of a SIC-graph from Sec.~\ref{sssec:chrome}. 
Given a SI-C set $\{\Pi_i\}_i$, we consider the associated SI-NC inequality presented in Eq.~\eqref{eq:nc_SIC_r1},
\begin{equation}\label{e:ncgen}
 \sum_i w_i p(\Pi_i=1) - \sum_{i} w_i \sum_{j\in \mathcal N(i)}
p(\Pi_i = \Pi_j=1)\le 1,
\end{equation}
such that the optimal classical assignment corresponds to one satisfying the exclusivity relations, and violated by the maximally mixed state, with a value $\chi_f(G)/d>1$.
This inequality can be transformed into the Bell inequality
\begin{align}
&\sum_i w_i p(\Pi_i^A=\Pi_i^B= 1) - \frac{1}{2}\sum_{i} w_i \times
 \\
&\times \sum_{j\in \mathcal N(i)}
\Big[p(\Pi_i^A = \Pi_j^B=1) + p(\Pi_i^B = \Pi_j^A=1)\Big]\le 1
\nonumber
\end{align}
by distributing a copy of all projectors $\{ \Pi_i\}_i$ to both Alice and Bob, i.e.,  $\Pi_i^A = \Pi_i$ and $\Pi_i^B=\Pi_i^t$, where on Bob's projectors the previously mentioned transposition has been applied. 

By Eq.~\eqref{eq:max_ent_trace}, we have the value on $\ket{\Psi}$,
\begin{equation}\label{e:spat_value}
\begin{split}
&\mean{\psi| \Pi_i \otimes \Pi_i^t |\psi}=\tr(\Pi_i \Pi_i )/d = \tr(\Pi_i)/d,\\
&\mean{\psi| \Pi_i \otimes \Pi_j^t |\psi}=\tr(\Pi_i \Pi_j )/d = 0, 
\end{split}
\end{equation}
if $i,j$ appear as a correlator in Eq.~\eqref{e:ncgen}, is the same as that of the lhs of Eq.~\eqref{e:ncgen} on the maximally mixed state $\openone/d$, namely, $\chi_f(G)/d > 1$. 

For the local hidden-variable bound, one can have an argument similar to that presented by \citet{CKB2015} for Eq.~\eqref{e:ncgen}. The main idea is that the weights $w_i$ are chosen such that the maximum deterministic value assignment is one that respects the orthogonality conditions among the projectors, i.e.,  if $\Pi_i \Pi_j = 0$, to the corresponding classical variables. We denote them by $\pi_i$ and $\pi_j$, which are assigned one $0$ and one $1$. Every time that we violate one of these constraints for just one of the parties, say on Alice's side, by flipping the value of $\Pi_i^A$ we get a factor $-w_i/2 \sum_{j \in \mathcal N(i)} \pi_j^B$, which decreases the total value (assuming we are violating an orthogonality constraint, so at least one of the $\pi_j^B$ is not $0$). Similarly, if we violate one orthogonality for both $\Pi_i^A$ and $\Pi_i^B$, we get a factor $w_i -w_i/2 \sum_{j \in \mathcal N(i)} (\pi_j^A+\pi_j^B)$, which is again negative. We therefore find that optimal classical assignments are those respecting the orthogonality relations on both Alice and Bob's side.

The previous construction is a simple one, but other constructions are possible (see the aforementioned corresponding references). In particular, we highlight the fact that some of these constructions, such as that in \citet{Aolita:2012PRA}  also enable one to construct Bell inequalities where the quantum and nonsignaling bounds coincide. 

The first experiments on what is now called Peres-Mermin Bell inequality were based on the encoding proposed in \citet{ChenPRL2003} and carried out in \citet{CinelliPRL2005,YangPRL2005}, and subsequently repeated by the group in Rome, Italy, to improve the violation and fix a conceptual problem with the first experiment \cite{BarbieriPRA2005,BarbieriOS2007}. \citet{Aolita:2012PRA} also reported the results of an improved experiment in Rome. On the basis of the proposal made by \citet{Cabello:2010PRL}, an experiment with sequential measurements on entangled photons was performed  \cite{Liu:2016PRL}.


\subsection{Classical simulation of quantum contextuality}\label{class_sim}
The fact that quantum mechanics results in different predictions 
than noncontextual theories leads to the question regarding which 
contextual theories can simulate the quantum mechanical 
behavior. More precisely, one can ask, which classical 
resources are needed in order to classically simulate 
the quantum behavior in a contextuality experiment. 

This question has some precedent in the analysis of Bell 
scenarios. There, one may ask how much communication between
the two parties is needed in order achieve a maximal 
violation of a Bell inequality. For the case of the simplest 
Clauser-Horne-Shimony-Holt inequality, this has been discussed
in detail and optimized simulation schemes have been designed 
\cite{Toner:2003PRL, Cerf:2005PRL}.

Concerning contextuality, several contextual models have 
been designed, such as the PM square 
\cite{LaCour:2009PRA, Blasiak:2015AOP}. In addition, there
are  general approaches for simulating quantum mechanics
with classical models, including contextuality \cite{Spekkens:2007PRA, 
vanEnk:2007FPH, Larsson:2012AIP}. These models, however, were not 
constructed to be resource efficient, and they do not allow 
for a clear estimate of the minimal necessary classical resources. 

In the following, we discuss models to simulate
quantum contextuality in sequential measurements. Such contextual models require some 
memory to work: For instance, for obtaining the maximal 
value $\mean{\mathsf{PM}} = 6$ in Eq.~(\ref{s2:PMineq})
one needs to remember the previous measurements in the 
measurement sequence. Thus, the question arises as to what minimal
memory is needed for the simulation. This depends on the 
underlying computational model and the quantum mechanical 
correlations that should be simulated. In the following, 
we  first discuss two concrete models and then we
mention some more general approaches for simulating temporal 
correlations.

\subsubsection{Simulation with Mealy machines}
A simple attempt to simulate contextual behavior in
sequential measurements is the following \cite{Kleinmann:2011NJP}.
One assumes a classical automaton with $k$ internal states. 
For a given internal state one can ask a certain question
(or, in physical terms, perform a certain measurement)
and obtains an answer (or, result). After providing the 
result, the automaton changes its internal state, depending 
on the measurement that was performed. Therefore, in this model, 
each internal state is characterized by two discrete functions: 
One function determines the output, depending on the measurement, 
and the other function determines the update of the internal 
state, again in dependence on the measurement. 

Such a model is called a Mealy machine \cite{Mealy:1955BSTJ}. 
Given this class  of models, one can easily define the memory cost required for 
a simulation as the minimal number of internal states that is 
required for a simulation. 

This concept is best explained with an example. We focus on the 
PM square as in Eq.~(\ref{eq:basicPM}). For a simulation, 
we assume that the automaton has three internal states $S_1$, $S_2$, and 
$S_3$. For each state, we define the automaton via the tables
\begin{align}
S_1\!:\!  & \left[\begin{matrix}
     +&+&(+,2)\\
     +&+&(+,3)\\
     +&+&+
    \end{matrix}\right]\!,\;\;
S_2\!:\! \left[\begin{matrix}
     +&(+,1)&+\\
     -&+&-\\
     -&(-,3)&+
    \end{matrix}\right]\!,\;\;
\nonumber \\
S_3\!:\!  & 
\left[\begin{matrix}
     +&-&-\\
     (+,1)&+&+\\
     (-,2)&-&+
    \end{matrix}\right]\!.
    \label{eq-simplemealy}
\end{align}
This defines the automaton as follows: Assume that the Mealy 
machine is in state $S_1$ and we measure the observable $\gamma$ of Eq.~\eqref{eq:basicPM}. 
We then consider the first table at the position of $\gamma$ 
(i.e., the last entry in the third row). The simple plus sign 
at this position indicates that the measurement result will 
be $+1$, while the system stays in state $S_1$. If we continue 
and measure $C$, we encounter the entry $(+, 2)$ which indicates 
the measurement result $+1$ and a subsequent change to the 
internal state $S_2$. Being in state $S_2$, the second table 
defines the behavior for the next measurement: For instance a measurement 
of $c$ now yields  the result $-1$, and the system stays in state 
$S_2$.

Thus, starting in  $S_1$ the measurement results for the sequence 
 $\gamma C c$ are $+1, +1, -1$, so the product is $-1$ in 
 accordance with the quantum prediction. It is straightforward 
 to verify that this model yields $\mean{\mathsf{PM}} = 6$. In 
 addition, the observables within each context (defined by a 
 column or row) are compatible in the sense 
 that in sequences of the form $A_1 A_2$, $A_1B_2A_3$, or 
 $A_1\alpha_2 a_3 A_4$ (here, the subindices indicate the 
 temporal ordering in the measurement sequence) the first 
 and last measurements of $A$ yield the same output.
 
One can show that this automaton is the smallest automation 
to reproduce these predictions; that is,  Mealy machines with 
two internal states cannot do that \cite{Kleinmann:2011NJP}.
In this example, however, one has to be careful about the 
correlations that one wants to simulate. For instance, while 
the previously mentioned Mealy machine reaches $\mean{\mathsf{PM}} = 6$, 
it does not reproduce all deterministic quantum predictions: 
Starting in $S_1$, the sequence $B_1{C_2\beta_3} B_4$ yields 
the sequence of results $(+1, +1, -1, -1)$, i.e., $B$ changes
its value. This is in contrast to quantum mechanics, since  
$C$ and $\beta$ are both compatible with $B$. Thus, while this 
machine reproduces compatibility constraints within one 
column or row, it does not reproduce all compatibility 
conditions. For incorporating more compatibility constraints 
one needs four internal states \cite{Kleinmann:2011NJP}.

In this example, Mealy machines were used only to simulate 
some outcomes of quantum mechanics in a deterministic manner. 
However, no quantum state gives deterministic outcomes for 
all measurements of the PM square. For this, one 
can consider probabilistic mixtures of different Mealy machines. 
For instance, \citet{Fagundes:2017JPA} considered a class of variations of 
the automaton in Eq.~(\ref{eq-simplemealy}), and then probabilistic
mixtures of these. They then showed that the predictions 
for any quantum state can  be reproduced as long as only 
compatibility relations within the columns and rows are 
considered. In an extension of this research line, it was also
shown that a Mealy-type machine with a single qubit cannot
simulate contextual correlations \cite{Budroni:2019NJP}.
In a significantly different approach, the problem of determining the initial state of a 
Mealy or More machine was connected to quantum logic \cite{schaller-96, 
dvur-pul-svo}.

\subsubsection{Simulation with $\varepsilon$ transducers}
A different approach to simulate contextuality or temporal 
correlations comes from the analysis of time series and can also
be used to quantify the memory needed for a simulation. 
Before starting the explanation of $\varepsilon$ transducers, 
it is useful to recall the notions of hidden Markov models 
\cite[HMMs,][]{Rabiner:1989IEEE} and $\varepsilon$ machines \cite{Crutchfield:1994PD}.

\begin{figure}
\begin{center}
\includegraphics[width=.95\columnwidth]{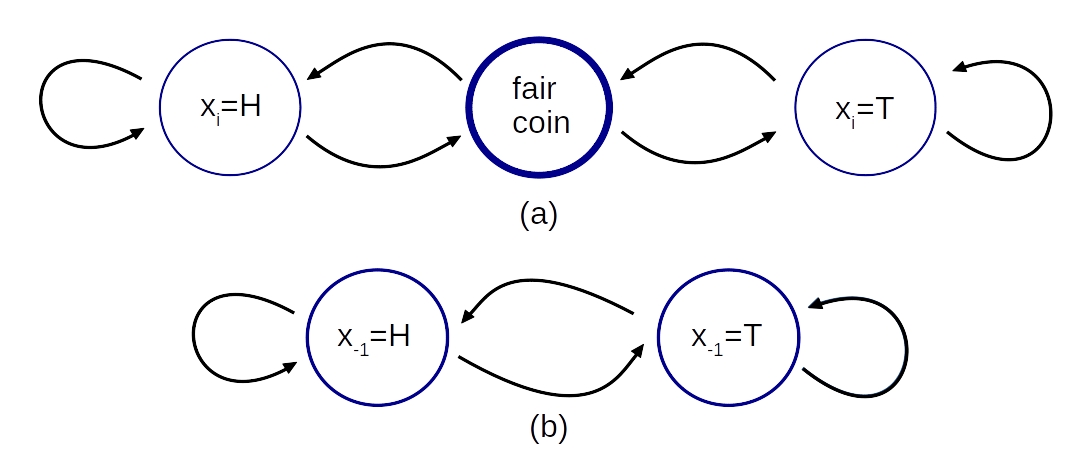}
\end{center}
\caption{\label{fig:hmm-scheme}%
Examples of a HMM and an $\varepsilon$-machine for the  simulation 
of a biased coin flip (also called a perturbed coin). The process is 
given by a coin, which flips with a certain probability, as given by 
$P(x_i=H)=P(x_i=T)={1}/{2}$ and 
$P(x_i=T|x_{i-1}=H)={1}/{2}-\delta$ 
and 
$P(x_i=H|x_{i-1}=T)={1}/{2}-\delta$. This can be seen as a fair 
coin ($\delta=0$) that is disturbed toward a constant process.
Detailed descriptions of this and the following models were given 
by \citet{Lohr:2009Proc}.
(a) A general HMM could simulate this with three internal states, 
 corresponding to a fair coin and two maximally biased coins, 
which always give heads or tails. The fact that the biased coin 
flip tends to reproduce the previous result is modeled by the 
rule that the automaton acts most of the times as a fair coin, 
but sometimes this is replaced by a deterministic coin. 
(b) When constructing the $\varepsilon$ machine, one first
observes that only the last output matters, so this defines
the causal states. Since the two outputs are equally probable, 
the statistical entropy of the process is 1 bit. The HMM in 
(a) requires less memory, but it contains oracular information: 
If one knows that the automaton is in a state corresponding 
to a maximally biased coin, the next output is foreseeable.}
\end{figure}

HMMs are probabilistic automata to simulate time series. 
The automaton contains a set of internal states $S_k$, and 
for each internal state, there is a probability distribution
$P_k$ of the outcomes and a probability distribution $T_k$ 
of the transitions. If the automaton is in a given state 
$S_k$, it will output an outcome drawn from $P_k$ and move
to another state, chosen according to $T_k$. For a
description of a given time series the HMM does not have to 
be in a definite state, instead one has a probability distribution
over all internal states. Two examples of HMMs are given in 
Fig.~\ref{fig:hmm-scheme}.

$\varepsilon$ machines can be seen as a special instance 
of HMMs. Consider an infinite time series 
$\overleftrightarrow{\mathcal{X}} 
= \{\dots,X_{-2},X_{-1},X_{0},X_{1},X_{2},\dots\}$
where the $X_i$ are random variables over some alphabet. 
One can split it into a past and a future, 
\begin{align}
\overleftarrow{\mathcal{X}} & = \{ \dots, X_{-3}, X_{-2}, X_{-1}\},
\nonumber
\\
\overrightarrow{\mathcal{X}} & = \{ X_{0}, X_{1}, X_{2}, \dots\}.
\end{align}
One can then define an equivalence relation on the set of all possible 
pasts, by calling two pasts equivalent if they both predict the same 
future. Mathematically, one defines 
$\overleftarrow{{x}} \sim \overleftarrow{{x}}'$
if and only if
$P(\overrightarrow{\mathcal{X}}|\overleftarrow{{x}})
=
P(\overrightarrow{\mathcal{X}}|\overleftarrow{{x}}').
$
The equivalence classes of this relation are then called the
causal states $\{S_k\}$. The causal state contains all 
information from the past that is relevant for the future;  
knowledge of the precise history does not add anything to it.

Given a causal state, one obtains a $x_0$ as a new output. This
additional output defines a new history, belonging to a potentially
different causal state and, consequently, the output $x_0$ defines
a transition to a new causal state. Note that in a general HMM 
the output does not determine the transition. Finally, one can 
consider the probability distribution of the causal states and 
its entropy, $H= -\sum_k p(S_k) \log[p(S_k)]$. This is the 
statistical complexity of the process and can be used to 
quantify the memory needed for a simulation. Recently it was 
shown that in this context quantum mechanics can help to reduce 
the memory required for a simulation of a time series \cite{Gu:2012NC, 
Palsson:2017SA}.

Before discussing the application to contextuality, it 
is useful to explain in more detail the difference between 
an $\varepsilon$ machine and general HMMs. In a HMM (or a 
Mealy machine) the simulation automaton 
may contain information about the future that cannot be derived 
from the past. Consider Alice and Bob, where Alice  
knows only the current internal state of the automaton and Bob
 knows only the past sequence of results. If the simulation 
works properly, Alice can predict the future as well 
as Bob. It is possible, however, that Alice could predict the
future better than Bob, such as if the given internal state 
$S_k$ predicts a deterministic outcome for the next 
measurement that cannot be deduced from the past. This 
difference is also illustrated in Fig.~\ref{fig:hmm-scheme}. 
Physically, such general HMMs with oracular
information may be excluded due to the demand that only 
the past observations should be used for simulating 
the future. This then leads  to $\varepsilon$ machines.

For simulating contextuality, one still has to extend
the scheme a bit, as different measurements in each time
step are possible. This, however, can easily be done by 
combining the sequence of measurement choices 
$\overleftrightarrow{\mathcal{Q}}$ and the sequence
of results $\overleftrightarrow{\mathcal{A}}$ to a 
single variable, 
$\overleftrightarrow{\mathcal{X}}
=
(\overleftrightarrow{\mathcal{Q}}, \overleftrightarrow{\mathcal{A}}).$
The corresponding $\varepsilon$-machine is 
then called the $\varepsilon$-transducer \cite{Barnett:2015JSP}.

The simulation of the PM square with 
$\varepsilon$-transducers was considered by \citet{Cabello:2018PRLEPSILON},
who found that the causal states are effectively
the 24 eigenstates occurring in the observables of the square. 
They occur with equal probability, so the required memory is 
$H= \log(24) \approx 4.585$ bits. For the Yu and Oh scenario, 
the causal states are more difficult to identify, but at least
$H \approx 5.740$ bits are required for a simulation. Although 
the Yu-Oh scenario is more difficult to
simulate, the scenario has a smaller degree of contextuality
according to several contextuality measures \cite{Kleinmann:2012PRL, Abramsky:2011NJP,Grudka:2014PRL}.

\subsubsection{Other related results}
Contextuality is relevant for quantum computation (see 
also Sec.~\ref{sec:contextualityincomputation}), so 
the simulation of both phenomena is connected. In quantum
computation, the so-called stabilizer operations contain an
important class, that is, however, not sufficient for 
universal computation. 
Recently, \textcite{Hindlycke:2022} 
provided a model where the correlations of all 
Pauli-stabilizer states, Clifford transformations, and 
Pauli-tensor-product measurements can be simulated in time 
and space quadratic in the number of qubits. 
Thus this contextual hidden-variable model gives an efficient
simulation of the stabilizer subtheory of quantum mechanics 
including its complete contextual behavior.

For making stabilizer operations universal, so-called 
magic states need to be added and an explicitly contextual 
(but not efficient)
hidden-variable model for these was found by \citet{Zurel:2020PRL}. 
Similarly, explicitly contextual
classical models that simulate quantum contextuality were
investigated by \citet{Bravyi:2018SCI,Bravyi:2020NP}. 
In this case, the cost of the classical simulation of contextual 
correlations is quantified in terms of the circuit depth, which
increases with the input size for classical models
that use gates with bounded fan-in, 
but remains constant for any input size for quantum models. 

The previous results lead to the question as to how general 
temporal quantum correlations can be simulated in a 
classical manner. If the dimension of the underlying
quantum system is not bounded, the space of all correlations 
forms a polytope \cite{Abbott:2016PRA, Hoffmann:2018NJP}, while
the correlation space becomes nonconvex for fixed dimension 
\cite{Mao:2020arxiv}. Mealy machines have been used
to characterize the memory cost for simulating correlations
\cite{Budroni:2019NJP,Budroni:2020PRR,Vieira:2022Q}, 
and the minimal dimensions for reaching
certain correlations have been characterized 
\cite{Spee:2020NJP, Mao:2020arxiv}.




\subsection{Resource theory of contextuality}\label{res_th}

The connection between quantum contextuality and practical applications in quantum information processing, most notably quantum computation (see Sec. \ref{sec:applications}), stimulated the development of various proposals for a resource theory of quantum contextuality, which we review in the following. 
In simple terms, the basic elements of a resource theory are the resourceful and resourceless objects, as well as the operations that do not increase the resource, i.e., the \textit{free operations}, and finally a quantifier of the resource, which should be monotonic with respect to the free operations~\cite{Coeke2016}. 
The first works in this area were the one by \citet{Grudka:2014PRL} and the follow-up paper by \citet{HorodeckiPRA2015}. They proposed a measure of contextuality, called {\it relative entropy of contextuality}, together with an abstract characterization of the axiomatic structure of the resource theory. The monotonicity of their measure, however, was proven only for a restricted set of operations, while a general characterization of the free operations was not provided. This gap was filled by \citet{AmaralPRL2018}, who provided a characterization of free operations, the so-called {\it noncontextual wirings}, and showed that the relative entropy of contextuality is indeed a monotone. In simple terms, noncontextual wirings can be understood as a preprocessing and postprocessing of each joint measurement associated with a context, regardless of the particulars of its experimental implementation (such as joint or sequential measurement). The preprocessing affects the choice of the inputs, i.e., the sequence to be measured, whereas the postprocessing, which may also depend  on the preprocessing operation, affects the measurement outputs. Preprocessing and postprocessing cannot be arbitrary, but they obey constraints that make their application consistent: restricting the measurements only to contexts, etc.

Similarly, \citet{Abramsky:2017PRL} explored the properties {\it contextual fraction}, which was introduced by \citet{Abramsky:2011NJP} and \citet{Amselem:2012PRL}) as a contextuality monotone. There, however, they proved the monotonicity with respect to a restricted set of operations, namely, a translation of measurements and coarse graining of outcomes. This work was then further expanded by \citet{AbramskyIEEE2019} and \citet{Barbosa2021XXX}. In particular,  \citet{AbramskyIEEE2019} introduced a new operation in the resource theory called {\it conditioning on a measurement}, namely, the possibility of choosing the current measurement to perform in a temporal sequence on the basis of the outcomes of the previously performed (compatible) measurements [i.e., a measurement protocol according to the definition of \citet{AcinCMP2015})]. The contextual fraction has been proven to be a monotone both for this extended set of operations \cite{AbramskyIEEE2019} and for the noncontextual wirings \cite{AmaralPRL2018}. One can show that the extended set of operation given by \citet{AbramskyIEEE2019} is strictly larger than the noncontextual wirings of \cite{AmaralPRL2018}. This is achieved \cite{Karvonen2022} by showing that any noncontextual wiring can be reproduced by the operations of \citet{AbramskyIEEE2019} and then providing a transformation, protocol 1 of \citet{Barrett:2005PRA}, that can be expressed in the resource theory of \citet{AbramskyIEEE2019}, but not as a noncontextual wiring.

Finally, within the framework of \citet{AbramskyIEEE2019}, \citet{Karvonen:PRL2021} proved that the resource theory of contextuality does not admit catalysis, meaning that there are no resources (in this case, correlations) that can enable an otherwise impossible resource conversion and still be recovered afterward.


\subsection{The so-called nullifications of the Kochen-Specker theorem}
\label{nullif}


The KS theorem was developed in the framework of ideal measurements and, as we saw in Sec.~\ref{ssec:compatible} and \ref{ssec:exper_imp}, several problems arise when one tries to map those ideal measurements to actual experimental implementations. Some of the first criticisms regarding the physical implications of the KS theorem
precisely involved this transition from ideal to actual measurements, in particular the impossibility of arbitrarily precise measurements, and were raised by \citet{Meyer:1999PRL}, \citet{Kent:1999PRL}, \citet{Clifton:2000PRSA}, and \citet{Barrett:2004SHPB}. These works played a fundamental role in the development of the modern approach to contextuality by stimulating the extension of the KS notion of contextuality from a logical to a probabilistic framework. In fact, they motivated the derivation of the KS inequalities, which appeared in those years~\cite{Larsson:2002EPL,Simon:PRL2001}. This transition from the logical to the probabilistic perspective in Kochen-Specker's contextuality, and in particular the subsequent theoretical and experimental effort in testing contextuality on physical systems, is the most interesting outcome of this debate. 
Notwithstanding the value of both the criticisms of Meyer, Kent, Clifton, and Barrett and the responses that they received \cite{Cabello:1999XXX, Mermin:1999XXX,Appleby:2000XXX, Havlicek:2001JPA,Simon:PRL2001,Appleby:2001XXX, Larsson:2002EPL, Appleby:2002PRA, Peres:2003XXX, Appleby:2005SHPS}, we decided in the interest of brevity not to present them in detail here.
Instead, after presenting the arguments by \citet{Meyer:1999PRL},  \citet{Kent:1999PRL} and the one by \citet{Clifton:2000PRSA}, we compare them to the broader perspective of probabilistic approaches to contextuality discussed in previous sections  and, in particular, the problem of designing and implementing valid experimental tests of contextuality (Secs.~\ref{ssec:compatible} and \ref{ssec:exper_imp}). These results provide the strongest argument against any possible claim of ``nullification'' of Kochen and Specker's contextuality.


\subsubsection{Meyer's ``nullification'' of the KS theorem}


What if not all sharp measurements are physically realizable?   \citet{Meyer:1999PRL} suggested an explicit way in which this can happen while being undetectable due to the unavoidable finite precision of actual measurements. 
Each direction in the three-dimensional Euclidean space can be represented by a unit vector 
\begin{equation}
\langle v_j |=\frac{1}{\sqrt{x_j^2+y_j^2+z_j^2}}(x_j,y_j,z_j),
\end{equation}
with 
\begin{equation*}
\frac{x_j}{\sqrt{x_j^2+y_j^2+z_j^2}},\frac{y_j}{\sqrt{x_j^2+y_j^2+z_j^2}},\frac{z_j}{\sqrt{x_j^2+y_j^2+z_j^2}}\in \mathbb{R}.
\end{equation*}

According to quantum theory, each direction in the three-dimensional Euclidean space corresponds to a sharp measurement on a three-dimensional quantum system. The corresponding $\pm 1$-valued observable is represented in quantum theory by the self-adjoint operator constructed from 
\begin{equation}
A_j = 2 |v_j\rangle \langle v_j|-\openone,
\end{equation}
where $\openone$ is the $3 \times 3$ identity matrix. The possible outcomes are the eigenvalues of $A_j$: $-1$ (doubly degenerate) and $1$ (nondegenerate).

If all directions $\ket{v_j}$ correspond to physically realizable sharp measurements, then at least four colors are needed to color every $\ket{v_j}$ respecting that orthogonal $\ket{v_j}$'s are colored differently \cite{Hales1982}. However, if 

\begin{equation}
frac{x_j}{\sqrt{x_j^2+y_j^2+z_j^2}},\frac{y_j}{\sqrt{x_j^2+y_j^2+z_j^2}},\frac{z_j}{\sqrt{x_j^2+y_j^2+z_j^2}}\in \mathbb{Q},
\end{equation} 
then only three colors are needed \cite{Godsil1988XXX}. This implies that noncontextual assignments of $-1$ and $1$ respecting the fact that, for each orthogonal trio, $1$ is assigned to only one vector are possible \cite{Meyer:1999PRL}. In fact, it is enough to take the three colors and assign the value $0$ to two of them and the value $1$ to the remaining one to obtain a valid KS assignment.  Moreover, the rational unit sphere is dense in the real unit sphere and thus there is no experimental way to distinguish between the two spheres. This result was derived by \citet{Pitowsky:1985Phil} several years earlier. Using just the continuum hypothesis, but a version with a weaker assumption (Martin's axiom), Pitowsky provided an assignment $s$ of values 
$\{-1,0,1\}$ to all triples of orthogonal vectors in the unit sphere $S^2$ such that the condition $s^2(x)+s^2(y)+ s^2(z)=2$ is satisfied by all orthogonal triples $x,y$ and $z$ except for a countable number of them. In other words, Pitowsky found a valid value assignment for   almost all  triples of orthogonal vectors in $S^2$; see also the discussion in \cite[][]{Fuchs:2011} p. 503.

According to Meyer, this shows that, despite the KS theorem, NCHV models can simulate the predictions of quantum theory within any fixed finite experimental precision.   \citet{Kent:1999PRL} generalized Meyer's result and showed a construction of KS colorable dense sets of projectors onto vectors with rational components in complex Hilbert spaces of arbitrary finite dimension.  \citet{Kent:1999PRL} claimed that this shows that ``noncontextual hidden variable theories cannot be excluded by theoretical arguments of the KS type once the imprecision in real world experiments is taken into account''.




A simple counterargument to  Meyer's and Kent's NCHV models is given by the fact that their models cannot  reproduce the probabilistic predictions of quantum theory. The following example is taken from \citet{Cabello:2010PLA}. Consider $d=3$ and the initial state
\begin{equation}
\langle \psi| = \frac{1}{527}(354,357,-158)
\end{equation}
and the sharp measurements associated with
\begin{subequations}
	\begin{align}
	 \langle v_1| & = ( 1,0,0),\\
	  \langle v_2| & = (0,  1, 0),\\
	   \langle v_3| & = \frac{1}{73}(48,0,-55),\\
	    \langle v_4| & = \frac{1}{3277}(1925,2052,1680),\\
	     \langle v_5| & = \frac{1}{221}(0,140,-171).
	\end{align}
\end{subequations}
This state and all these ideal measurements are allowed according to Meyer. For this state and these measurements, quantum theory predicts
\begin{equation}
\kappa = 3.941
\end{equation}
for 
\begin{equation}
\label{kappa}
\kappa= -\langle A_1 A_2 \rangle-\langle A_2 A_3 \rangle-\langle A_3 A_4 \rangle-\langle A_4 A_5 \rangle-\langle A_5 A_1 \rangle.
\end{equation}
However, for any NCHV model \cite{Klyachko:2008PRL,Klyachko:2007NATO},
\begin{equation}
\label{KCBS}
\kappa \leq 3.
\end{equation}
Therefore, Meyer's NCHV models fail to simulate the predictions of quantum theory. 
Notice that the set of inequalities of the form of \eqref{KCBS} with $\kappa$ of the form  of Eq.~\eqref{kappa} with an odd number of minus signs [such as Eq.~\ref{kappa}, which has five minus signs] provides a necessary and sufficient condition for the existence of a NCHV model \cite{Araujo:2013PRA}.


\subsubsection{Clifton and Kent's nullification of the KS theorem}


\citet{Clifton:2000PRSA}, starting with similar ideas, adopted a different approach. They asked the following question: What if every sharp measurement  belongs only to one context?  They showed that there is a set of directions in the three-dimensional Euclidean space that is dense in the real unit sphere and consists of directions such that none of them are  orthogonal to any of the others. Therefore, one can assign any predetermined outcome to any of these directions \cite{Clifton:2000PRSA}. In other words, they substituted the measurements defining the set of contexts (compatible ideal measurements) with other measurements for which contexts consist of a single measurement; i.e., all measurements are mutually incompatible. In this way, as noted by \citet{Kochen:1967JMM}, no constraint is imposed on the NCHV model except the reproduction of single-measurement marginals, hence a NCHV model can be constructed as a product of all single-measurement distributions.

This can be formulated as a problem of imperfect compatibility, in analogy with the one addressed by~\citet{Larsson:2002EPL}, \citet{WinterJPA14}, and \citet{Kujala:2015PRL}. Possible solutions to this problem within the probabilistic framework of contextuality and involving modifications to the standard NC inequalities were extensively discussed in Sec.~\ref{ssec:exper_imp}; see that section for further details.

We now comment on the relation among Bell nonlocality, contextuality, and imperfect compatibility. It is true, as \citet{Barrett:2004SHPB} claimed,  that the original KS contextuality and Bell nonlocality are logically independent concepts. However, in the probabilistic framework for contextuality, developed precisely after the entire nullification debate, Bell nonlocality can be seen as an special case of contextuality. We have a notion of contexts, compatible measurements, and the goal of reproducing observed correlations, associated with single contexts, from a global probability distribution; see Sec.~\ref{ssec:compatible}. In a Bell scenario, however, perfect compatibility is always guaranteed by the spacelike separation of measurement events. Hence, by the locality condition of special relativity no disturbance is allowed between them. Moreover, imperfect measurements can always be ``dilated'' to projective measurements using Neumark's theorem \cite{Neumark:1940IANa,Holevo:1982,Peres:1993,Neumark:1940IANb,Neumark:1943CRA}.

Finally, imperfect measurements seem to forbid contextuality by another mechanism, namely, the transformation of degenerate observables into nondegenerate ones. The degeneracy property of quantum observables is fundamental in creating a nontrivial structure of measurement contexts (see the discussion of Vorob'ev's theorem in Sec.~\ref{graphth}). In fact, for nondegenerate observables, commutativity becomes a transitive property (i.e., $[A,B]=[B,C]=0$ implies that $[A,C]=0$), which guarantees the existence of a NCHV model~\cite{Correggi2002}. In our jargon, the compatibility graph is a collection of disconnected cliques; see Sec.~\ref{graphth}. For physically relevant observables of a single system, the degeneracy is a consequence of some symmetry of the system (consider the rotational symmetry of the hydrogen atom) that is removed when the symmetry is no longer exact (consider the level splitting due to a small electric or magnetic field). Arguably, this is the case of imperfect experimental realization (for instance, it is impossible to completely remove any electric and magnetic field). In contrast, the symmetries related to the space-time structure are robust, i.e., never removed by small imperfections, and thus preserve the degeneracy of the relevant physical observables (for instance, consider observables of the form $A_x\otimes \openone$ and $\openone\otimes B_y$ for a bipartite system). This problem can be analyzed from the perspective of experimental imperfections presented in Sec.~\ref{ssec:compatible}.


\section{Applications of quantum contextuality}
\label{sec:applications}
Since contextuality is a fundamental phenomenon of quantum 
mechanics, it is not surprising that some studied 
its applications and its relevance to quantum information 
processing. In the following, we  discuss three examples: First, the role of contextuality in quantum 
computation; second, potential applications in quantum cryptography;
and third, an application of contextuality in randomness generation.
Finally, we mention some further connections to information 
processing.

\subsection{Contextuality and quantum computation}
\label{sec:contextualityincomputation}
The first works to investigate the relation between quantum contextuality and computation appeared in the framework of measurement-based quantum computation \cite[MBQC,][]{Raussendorf:2001PRL,BriegelNatPhys2009}, where computation is performed by adaptively measuring single qubits prepared in a large entangled state. Thus, in each experimental run a set of compatible measurements (i.e., measurements on different qubits) are performed. It is natural to interpret the entire experiment as a contextuality experiment (notice that it cannot be interpreted as a Bell experiment since the systems are not far apart) and ask whether the computational power arises as a consequence of quantum contextuality. The first result in this direction was presented by  \citet{Anders:2009PRL} who showed that GHZ-type correlations enable the deterministic computation of the NAND gate, effectively promoting a classical parity computer into a universal (classical) one; see also the non-adaptive case  by \citet{Hoban:2011NJP}. Starting with this observation,  \citet{Raussendorf:2013PRA} proved that all MBQC with mod 2 linear classical processing that compute a nonlinear Boolean function with a sufficiently high success probability are contextual. Raussendorf's result was then further generalized. Specifically, \citet{Abramsky:2017PRL}  provided an explicit lower bound to the failure probability in terms of the noncontextual fraction and the distance from the set of linear functions. \citet{OestereichPRA2017} extended the result to include reliable computation, i.e., with success probability strictly greater than $1/2$; and \citet{FrembsNJP2018} considered the case beyond qubits.

A fundamental result showing a strong interplay between contextuality 
and computation was that of \citet{Howard:2014NAT}. More precisely, 
the result connected contextuality in the framework of NCHV models
with additional exclusivity (see Sec.~\ref{sssec:NCHVE}), and 
quantum computation in the framework of quantum computation 
via magic state distillation for qudit systems with $d$ as an 
odd prime number. 
Finally, another important result is the one obtained by 
\citet{Bravyi:2018SCI,Bravyi:2020NP}, who showed an
example of a problem that can be solved with quantum circuits of 
constant depth, regardless of the input size (shallow circuits),
but requires classical circuits to increase in depth logarithmically 
with the input size. This result can be directly connected to the 
problem of classical simulation of contextual correlations.

Given the relevance of such results in the quantum information 
community and their direct connection with topics discussed in this
review, namely, the graph-theoretical approach to contextuality presented in 
Sec.~\ref{ssec:graph_CSW} for \citet{Howard:2014NAT} and the cost of 
classical simulation of contextuality presented in Sec.~\ref{class_sim}
for \citet{Bravyi:2018SCI,Bravyi:2020NP},
we summarize these two results in the following.

\subsubsection{Contextuality and magic states}
One of the basic building blocks of the paradigm of computation 
via magic state distillation is stabilizer codes, which provide 
a fault-tolerant implementation of a subset of preparations, 
measurements, and unitary transformations. This subset of 
operations, however, is not only not-universal for computation, 
but also efficiently classical simulable, as shown by 
the Gottesman-Knill theorem~\cite{GottesmanPhD1997}. 
An additional resource that provides universal quantum computation 
is nonstabilizer states, called {\it magic states}, possibly 
provided in a noisy form, but distillable to some target magic 
state~\cite{BravyiPRA2005}. With these states non-Clifford gates, 
such as the $\pi/8$ gate or its qudit generalization, 
can be implemented, thus promoting stabilizer computation 
to universal quantum computation. Not all magic states are 
useful, as a large class of them cannot be distilled to 
pure states and some can even be efficiently classical simulable~\cite{AaronsonPRA2004,VeitchNJP2012,MariPRL2012,VeitchNJP2014}. 

\citet{Howard:2014NAT} showed that a state is contextual, with respect to stabilizer measurements, 
if and only if the state is outside the polytope of efficiently simulable 
states $\mathcal{P}_{\rm sim}$. They proved the statement for qudits 
with $d$ being an odd prime and for the special case of qubits. This result 
identified contextuality as a necessary condition for universal quantum 
computation via magic state distillation. The proof of the sufficiency 
requires one to show that any state $\rho \notin \mathcal{P}_{\rm sim}$ 
can be distilled to a sufficiently pure magic state.

In the following, we explain the stabilizer formalism 
and how the set of efficiently simulable states can be identified 
with the noncontextual states, with respect to stabilizer measurements, 
defined as all projective measurements
consisting of rank-1 projectors onto stabilizer states. We consider 
a system of dimension $p$, where $p$ is an odd prime. The qubit 
case has also been discussed by \citet{Howard:2014NAT}. 

We first recall the following definition of the displacement 
operators in the discrete phase space~\cite{Vourdas2004, 
GibbonsPRA2004, Gross2006}  
\begin{equation}
D_{l,m} :=\omega^{2^{-1} l m} X^l Z^m,
\end{equation}
where the generalized $X$ and $Z$ are defined in the 
computational basis by  $Z\ket{k}=\omega^k \ket{k}$, $X\ket{k}=\ket{k+1}$, 
and $\omega:=e^{i2\pi/p}$, and $2^{-1}$ is the multiplicative 
inverse of $2$ in the field $\mathbb{Z}_p$, i.e., $\omega^{2^{-1}} = e^{i\pi/p}$.

In analogy with the continuous-variable case, the discrete Wigner function of 
a state $\rho$ can then be defined as the expectation values of displacement 
operators on it. Since the dimension is finite and some of the displacement 
operators commute (since $D_{l,m} D_{l',m'} = \omega^{2^{-1} (l m'-l'm)} D_{l+l',m+m'} $), 
it is sufficient to consider only $p+1$ of them, for instance,
 $L=\{D_{0,1},D_{1,0},D_{1,1},D_{1,2},\ldots,D_{1,p-1}\}$, as their 
 eigenvectors form a complete set of mutually unbiased bases~\cite{ApplebyJMP2008}. 
 Now denote by $\Pi_j^{q_j}$ the projector onto the eigenvector with eigenvalue 
 $\omega^{q_j}$ for the $j$-th operator in $L$. For a vector 
 $\mathbf{q}\in \mathbb{Z}_p^{p+1}$ define the operator 
 $A^{\mathbf{q}}:= -\openone + \sum_{j=1}^{p+1}\Pi_j^{q_j}$ 
 and finally the discrete Wigner function as~\cite{GibbonsPRA2004, Gross2006}
 \begin{equation}
 W_\rho(\mathbf{q})=\tr(\rho A^\mathbf{q}).
 \end{equation}

Following \citet{VeitchNJP2012,MariPRL2012}, the efficently simulable 
states are identified with those with a positive Wigner function, 
namely
\begin{equation}\label{eq:Psim}
\mathcal{P}_{\rm sim} := \
\{ \rho \ |\ \tr(\rho A^{x\mathbf{a}+z\mathbf{b}})\geq 0,\ x,z\in \mathbb{Z}_p \},
\end{equation}
with $\mathbf{a}:= (1, 0, 1, \ldots, p - 1 ) $ and $\mathbf{b} := - (0, 1, 1, \ldots , 1)$.
The connection between simulability and contextuality was then obtained 
by associating an exclusivity graph to a collection of projectors [specifically, 
those entering into the definition of $A^{x\mathbf{a}+z\mathbf{b}}$ in 
Eq.~\eqref{eq:Psim}], computing its independence number (i.e., the noncontextual 
bound) and showing that the condition $\tr(\rho A^{x\mathbf{a}+z\mathbf{b}}) < 0$ 
amounts to a violation of the noncontextual bound. More precisely, for a 
two-qudit system, with the appropriately chosen set of projectors 
$\{\Pi_j^{s_j}\}$ and their sum $\Sigma^\mathbf{r}$, they showed that 
\begin{equation}\label{eq:nchv_pos}
\tr[\Sigma^\mathbf{r} (\rho\otimes \sigma)] \leq p^3 \Leftrightarrow \tr(\rho A^{\mathbf{r}})\geq 0
\end{equation}
for any state $\sigma$ on the second qudit. Finally, the bound $p^3$ was proven to be the independence number of the graph $\Gamma^{\mathbf{r}}$ associated with the set of projectors appearing in $\Sigma^\mathbf{r}$, i.e., $\alpha(\Gamma^{\mathbf{r}})=p^3$, showing that the contextual states ($\rho\otimes \sigma$ for all $\sigma$) are precisely those associated with a negative Wigner function for $\rho$. This connection between contextual states and the negativity of their Wigner function has been proven to be general~\cite{DelfosseNJP2017} for $n$-qudit 
systems with $n>1$ and $d$ an odd prime, without requiring the construction of the tensor product $\rho\otimes \sigma$ as in Eq.~\eqref{eq:nchv_pos}.

The result obtained by \citet{Howard:2014NAT} was then extended to rebits, i.e., 
a restriction of qubits to real-valued density matrices and operators
\cite{Delfosse:2015PRX}, and finally to qubits 
\cite{Bermejo-Vega:2017PRL,Raussendorf:2017PRA}. 
Finally,  a  different notion of contextuality, called
sequential contextuality has been investigated from the perspective of
computation~\cite{MansfieldPRL2018} and quantum information processing tasks such as
quantum random access codes for systems with bounded memory~\cite{Emeriau:2020XXX}.

\subsubsection{Contextuality and shallow quantum circuits}
The name {\it shallow circuit} refers to circuits that are of constant depth, regardless of the input size. Constant depth implies that the corresponding operations can be run in a constant time, as operations on different sets of qubits can be run in parallel.

As a starting point, \citet{Bravyi:2018SCI} showed that there are problems that can be solved by a quantum algorithm with certainty and in constant time for any input size, i.e., with a shallow circuit, but require a time logarithmic in the length of the input for any classical circuit that solves them with a sufficiently high probability. This work was further extended \citet{Bravyi:2020NP} to account for what happens if noise is explicitly modeled. \citet{Bravyi:2020NP} presented  an alternative, and arguably simpler, argument to show the gap between quantum and classical shallow circuits. 

Note that the arguments presented  by \citet{Bravyi:2018SCI,Bravyi:2020NP} do not use the oracular paradigm. This is important, since a speedup proven in the oracular paradigm may not translate into a real-world advantage, as a classical algorithm may take advantage of the internal structure of the oracle to solve the problem more efficiently~\cite{Johansson:2017QIP,Johansson:2019ENT}. 

Two different arguments were presented by \citet{Bravyi:2018SCI} and \citet{Bravyi:2020NP}, but they are based on  similar reasoning. An intuitive understanding of them, based on nonlocal games, can be obtained as follows. We first consider the case of a circuit that is fed with a fixed entangled state plus classical bits as input, but implements only local (such as single-qubit) operations. This means that each gate has only a wire coming in and one coming out. Recall that the number of wires coming in each gate, not necessarily among nearest neighbors, is  called the {\it degree} or {\it fan-in} of the gate. If the input-output relation of the circuit is modeled after that of a nonlocal game, then the observation of a probability of success for the correct output above a certain threshold can be interpreted as the violation of a Bell inequality. Now imagine we allow for fan-in $2$ with nearest-neighbor interactions.  Thus, in the analogy of the nonlocal game, any two parties can collaborate to win, but only if their distance is less than $2^D$, where $D$ is the circuit depth. To visualize this, one can imagine that each output of the circuit has some ``past lightcone'' indicating the initial inputs that could have influenced it. 

The number of parties in the game corresponds to the input size of the circuit, hence, to allow for a collaboration among distant parties, the depth of the circuit must grow logarithmically with the input size.  In other words, to win with the aforementioned classical communication strategy, the depth of the circuit must grow (logarithmically) with the input size. The argument is then extended to the case of gates of fan-in $K$, with $K$ arbitrary but fixed for all possible input sizes  and the condition of a nearest-neighbor interaction removed. Finally, the nonlocal game is chosen such that it can be won by quantum players without any communication, corresponding to a fixed depth circuit necessary to prepare the correct initial entangled state. Notice that even if the number of operations needed for preparing this entangled state grows with the input, they can be performed in parallel; hence, the depth of the circuit remains constant.

A more detailed description can be provided by explicitly considering the game given by \citet{Bravyi:2020NP},  which is based on the PM-square~\cite{Peres:1990PLA,Mermin:1990PRL} nonlocal game \cite{Aravind:2004AJP,Cabello:2001PRLb, CHTW04}, and called the 1D magic square problem. \citet{Bravyi:2018SCI} used a similar game based on a GHZ-type contradiction, but the proof is more elaborate as the game is based on a two-dimensional qubit architecture. \citet{Bravyi:2020NP} had the $N$ input wires of the circuit represent the $N$ players and additional classical inputs were provided to specify the game. In each round, only two of them play the 
PM-square game, whereas the other players collaborate to create the right correlations. The different roles are assigned at random at the beginning of each round, such that a perfect winning (classical) strategy would necessarily require a collaboration between any pair of players. 

The nonlocal game can be straightforwardly translated into an abstract {\it relation problem}: a relation is simply a set of valid input-output pairs, $(z_{\rm in}, z_{\rm out})$, defined by a function $R(z_{\rm in}, z_{\rm out})$ taking value $0$ or $1$. A circuit is said to solve the relation problem $R$ if for each input $z_{\rm in}$ it produces an output $z_{\rm out}$ such that $R(z_{\rm in}, z_{\rm out})$ equals $1$. The pairs $(z_{\rm in}, z_{\rm out})$ can then be derived from the input-output pairs of the nonlocal game. The problem is said to have $n$ input-output bits if $|z_{\rm in}| + |z_{\rm out}| =n$.

We can then summarize the result by \citet{Bravyi:2020NP} as follows. They showed that for each $n$ there is a relation problem $R$ with $n$ input-output bits and a set of inputs $S$, with $|S|={\rm poly}(n)$ such that (i) $R$ can be solved for all inputs with a 
constant-depth  quantum circuit and (ii) any classical circuit with constant fan-in that solves $R$ with probability $p_{\rm success}\geq 90\%$, for $S$ uniformly distributed, has a depth growing at least as fast as $\log(n)$.

\citet{Bravyi:2020NP} generalized the result to the case of quantum circuits with local stochastic noise, namely, a random Pauli error is applied at each time step to the ideal quantum circuit. More precisely, they showed that a constant-depth noisy quantum circuit, this time with a 3D geometric structure, can solve the $n$-bit problem $R$ with high probability ($p_{\rm success} \geq 99\%$), whereas any classical noiseless circuit that solves it with $p_{\rm success} \geq 90\%$ has depth growing at least as fast as $ {\log(n)}/{\log[\log(n)]}$.

Beyond the technical details of the results and from a contextuality perspective, an interesting observation is the following. The quantum circuit can solve the problem with a constant depth due to its ability to generate contextual correlations. In fact, any classical simulation of these contextual correlations requires communication among the parties, which in turn requires the depth of the circuit to grow with the number of parties (or inputs of the game). The results obtained by \citet{Bravyi:2018SCI} and \citet{Bravyi:2020NP} can therefore be interpreted in the framework of {\it memory cost} (or, {\it communication cost}) for the classical simulation of quantum contextuality (see Sec.~\ref{class_sim}). Finally, notice that even if we are discussing nonlocal games and Bell inequalities, in any realization of the considered circuit the single measurements are not far apart. It is therefore more appropriate to identify the corresponding phenomenon as quantum contextuality  than as Bell nonlocality.


\subsection{Contextuality and quantum cryptography}


\subsubsection{Svozil's quantum key distribution protocol}


The possibilities of contextuality for quantum key distribution (QKD) were first devised by \citet{BechmannPasquinucci:2000PRL}. Here, we review a QKD scheme introduced by \citet{Svozil:2010}, which provides a good example of how contextuality adds features beyond those provided by measurement incompatibility. Specifically, it shows how contextuality can be used to counteract a possible attack [described by \citet{Svozil:2006AJP}] that may be used to 
attack the standard BB84 QKD protocol \cite{Bennett:1984}. 

Recall how the BB84 protocol works. There are two separate parties Alice and Bob who want to obtain a shared secret key (i.e., a sequence of bits known only to them). For that, they send physical systems from Alice to Bob and share classical information over a public channel. The protocol goes as follows: (i) Alice randomly picks one from two basis of qubit states (the computational basis and the Hadamard basis) and sends Bob over a public and authenticated quantum channel a randomly chosen state of that basis. (ii) Bob picks a basis at random from the two and measures in this basis the system received from Alice. (iii) Over a public channel, Bob announces his bases and Alice announces those events in which the state sent belongs to the measured basis. (iv) Alice and Bob repeat steps (i)-(iii) many times and, from the bits where both Alice and Bob used the same basis, Alice randomly chooses half of them and discloses her choices over the public channel. Both Alice and Bob announce these bits publicly and run a check to see whether more than a certain number of them agree. If this check passes, then Alice and Bob use the remaining
undisclosed bits to create a shared secret key via additional techniques like 
error correction and privacy amplification.

We now describe Svozil's attack to the BB84 protocol \cite{Svozil:2006AJP}. The adversary replaces the preparations and measurements of quantum states by classical preparations and measurements. Thus, in step (i), Alice is actually picking one of two differently colored eyeglasses (instead of one of the two bases) and picking a ball from an urn (instead of picking one quantum state) with two color symbols in it (corresponding to the basis that the state belongs to). Each one of the two differently colored eyeglasses allows her to see only one of the two colors. Svozil 
observed that the adversary can mimic the quantum predictions if: (a) each of the balls has one symbol $S_i \in \{0,1\}$ written in two different colors chosen from the two
possible pairs. Her choice of eyeglasses decides which symbols Alice can see. (b) All colors are equally probable, and for a given color the two symbols are equally probable. Therefore, in step (ii) Bob is actually picking one of two differently colored eyeglasses and reading the corresponding symbol. Since the requirements (a) and (b) can be satisfied simultaneously, the strategy can successfully imitate the quantum statistics of the BB84 protocol. Therefore, if the replacement remains unnoticed to Alice and Bob, just by checking the statistics they cannot realize that the adversary may have full knowledge of their ``secret'' key.

However, \citet{Svozil:2010} also noticeed that if one replaces the quantum states and basis of the BB84 protocol by those used in a proof of the KS theorem, then in the classical attack requirements (a) and (b) cannot be satisfied simultaneously and  the adversary cannot simulate the quantum statistics. Specifically, Svozil's protocol uses the 9-basis 18-state KS proof in dimension $4$ in Fig.~\ref{fig:cab18}. In step (i), Alice randomly picks a basis from the nine bases in Fig.~\ref{fig:cab18} and sends Bob a randomly chosen state of that basis. (ii) Bob picks a basis at random from the nine and measures the system received from Alice. The remaining steps are as in BB84, but one notices that each measurement now has four outcomes (rather than two).


\subsubsection{Contextuality offers device-independent security}


The core idea behind BB84 is that information gain for one quantum observable must cause a disturbance to another incompatible observable. This does not require entanglement or composite systems. However, there is a second generation of QKD protocols, offering much higher levels of security that relies on nonlocality. These protocols were initiated by \citet{Ekert:1991PRL}, were advanced through the work of  \citet*{Barrett:2005PRL}, and lead to the schemes for device-independent QKD \cite{Acin:2007PRL}. Security in the protocols is verified solely through the statistics of the measurement outcomes, with no assumptions about the inner working of the devices (except those that are standard in cryptography). Since nonlocality can be seen as contextuality produced by local measurements on composite systems, all these schemes may be considered as applications of contextuality. However, in most of them the exact role of contextuality is difficult to follow. 

Here we review a result and a corresponding QKD scheme presented by \citet{Horodecki:2010XXX}, in which local contextuality plays a crucial role. The result can be summarized as follows: if two parties share systems that, locally, show a KS contradiction and, in addition, exhibit perfect correlations, then they can use them to extract a secure key in a device-independent manner. As an application, they introduced a QKD protocol exploiting the properties of the PM magic square; see Sec.~\ref{PM}. Here we do not provide the details of the QKD protocol: we instead simply describe the resources used and the steps to prove device-independent security.

A distributed PM box (shared by Alice and Bob) is defined as follows \cite{Cabello:2001PRLb, Aravind:2004AJP, CHTW04}. 
Both Alice and Bob have a PM set of observables. Alice measures columns of the PM magic square, while Bob 
measures rows, as first proposed by \cite{Cabello:2001PRLb}. That is, a distributed PM box is a set 
of nine conditional distributions $p(a, b|x, y)$, where $x$ labels the columns of the PM table, 
$y$ labels the rows, and $a = (a_1,a_2,a_3)$ and $b = (b_1,b_2,b_3)$ are the outcomes of 
the joint measurement of the three observables in the respective column or row, where 
$a_i,b_j \in \{+1,-1\}$. The outcomes are assumed to satisfy the corresponding quantum 
predictions. That is, $a_1 a_2 a_3=+1$ for all columns of the PM table except the 
last one (for which $a_1 a_2 a_3=-1$), and $b_1 b_2 b_3=+1$ for all rows. In addition, 
there are perfect correlations between the outcomes of the same observables on Alice's 
side and Bob's side. Finally, nonsignaling holds. That is, Alice's (Bob's) local 
distributions do not depend on the choice of measurement by Bob (Alice). In quantum 
mechanics, such a distributed PM box can be realized, if both parties share two singlet 
states.

Consider a PM distributed box such that the parties do not know how it is implemented, 
i.e., what observables are measured and in which quantum state. However, if they assume 
the validity of quantum mechanics, as is usually done in the device-independent paradigm 
\cite{Acin:2007PRL}, it can be shown \cite{Horodecki:2010XXX} that the outcomes of a 
fixed row or column possess about $0.44$ bits of intrinsic randomness, and hence that the 
correlations offer security. It can also be proven \cite{Horodecki:2010XXX} that a 
secure key can be obtained both in the noiseless case and when assuming a small amount 
of noise in the state.


\subsection{Random number generation}


The inherent randomness of quantum mechanics together with the impossibility 
of a classical simulation of some of its aspects suggests the possibility of using it for the generation of random numbers. For instance,  protocols of randomness expansion based on Bell nonlocality and with minimal assumptions on the measuring devices have been proposed \cite{Colbeck:2006XXX}. 

In the following, we review a protocol of {\it randomness generation} based on quantum contextuality 
proposed by  \citet{Abbott:2012PRA}. The main idea is to exploit a Kochen-Specker-type contradiction, 
namely, the impossibility of a preassigned value for certain quantum properties of a system to claim that 
the outcomes generated by the measurement of such properties are genuinely random. In contrast to previous 
approaches, \citet{Abbott:2012PRA} not only used the impossibility of a simultaneous assignment to {\it all} 
variables (NCHV) but also precisely localized which variable cannot have a definite value. The intuition is 
similar to the one at the basis of the bug graph in Fig.~\ref{fig:KS8-Yu-Oh} on Pag.~\pageref{fig:KS8-Yu-Oh}: 
if $A$ is assigned the value $1$, then $B$ must be assigned the value $0$. \citet{Abbott:2012PRA} extended 
this idea by proving a stronger result: they found a graph in $d=3$ such that, whenever $A$ is assigned the 
value $1$,  the assignments $B=0$ and $B=1$ both generate some contradiction. In this case $B$ is said 
to be value indefinite. Moreover, they showed that this graph can be constructed for any two projectors 
$P_A=\ketbrac{a}$ and $P_B=\ketbrac{b}$ such that $\sqrt{5/14}\leq |\mean{a|b}|\leq 3/\sqrt{14}$. These numbers
were subsequently improved by \citet{Abbott:2015JMP}, to the general condition $0< |\mean{a|b}| <1$.  This 
construction relies on the assumption that value assignments respect QM predictions for one-dimensional 
projectors, in particularly orthogonality ($\mathbf{O'}$) and completeness ($\mathbf{C'}$) for an orthogonal basis; see Sec.~\ref{sec:KS-sets}. 
Moreover, the assumption of a definite value for $P_A$ is translated, through the {\it eigenstate 
assumption} \cite{Abbott:2012PRA}, to the assumption of the preparation of an eigenstate of $P_A$. 

This argument can be translated to a practical random number generation protocol that consists of 
preparing a three-level quantum system in the pure state $\ket{\psi}$ and then measuring it in a basis 
containing vectors $\ket{\phi_+},\ket{\phi_-}$ such that 
$0 < |\mean{\psi|\phi_\pm}|< 1$. An explicit implementation is given in terms of 
the spin operators for a spin-1 system with $\ket{\psi}=\ket{S_z = 0}$ and 
$\ket{\phi_\pm} = \ket{S_x=\pm 1}$. In other words, the system is prepared in the eigenstate associated with 
$0$ for the spin along the $z$ direction, and a measurement in the $S_x$ basis is performed. 
By the geometry of the problem $\mean{S_z=0|S_x=0}=0$ and  $\mean{S_z=0|S_x=\pm 1}=1/\sqrt{2}$, which implies 
that the outcome $0$ never appears in the measurement of $S_x$ and that the two value indefinite outcomes 
$\pm 1$ appear with equal probability. In addition to the realization with spin-1 systems, Abbott {\it et al.} discussed an implementation based on photon interferometry.

This approach was explored experimentally by \citet{Kulikov:2017PRL}, and the quality of the
randomness produced in the experiment has been further analyzed by \citet{Abbott:2019PS}. See also \cite{Abbott:2014MSCS,Aguero2021} for more recent theoretical developments. Experimental
random number generation based on contextuality was also explored by \citet{UmSR2013,UmPRA2020}, 
but not in the framework developed by \citet{Abbott:2012PRA}.


\subsection{Further applications}


Finally, we want to mention some other applications where 
contextuality has proven useful.

\subsubsection{Zero-error channel capacities}\label{sec:zero_err}
In general, a classical channel $\mathcal{N}$ transforms 
inputs $x \in X$ on Alice's side to outputs $y \in Y$ on
Bob's side, so it can be considered a conditional 
probability distribution $p(y|x).$ If only a single use
of the channel is allowed, Bob may not be able to uniquely
determine Alice's input from his output. One may therefore ask what
the largest subset of inputs is  that can be perfectly distinguished. 
This is also called the one-shot, zero-error capacity 
$c_0(\mathcal{N})$ of the channel. 

This quantity can be interpreted in a graph-theoretical 
manner, using the so-called confusability graph 
$G(\mathcal{N})$. The vertices of this graph are the input
symbols $x \in X$ and two vertices $x_1$ and $x_2$ are connected 
if and only if the probability distributions $p(y|x_1)$ and 
$p(y|x_2)$ overlap. This means that there is a possible 
output $y$ that may originate from the two $x_i$, so these 
two inputs are confusable. In other words, the one-shot 
zero-error capacity $c_0(\mathcal{N})$ corresponds to 
the maximum independent set of the confusability graph.
In this jargon, Bob's perspective can be described as 
follows: He receives an output $y$ that may originate from 
several $x_i$. Any two of these $x_i$ are confusable, so the 
possible inputs $x_i$ form a clique in the confusability 
graph. 

How does the capacity of a channel change if Alice and Bob 
also have  some access to additional resources such as shared
randomness or entangled states? 
\citet{Cubitt:2010PRL,Cubitt:2011TIT} showed 
that from any Kochen-Specker set of vectors one can construct
an example of a channel, where the one-shot zero-error capacity 
in the presence of a shared entangled state, denoted by 
$c_{\rm E}(\mathcal{N})$, is strictly larger than the capacity 
without shared entanglement $c_0(\mathcal{N}).$

This connection is best explained with an example. Consider the nine 
orthogonal bases in four-dimensional space, which came out of the 
18-vector proof in Fig.~\ref{fig:cab18} in Sec.~\ref{sec:KS-sets}. The 36 overall  vectors can be
organized in a $9 \times 4$ array $\ket{\psi_{ij}}$ and the indices 
$(ij)$ constitute the input space $X$ of the channel $\mathcal{N}$.
One then constructs the channel such that two inputs $(ij)$ and $(kl)$
are confusable if the vectors $\ket{\psi_{ij}}$ and $\ket{\psi_{kl}}$
are orthogonal. This can be achieved in different ways: for instance
one can simply take  $Y=X$ as an output space and start with 
$p(y|x) = \delta_{xy}$. Small disturbances are added in order to
build the desired confusability graph.

The resulting channel has $c_0(\mathcal{N}) \leq 8.$ To see this, 
assume that $c_0(\mathcal{N}) = 9$ (or larger). The nine
distinguishable $x_m$ have to belong to the nine different bases 
(or rows in the array) since inputs within a row are by construction 
not perfectly distinguishable by Bob. Moreover, if the same vector 
appears on two positions in the array ($\ket{\psi_{ij}} = \ket{\psi_{kl}}$)
and $(ij)$ belongs to the set $\{x_m\}$, then   $(kl)$  also
belongs to the set since $\ket{\psi_{ij}}$ is orthogonal 
to all other vectors in the row $k$. Thereofre, one arrives at an 
assignment of values to the 36 vectors that obeys the rules of noncontextuality, and this is by construction not possible. 

On the other hand, it is evident that with the help of entanglement
$c_{\rm E}(\mathcal{N}) \geq 9.$ Assume that Alice and Bob share a
maximally entangled state in a $4 \times 4$-system. To 
send the row index $i$ to Bob, Alice then simply performs a projective 
measurement of the corresponding basis on her part of the state. She 
obtains the random result $j$ and sends $(ij)$ through the channel. 
From the channel output $y$, Bob can identify a clique of four possible 
inputs $(kl).$ The corresponding states $\ket{\psi_{kl}}$  are orthogonal, 
so he can identify Alice's input by performing a projective measurement
on his reduced state, which is given by $\ket{\psi_{ij}}.$


\subsubsection{Dimension witnesses} 
As mentioned in 
Sec.~\ref{sec:KS-sets}, the Kochen-Specker theorem 
requires at least a three-dimensional Hilbert space. It 
is therefore natural to connect the violation of contextuality 
inequalities to the dimension.

For the case of the PM square, this was done by
\citet{Guhne:2013PRA}. They considered, the contextuality inequality found
in Eq.~(\ref{eq:chshjocord}) and 
studied how the violation depends on the underlying dimension. 
It has been shown that 
\begin{equation}
\mean{\mathsf{PM}} 
\stackrel{\rm 2D, com.}{\leq} 2 
\stackrel{\rm 3D, com.}{\leq} 4(\sqrt{5}-1) \approx 4.94,
\end{equation}
where the bounds hold for the respective dimensions under the 
assumption that the measurements are projective and obey the 
compatibility (or commutation) relations of the PM 
square. These bounds can be generalized to certain POVMs and 
also to the KCBS inequality \cite{Guhne:2013PRA}. More recently 
a general method on dimension witnesses using the graph-theoretic approach was
 introduced \cite{Ray:2021NJP}.


\subsubsection{Self-testing} 
Quantum self-testing \cite{Yao:2004} is the art of certifying quantum states, quantum 
measurements, and other quantum features from the input-output statistics of measurement experiments and some 
minimal assumptions, which do not include assumptions about the quantum system. The method is based on the 
observation that some input-output statistics corresponding to extremal points in the corresponding sets of 
quantum correlations can be achieved, up to isometries, only with specific states and measurements. The idea was 
initially used for self-testing quantum states and measurements in Bell scenarios \cite{Yao:2004}, then 
extended to other features and scenarios. Here we review its application for self-testing states and 
measurements in contextuality experiments with sequences of ideal measurements \cite{Bharti:2019PRL,Bharti:2019XXX} and for self-testing states and measurements in Bell scenarios by exploiting the connection between 
quantum contextuality and graph invariants \cite{Bharti:2021XXX}.

In the first case, the distinctive assumption is that measurements are ideal. Under this assumption, it was
proven \cite{Bharti:2019PRL} that the quantum violations of the KCBS inequality and all the tight inequalities 
for the odd $n$-cycle scenarios (with $n \ge 5$) \cite{Araujo:2013PRA} allow for self-testing. It was also 
proven \cite{Bharti:2019XXX} that the quantum violations of the antihole noncontextuality inequalities 
\cite{Cabelloetal:2013PRA} allow for self-testing. The interest in the latter result relies on the fact that it allows 
for self-testing quantum states and measurements of any odd dimension $d \ge 3$.

\cite{Bharti:2021XXX} showed that the connection between quantum contextuality and graph invariants 
permits one to simplify the proofs of self-testability of certain Bell nonlocal correlations that were known to 
allow for self-testing, identify new Bell nonlocal correlations that allow for self-testing, and prove a 
conjecture about the closed form expression of the Lov\'asz theta number for a family of graphs.


\subsubsection{Applications of Spekkens' contextuality}

{\bf Parity-oblivious multiplexing.} 
This is an information processing
task for two parties, where preparation contextuality, as explained in 
Sec.~\ref{ssec:other_cont}, is useful \cite{Spekkens:2009PRL}. We 
first describe the problem. Consider a two-party system where Alice
receives a bit string $x \in \{0,1\}^{n}$ of length $n$. Bob 
receives a number $y \in\{1, \dots n\}$ and has to predict the 
bit $x_y$ of Alice's string. To succeed, Alice can send
Bob some information about her string. Thus far, this is a general 
scenario that also occurs in random access codes \cite{Ambainis:2002ACM}. The 
interesting point is to put constraints on the information Alice 
is allowed to send to Bob and then investigate the physical consequences.

In the scenario considered by \citet{Spekkens:2009PRL} one adds the 
constraint that the information that Alice is allowed to send to Bob 
should not give any information about the parity of her string on 
any subset containing two or more bits. Using a mathematical formulation, 
let $s\in \{0,1\}^{n}$ be an arbitrary bit string with at least two 
entries $1$. No information on $\sum_i x_i s_i$ should then be 
revealed, where addition is modulo $2$. This constraint makes the
information transmission from Alice to Bob ``parity oblivious.'' 

One can first ask what the optimal classical success probability  is 
for this game. In a classical system, the constraint effectively ensures
that Alice can  transfer only one single bit of the string $x$; without
losing generality one can assume that this is the first bit. Bob
can then predict the bit correctly for $y=1$, and he has to guess for all other
values of $y$. This leads to a success probability of 
$p(b=x_y) = 1/n + 1/2 \times (n-1)/n = (n+1)/2n.$ In ontological 
models obeying the constraint of preparation contextuality, one also
cannot exceed this value. The reason is that in these models 
parity obliviousness at the level of Alice's preparations and Bob's 
measurements already implies the parity obliviousness at the level of 
hidden variables; see \cite{Spekkens:2009PRL} for a 
detailed argumentation. 

In quantum mechanics, however, this bound does not hold. Consider 
the case $n=2.$ The four possible strings for Alice can be encoded 
in four single-qubit states with the Bloch vectors lying in the 
$x$-$y$ plane via 
$\vec{r}_{x_1, x_2} = \left((-1)^{x_1}, (-1)^{x_2}, 0\right)/\sqrt{2},$ 
and the states are 
$\varrho_{x_1, x_2} =(\openone + \vec{r}_{x_1, x_2} \vec{\sigma})/2.$
Since $\varrho_{11} + \varrho_{00} = \varrho_{10} + \varrho_{01}$,
no quantum measurement can give information on the parity of $x.$
If Bob wants to know $x_1$ he measures $\sigma_x$ and for predicting
$x_2$ he measures $\sigma_y$. This gives the right bit with probability
$\cos^2(\pi/8)\approx 0.8536$, which is larger than the classical optimum
of $3/4.$

The choice of the signal states is closely related to
the examples of inequalities for preparation noncontextuality;
see Eq.~(\ref{eq-spekkens-constraint1}) in Sec.~\ref{ssec:other_cont}. 
The connection of parity-oblivious communication with preparation contextuality 
has been further generalized in several directions 
\cite{SahaNJP2019, SahaPRA2019, Hameedi:2017PRL, Ghorai:2018PRA, Ambainis:2019QIP, Chailloux:2016NJP, Banik:2015PRA}.


{\bf State discrimination.} 
The task of minimum error state discrimination has been a well-studied
problem since the early days of quantum information processing; 
see \cite{Barnett:2009AOP} for review. In
the simplest scenario, two nonorthogonal states $\ket{\psi}$ and 
$\ket{\phi}$ are given with equal probability. The task is then
to make a measurement and identify the state. As the states are 
nonorthogonal this cannot be done perfectly, so the task is to 
minimize the error probability of the guess. Note that there is also
a different notion of unambiguous state discrimination in which no
error is allowed, but it is possible to pass as a third option. 

This can be connected to preparation noncontextuality, as shown by 
\citet{SchmidPRX2018}. Consider two single-qubit states $\ket{\psi}$ 
and $\ket{\phi}$ with overlap $c=|\braket{\psi|\phi}|^2$. We can 
assume without losing generality that their Bloch vectors are of 
the form $\vec{r}_{\psi/\phi} = \left(\cos(\alpha), 0, \pm \sin(\alpha)\right).$ 
The optimal measurement is then given by $\sigma_z$, leading 
to a success probability of $s = (1+\sqrt{1-c})/2.$  One can consider 
in addition the orthogonal vectors $\ket{\psi^\perp}$ and 
$\ket{\phi^\perp}$ with the Bloch vectors $\vec{r}_{\psi^\perp/\phi^\perp} 
= \left(-\cos(\alpha), 0, \mp \sin(\alpha)\right).$ They lead to essentially the 
same state discrimination problem, with the same success probability. 

From the perspective of preparation contextuality  
it is important that 
$
\ketbra{\psi}{\psi} + \ketbra{\psi^\perp}{\psi^\perp} 
=
\ketbra{\phi}{\phi} + \ketbra{\phi^\perp}{\phi^\perp} = \openone.
$ 
This puts constraints on the hidden-variable distributions
describing these four states in preparation-noncontextual 
theories, see also Eqs.~(\ref{eq-spekkens-constraint1} and \ref{eq:pnc_conv}).
Under these constraints and under the assumption that the 
relations and symmetries between the four states are preserved, 
one can prove that the success probability for the two-state problem
given by $\ket{\psi}$ and $\ket{\phi}$ is bounded by $s \leq 1-c/2$. 
For any $c$ this is strictly lower than the quantum mechanical value 
previously provided \cite{SchmidPRX2018}. This result can also  be shown to hold 
for states affected by noise.


\subsubsection{Further applications on the horizon}
Recently,  other works considered potential applications of 
contextuality, but it is currently difficult to predict the 
future impact of these research lines. The novel applications contain
machine learning \cite{Gao:2022PRX}, postselected metrology \cite{Arvidsson-Shukur:2020NC},
and state-dependent cloning \cite{Lostaglio:2020Q}.



\section{Summary and outlook}

Since the discovery of quantum contextuality more than 50 years ago, the topic 
has received increasing attention, with the largest number of significant 
contributions occurring only during the last decade. This development parallels 
the increased interest in Bell nonlocality and has been partially driven by the 
fast-growing community of quantum information scientists. A key breakthrough 
for quantum contextuality was the transformation of the logical contradiction 
that underlies the original theorem by \textcite{Kochen:1967JMM} to 
experimentally accessible noncontextuality inequalities; see 
Sec.~\ref{section4}. A recent key development has been the establishment of the 
connection between computational resources and the presence of contextuality;
see Sec.~\ref{sec:applications}. As the field of quantum contextuality is 
evolving faster then ever before, we identify three key topics that are 
essential for the consolidation of our current understanding of 
quantum contextuality and the further development of the field.

First, the mathematical structure of quantum contextuality has not yet been 
fully revealed, despite the numerous seminal results. For example, the smallest 
scenario for state-independent contextuality is not known. It is likely that it 
will be the scenario by \citet{Yu:2012PRL}, but a conclusive proof 
has not yet been provided. In addition, scenarios that are maximally contextual in 
certain ways have yet to be identified. For example, it is known 
\citet{AmaralPRA2015} that the quotient $\vartheta(G)/\alpha(G)$ for an 
exclusivity graph $G$ tends to the number of vertices of $G$, but a family 
of graphs with this property is not yet known.

Second, although there are a large number of convincing experiments that have 
confirmed quantum contextuality in physical systems (see 
Sec.~\ref{ssec:experiments}) the handling of experimental imperfections, or 
``loopholes,'' has not reached the thoroughness that has been achieved for Bell 
nonlocality \cite{Larsson:2014JPA,Brunner:2014RMP}. There are several (partially 
competing) methods to handle experimental imperfections (see 
Sec.~\ref{ssec:exper_imp}) but a comprehensive description in a unified 
framework is missing. Even if a truly loophole-free experiment might be
fundamentally impossible, this does not lessen the need for comprehensive 
treatment. Some of these difficulties can be traced back to the fact that 
quantum contextuality (with the notable exception of Spekkens's notion of contextuality, see 
Sec.~\ref{ssec:other_cont}) is based on the notion of ideal measurements and, 
in the case of implementations with sequential measurements, on the role of 
L\"uders rule for ideal measurements. Our understanding of both concepts within 
the foundations of quantum theory is not fully developed and might be a source 
of our struggle with the design of loophole-free contextuality experiments.
See \citet{WangSAdv2022} for  recent developments in this direction.

Finally, we mention the role of contextuality in quantum 
computation and communication; see also Sec.~\ref{sec:applications}. There 
are still no strong methods that would allow one to quantify the 
memory cost of quantum contextuality. Specifically, it is not known whether 
there is a quantum advantage regarding the cost when simulating a sequential 
implementation of quantum contextuality by means of a classical finite state 
machine. Current affirmative results (see Sec.~\ref{ssec:compatible}) are based 
on sequences of incompatible observables, which is an alien concept to quantum 
contextuality. Much broader and more general questions regard whether and how 
quantum contextuality plays a role in universal quantum computation. To date
this has been answered  in the case of measurement-based quantum computation,
quantum computation via magic states,
and shallow quantum circuits; see Sec.~\ref{sec:contextualityincomputation}. 
But whether and in what sense contextuality plays a role in the 
circuit model are widely open questions. Besides those more specific questions, 
we expect various new key applications of contextuality in quantum information 
science to emerge in the near future.

In conclusion, quantum contextuality plays a central role in quantum theory, encompassing both \textit{measurement incompatibility} at a fundamental level and \textit{Bell nonlocality} and \textit{entanglement} when subsystems are spatially separated. It is also strongly connected to new developments in quantum technology. 
Quantum contextuality is at the heart of the matter, more so than quantum uncertainty or quantum interference. Both of them could in principle be present in a classical model, whereas quantum contextuality cannot, as shown by \citet{Kochen:1967JMM}.
Paraphrasing their conclusion ``This way of viewing the results [presented here] seems to us to display a new feature of quantum mechanics in its departure from classical mechanics.'' 
Quantum contextuality is what makes quantum theory fundamentally nonclassical, and will indubitably play an important role in future developments of quantum physics.

\begin{acknowledgments}
We thank 
Samson Abramsky,
Antonio Ac\'in,
Evelyn Acu\~{n}a,
Joseba Alonso,
B\'arbara Amaral,
Elias Amselem,
Leandro Aolita,
Marcus Appleby,
Mateus Ara\'ujo,
Mauricio Arias,
Ali Asadian,
Flavio Baccari,
Guido Bacciagaluppi,
Piotr Badzi\c{a}g,
Jos\'e P.\ Baltan\'as,
Johanna F.\ Barra,
Hannes Bartosik, 
Ingemar Bengtsson,
Marco Bentivegna,
Kishor Bharti,
Kate Blanchfield,
Rainer Blatt,
Naresh Goud Boddu,
Gilberto Borges,
Mohamed Bourennane,
Harvey Brown,
\v{C}aslav Brukner,
Jeffrey Bub,
Gustavo Ca\~{n}as,
Jaime Cari\~{n}e,
Gonzalo Carvacho,
Marcos Carvalho,
Daniel Cavalcanti,
Rafel Chaves,
Jiang-Shan Chen,
Jing-Ling Chen,
Giulio Chiribella,
Andrea Chiuri,
Sujit K. Choudhary,
Rob Clifton,
Andrea Crespi,
Jin-Ming Cui,
Vincenzo D'Ambrosio,
Lars E.\ Danielsen,
Pierre-Louis de Assis,
Dong-Ling Deng,
Ehtibar Dhzafarov,
Cristhiano Duarte,
Joseph Emerson,
Paul Erker,
Jos\'e M.\ Estebaranz, 
Sebasti\'an Etcheverry,
Gabriel Fagundes,
Armando Fern\'andez Prieto,
Jos\'e Ferraz,
Stefan Filipp,
Manuel J.\ Freire,
Tobias Fritz,
Diego Frustaglia,
Chris Fuchs,
Rodrigo Gallego,
Ernesto Galv\~{a}o,
Ren\'e Gerritsma,
Robert B. Griffiths,
Philippe Grangier,
Helena Granstr\"om,
Esteban S.\ G\'omez,
Mile Gu,
Guang-Can Guo,
Yong-Jian Han,
Lucien Hardy,
Yuji Hasegawa,
Teiko Heinosaari,
Joe Henson,
Isabelle Herbauts,
Jonathan P.\ Home,
Michael Horne,
Pawel Horodecki,
Mark Howard,
Xiao-Min Hu,
Yun-Feng Huang,
Greg Jaeger,
Martti Karvonen,
Dagomir Kaszlikowski,
Adrian Kent, 
Michael Kernaghan,
Andreii Khrennikov,
Kihwan Kim,
Gerhard Kirchmair,
J\"urgen Klepp, 
Alexander Klyachko,
Simon Kochen,
Ravi Kunjwal,
Pawel Kurzy\'{n}ski,
Leong-Chuan Kwek,
Brian La Cour,
Raymond Laflamme,
Radek {\L}apkiewicz,
Matthew Leifer,
Florian M.\ Leupold,
Gustavo Lima,
Chuan-Feng Li,
Qiang Li,
Petr Lison\v{e}k,
Bi-Heng Liu,
Zheng-Hao Liu,
Antonio J.\ L\'opez Tarrida,
Vicente Losada,
Aintzane Lujambio,
Maciej Malinowski,
Shane Mansfield,
Breno Marques,
Paolo Mataloni,
Hui-Xian Meng,
David Mermin,
David A.\ Meyer,
Giovanni Morchio,
Osama Moussa,
Eleonora Nagali,
Miguel Navascu\'es,
Mohamed Nawareg,
Vlad Negnevitsky,
Choo Hiap Oh,
Roberto Osellame,
Sebasti\~{a}o P\'adua,
Jian-Wei Pan,
Mladen Pavi\v{c}i\'{c},
Marcin Paw{\l}owski,
Asher Peres,
Itamar Pitowsky,
Ioannis Pitsios,
Michel Planat,
\'Angel R.\ Plastino,
Davide Poderini,
Jos\'e R. Portillo,
Matt Pusey,
Marco T\'ulio Quintino,
Rafael Rabelo,
Magnus R{\aa}dmark,
Ravishankar Ramanathan,
Helmut Rauch,
Robert Raussendorf,
Maharshi Ray,
Renato Renner,
Rolf Riklund,
Christian Roos,
Davide Rusca,
Carlos Saavedra, 
Muhammad Sadiq,
Debashis Saha,
Ana Bel\'en Sainz,
Emilio Santos,
Valerio Scarani,
R\"udiger Schack,
Claus Schmitzer,
Fabio Sciarrino,
Simone Severini,
Abner Shimony,
Rui Soares Barbosa,
Alberto Sol\'{\i}s,
Adrian Specker,
Ernst Specker,
Susan Specker,
Rob Spekkens,
Stephan Sponar,
Allen Stairs,
Hong-Yi Su,
Kai Sun,
Karl Svozil,
Jochen Szangolies,
Marcelo Terra Cunha,
Stefan Trandafir,
Giuseppe Vallone,
Antonios Varvitsiotis,
Mar\'{\i}a C.\ Vel\'azquez Ahumada,
Giuseppe Vitagliano,
Mordecai Waegell,
Naqueeb Ahmad Warsi,
Harald Weinfurter,
Marcin Wie\'{s}niak,
Andreas Winter,
Elie Wolfe,
Chunfeng Wu,
Guilherme B.\ Xavier,
Ya Xiao,
Jin-Shi Xu,
Zhen-Peng Xu,
Bin Yan,
Sheng Ye,
Sixia Yu,
Xiao-Dong Yu,
Florian Z\"ahringer,
Anton Zeilinger,
Chi Zhang,
Jie Zhou,
Zong-Quan Zhou, 
Marek \.Zukowski, and
Wojciech \.Zurek
for interesting discussions on quantum contextuality over the years.
We also thank 
Alastair Abbott, 
Manik Banik,
Kishor Bharti,
Cris Calude,
Hyppolite Dourdent,
Pierre-Emmanuel Emeriau,
Robert B. Griffiths,
Alexei Grinbaum,
Martti Karvonen,
Zheng-Hao Liu,
Shane Mansfield,
Gaël Massé,
Karl Svozil, 
Armin Tavakoli,
and three anonymous referees
for helpful comments on the manuscript.
This work has been supported by 
Project Qdisc (Project No.\ US-15097) with FEDER funds, Projcet No.\ FIS2017-89609-P (MINECO, Spain) with FEDER funds, QuantERA grant SECRET, by MINECO (Project No.\ PCI2019-111885-2),
the Deutsche Forschungsgemeinschaft 
(DFG, German Research Foundation, Project No.\ 447948357 and No.\ 440958198), 
the
FQXi Fund (Silicon Valley Community Foundation) through the
projects ``The Nature of Information in Sequential Quantum Measurements'' and
``The Observer Observed: A Bayesian Route to the Reconstruction of Quantum Theory'',
the Sino-German Center for Research Promotion (Project M-0294),
the Austrian Science Fund (FWF) through Projects No. ZK 3 (Zukunftskolleg)
and No. F7113 (BeyondC), and the ERC (Consolidator Grant 683107/TempoQ).
\end{acknowledgments}

\appendix
\section{Quantum contextuality from a historical perspective}
\label{app:history}
Here we present a historical introduction to quantum contextuality from its 
origins to the time when the basis for experimentally testing Kochen-Specker 
contextuality was settled. The aim of this section is to frame the results 
presented in the review within a historical context and trace the connections 
between them that may help us to understand the evolution and ramifications 
of the field.


\subsection{The problem of hidden variables}


The discussion, in the late 1920s, of whether quantum mechanics can be supplemented by ``hidden variables'' was motivated by two results: Born's probabilistic interpretation of Schr\"odinger's wave function \cite{Born:1926ZP,Born:1926ZPb}, which expresses the fundamentally probabilistic character of the predictions of quantum mechanics, and Heisenberg's uncertainty principle \cite{Heisenberg:1927ZP}, which asserts a fundamental limit to the precision with which the values of position and momentum can be predicted in quantum mechanics. While Heisenberg, Born, Pauli, and, notably, Bohr made strong claims that quantum mechanics provides a complete framework for physics and manifested their skepticism about the possibility of completing it with hidden variables, Schr\"odinger, de Broglie, and especially Einstein hoped 
to recover quantum mechanics from a deeper nonprobabilistic theory and viewed the quantum state as an incomplete description in need of supplementation by hidden variables \cite{Lorentz:1928, Fine:1990FOP}.

At the Solvay Conference in 1927, de Broglie presented an explicit hidden-variable theory \cite{Lorentz:1928}. However, the criticisms received, particularly those from Pauli \cite{Lorentz:1928}, persuaded de Broglie to abandon his theory. 

In 1931, the skepticism of Bohr received support from a proof of impossibility of hidden variables presented by \citet{vonNeumann:1931AM} and included in his book \cite[see Sec.~IV.~2]{vonNeumann:1932SPR}. This proof was  soon after shown to be inconclusive by \citet{Hermann:1935}, but her work was mostly ignored for many years~\cite{Mermin:2018FOP}. The influence of von Neumann's book, then, strongly discouraged any discussion of hidden-variable theories for decades.

Paradoxically, at that time, \citet{Wigner:1932PR} found something that could have been used against hidden variables: when attempting to link Schr\"odinger's wave function to a distribution on phase space (which would be the analog in quantum mechanics to the distribution function of classical statistical mechanics), Wigner found that such a distribution has negative values and cannot be made non-negative. The importance of this discovery was not recognized until much later.

In \citeyear{Einstein:1935PR}, \citeauthor*{Einstein:1935PR} (EPR) showed that quantum mechanics is incomplete, in the sense that it does not assign definite outcomes to measurements whose results can be predicted with certainty from the outcomes of spacelike separated measurements. Several years later \citet{Bell:1964PHY} showed that EPR's hidden variable-theories collide with quantum mechanics, but at that time the EPR argument reinforced the resistance of Einstein (and many others) toward accepting quantum mechanics as a final theory. 

Meanwhile, von Neumann observed that the two-valued observables, represented in quantum mechanics by projection operators, constitute a sort of ``logic'' of experimental propositions and, together with Birkhoff \cite{Birkhoff:1936AM}, developed a ``quantum logic'', a set of algebraic rules governing operations to combine, and predicates to relate propositions associated with physical events. This logic would eventually provide a new basis for discussing the problem of hidden variables.

In 1952, \citet{Bohm:1952PR,Bohm:1952PRb} presented a hidden-variable theory that is a further elaboration of de Broglie's theory of 1927. Bohm's theory is deterministic and explicitly nonlocal at the level of hidden variables.
 
In parallel, \citet{Mackey:1957AMM} asked whether every measure on the lattice of projections of a Hilbert space can be defined by a positive operator with unit trace. A positive answer would show that the Born rule follows from a particular set of axioms (framing a generalized probability theory) for quantum mechanics \cite{Mackey:1957AMM,Mackey:1963}. Although Kadison \cite{Chernoff:2009NAMS} [and later \citet{Bell:1966RMP} and  \citet{Kochen:1967JMM}] proved that this was false for two-dimensional Hilbert spaces, \citet{Gleason:1957JMM} showed it to be true for higher dimensions. Gleason's theorem is going to play a crucial role in the discussion of hidden variables. Mackey's program \cite{Mackey:1957AMM} was further developed in several directions by \citet{Ludwig:1964ZP,Ludwig:1967CMP,Ludwig:1968CMP,Ludwig:1972CMP,Piron:1964HPA,Piron:1976,Foulis:1972JMP,Foulis:1974SYN,Randall:1970AMM,Randall:1973}. All these works provided the basis of what is now called the framework of generalized probabilistic theories; see \cite{Hardy:2001XXX,Chiribella:2010PRA}, which views quantum probability theory as one possibility in a landscape of probability theories and asks what is special about it. 


\subsection{The Kochen-Specker theorem}


In \citeyear{Specker:1960D}, \citeauthor{Specker:1960D}, a mathematician with theological concerns who inspired by ``the question whether the omniscience of God also extends to events that would have occurred in case something would have happened that did not happen'' \cite{Specker:1960D} and by the logic of Birkhoff and von Neumann, reformulated the question of hidden variables as follows: ``Is it possible to extend the description of a quantum mechanical system through the introduction of supplementary ---fictitious--- propositions in such a way that in the extended domain the classical propositional logic holds?'' Specker found that ``the answer to this question is negative, except in the case of Hilbert spaces of dimension 1 and 2'' as ``an elementary geometrical argument shows'' \cite{Specker:1960D} [quotation from the English translation of \citet{Seevinck:2011arxiv}]. 

In fact, according to Specker \cite{Specker1990}, ``the basic theorem of the paper was proved shortly [after a seminar on the foundations of quantum theory],'' a seminar that probably took place during the summer semester of 1948; see \cite{Enz:1997}.  The geometrical argument was not fully presented
until the collaborative paper of 1967 with Kochen \cite{Kochen:1967JMM}, although the fundamental building block for it discussed by \citet{Kochen:1965a} \cite[Fig.~1, p.~182; see also][]{Kochen:1965b}. The Kochen-Specker (KS) theorem shows the incompatibility between some predictions of quantum mechanics and a type of hidden variables that later came to be called noncontextual.

In 1963 (although it was not published until 1966) Bell developed a similar geometrical argument but one using a more complex building block \cite{Bell:1966RMP}. Bell also used an infinite set of quantum observables. In contrast, Kochen and Specker managed to prove their theorem using $117$ observables by concatenating their building block $15$ times. In his paper, Bell seemed to have found this geometrical argument after Jauch draw his attention to the consequences of Gleason's theorem to the problem of hidden variables \cite{Bell:1966RMP}. In fact, Bell later referred to this proof as ``observed by Jauch'' \cite{Bell:1971XXX} and ``subsequently set out by S.\ Kochen and E.\ P.\ Specker'' \cite{Bell:1971XXX} and even later Bell wrote that he ``was told of it by J.\ M.\ Jauch in 1963'' \cite{Bell:1982FP} and that ``the idea was later rediscovered by Kochen and Specker'' \cite{Bell:1982FP}. 
As we pointed out, the idea was already in print in 1960.

Notably, Bell was not convinced that the proof was compelling. His source of discomfort was the observation that measuring the same observable in different contexts  ``require[s] different experimental arrangements; [and thus] there is no {\em a priori} reason to believe that the results $\ldots$ should be the same'' \cite{Bell:1966RMP}. Bell added: ``The result of observation may reasonably depend not only on the state of the system (including hidden variables) but also on the complete disposition of the apparatus'' \cite{Bell:1966RMP}. Both \citet[see p.~73]{Kochen:1967JMM} and \citet[see p.~451]{Bell:1966RMP} seemed to believe that the only way to measure the same observable in two contexts is by measuring two maximal (and incompatible) quantum observables, one for each context. They did not consider the possibility, also offered by quantum mechanics, of measuring each observable using the same apparatus such that in each context one measures sequentially the observables of the context, as is done in modern sequential contextuality experiments \cite[e.g.,][]{Kirchmair:2009NAT}.

In addition, Bell noticed the nonlocality in Bohm's theory of 1952 [he writes, ``[I]n this theory an explicitly causal mechanism exists whereby the disposition of one piece of apparatus affects the results obtained with a distant piece,'' \citealp{Bell:1966RMP}] and how this is an unwanted feature, as it solves the EPR paradox ``in the way Einstein would have liked least'' \cite{Bell:1966RMP}. Finally, Bell pointed out that ``there is no {\em proof} that {\em any} hidden variable account of quantum mechanics {\em must} have this extraordinary character. It would therefore be interesting $\ldots$ to pursue some further `impossibility proofs,' replacing the arbitrary axioms objected to above by some condition of locality , or separability, of distant systems'' \cite{Bell:1966RMP}. This led to Bell's famous proof of impossibility of ``local'' hidden variables \cite{Bell:1964PHY}.


\subsection{The origin of the word contextuality}


The term contextuality in association with quantum mechanics derives \cite{Shimony:2009CQP,Jaeger:2019PTR} from the term introduced by Shimony \cite{Shimony:1971}  
to designate the hidden-variable theories ``in which the value of an observable $O$ is allowed to depend not only upon the hidden state $\lambda$, but also upon the set $C$ of compatible observables measured along with $O$'' \cite{Shimony:1971}. Shimony called them ``contextualistic'' hidden-variable theories. The shortening to ``contextual'' was made by \citet{Beltrametti:1981} and then adopted by Shimony and others. In the 1990s, ``Contextuality'' became the title of a chapter of Peres's influential book on quantum theory \cite{Peres:1993}.


\subsection{The relation between the KS and Bell theorems and the need for a theory-independent notion of noncontextuality}


While Bell's theorem gained prominence among physicists and the general public after the experiment of \citet{Freedman:1972PRL}, \citet{Clauser:1976PRL,Clauser:1976NCB}, 
\citet{Aspect:1982PRL}, and others and its applications to cryptography \cite{Ekert:1991PRL} and quantum information, the KS theorem was for a long time a subject that interested primarily philosophers of science and a few physicists concerned about the foundations of quantum mechanics.

The situation began to change in the 1990s. On the one hand, \citet{Peres:1990PLA,Peres:1991JPA,Peres:1992FP,Peres:1993} and \citet{Mermin:1990PRL,Mermin:1993RMP} simplified the proof of the KS theorem using a small number of two- and three-qubit observables, making the KS theorem accessible to a wider audience. On the other hand, Mermin's Bell inequality \cite{Mermin90b} and his ``unified form for the major no-hidden-variables theorems'' \cite{Mermin:1990PRL,Mermin:1993RMP} connected the proof of \citet*[GHZ,][]{GHZ89} to the Bell inequalities and the KS theorem, respectively. Similar connections between the Bell and KS theorems had been found before by Kochen in a private communication with Shimony \cite{Stairs:1983PS,HR83}; \citet{Stairs:1983PS}, to whom Mermin acknowledges input \cite{Mermin:1993RMP}, and \citet{HR83}; see also \citet{Brown:1990FPH}. 

There was still something that blocked unification of the KS theorem and Bell's theorem of impossibility of local hidden-variable theories. While Bell's theorem leads to experimental tests of whether the world can be explained with theories that can be defined without any reference to quantum mechanics, the KS theorem is deeply attached to quantum mechanics. This attachment is triple.

First, the KS theorem does not refer to general measurements, but to those that are represented in quantum mechanics by the spectral projectors of a self-adjoint operator. What does this restriction mean from a theory-independent point of view? Moreover, in quantum mechanics there are measurements that are not represented by projective measurements but by POVMs. 

Second, the proof of the KS theorem includes constraints that are specific to quantum systems. Examples of these constraints are that the values of the squared spin components of spin-$1$ particles for any orthogonal triad $\{x,y,z\}$ should satisfy the equation $v(S_x^2) + v(S_y^2) + v(S_z^2) = 2$ \cite{Kochen:1967JMM} and that the values for the Pauli observables of two spin-$1/2$ particles should satisfy the equation $v(\sigma_x^{(1)}) v(\sigma_x^{(2)}) v(\sigma_x^{(1)} \otimes \sigma_x^{(2)})=1$ \cite{Peres:1990PLA,Mermin:1990PRL}. 
Third, the experimental translation of the KS theorem (as proposed by KS and Bell) assumes quantum mechanics, as it is assumed that coarse grainings of two different (and incompatible) measurements represent the same observable based on the fact that in quantum mechanics both yield the same outcome statistics. 

Therefore, the problem was how to translate the KS theorem into experimental tests of contextuality in nature \cite{Cabello:1998PRL}. For that, what was needed was a theory-independent notion of contextuality that removes all the quantum constraints, includes a theory-independent definition of the type of measurements for which the assumption of outcome noncontextuality is made (similar to Bell's theorem's focus on local measurements), of the sets of measurements (contexts) whose correlations are considered (similar to Bell's theorem's focus on spatially separated local measurements), and a physical motivation for assuming outcome noncontextuality for these measurements and contexts (that plays the same role as the impossibility of communication between spacelike separated events in Bell's theorem). 

Nevertheless, the lack of such a formal framework did not impede experimental progress and the first ``experiments towards falsification of noncontextual hidden variable theories'' on single systems \cite{Michler:2000PRL}, which took advantage of the analogy between two two-dimensional separated subsystems and $2$ dichotomic degrees of freedom of a single photon and tested the violation of the single-particle equivalent of the Clauser-Horne-Shimony-Holt extension of the Bell inequality \cite{Clauser:1969PRL}; see also \citealp{Hasegawa:2003NAT} for a similar experiment with neutrons. 

However, it was the criticisms of \citet{Meyer:1999PRL}, \citet{Kent:1999PRL}, and \citet{Clifton:2000PRSA} toward the idea of giving the KS theorem a similar experimental status as Bell's theorem \cite{Cabello:1998PRL} that gave a definitive push to the transformation of contextuality into an experimentally testable property with no reference to quantum mechanics. 
These criticisms boosted vivid discussions \cite{Appleby:2000XXX,Appleby:2001XXX,Appleby:2002PRA,Appleby:2005SHPS,Barrett:2004SHPB,Cabello:1999XXX,Cabello:2002PRA,Havlicek:2001JPA,Mermin:1999XXX,Peres:2003XXX} and stimulated new developments. On the one hand, they stimulated the attempt to obtain experimentally testable ``KS inequalities'' \cite{Simon:PRL2001,Larsson:2002EPL}. However, these first inequalities still made assumptions that hold only in quantum mechanics. 

On the other hand, they stimulated a new notion of noncontextuality \cite{Spekkens:2005PRA}. 
This notion implicitly assumes that the hidden variables (or ontological models) merely provide a classical description of the same operations as those allowed in quantum mechanics, without any possibility of predictions deviating from those of quantum mechanics or even redundancy in the description, i.e., a different description at the level of the formalism for physically equivalent situations, as it happens for gauge symmetries. 


\subsection{Noncontextuality for ideal measurements}


The final boost for a general theory-independent framework for contextuality rooted in the notion of noncontextuality used by KS (but free of the assumptions that hold only  in quantum mechanics) was the discovery of the quantum violation of the KCBS inequality \cite{Klyachko:2008PRL} by single qutrits in a specific quantum state, followed by the discovery of similar inequalities that are violated by any quantum state (of a given dimension) \cite{Cabello:2008PRL,Badziag:2009PRL}. Unlike previous inequalities, the bounds of these inequalities are derived only from the assumption of outcome noncontextuality, without extra constraints inspired by quantum mechanics.

The KCBS inequality was introduced in an earlier paper \cite{Klyachko:2007NATO} as a way of showing that single spin-$1$ particles can exhibit a form of ``single-particle entanglement,'' defined as maximal uncertainty of a set of observables associated with a Lie algebra.
This led to the question: For what type of measurements and contexts  is there  an ``{\em a priori} reason to believe that the results for should be the same'' \cite{Bell:1966RMP}?
One possible answer is, for those measurements that yield the same result when performed repeatedly on the same physical system and do not produce any change in the outcomes of any jointly measurable observable, and for contexts made of compatible sets of them. These measurements are called ideal \cite{Cabello:2019PRA} or sharp \cite{Chiribella:2016IC}. Intuitively, ideal measurements reveal preexisting context-independent ``properties'' of the measured system that are preserved after the act of measuring. However, in general, this may not be the case. 

The focus on contexts made of compatible ideal measurements allows us to formulate a notion of contextuality in the operational framework of generalized probabilistic theories without any reference to quantum mechanics.
This theory-independent notion of contextuality is referred to as contextuality for ideal measurements or KS contextuality, as it is inspired by the work of Kochen and Specker. This notion allows us to replace or remove the two assumptions of the KS theorem that refer to quantum mechanics. Namely, (I) that measurements represented in quantum theory by self-adjoint operators reveal preexisting values that are independent of the ``context,'' where context meant set of measurements represented by mutually commuting self-adjoint operators, and (II) that measurement outcomes must satisfy the same functional relations that quantum mechanics predicts for commuting measurements on quantum systems of a given dimension. Instead of that, in KS contextuality, (I) is replaced by the assumption of outcome noncontextuality for ideal measurements and (II) is completely removed. Notably, the new notion provides a basis for experimentally testing KS contextuality in nature.


\subsection{The hidden history of noncontextuality inequalities}


The mathematical tools needed for studying contextuality were developed independently of physics and long before quantum mechanics. We call a contextuality scenario a set of abstract ideal measurements, each of them having a number of possible outcomes, and their relations of compatibility. For example, the scenario considered by KCBS \cite{Klyachko:2008PRL} has five measurements $M_i$, $i=0,\ldots,4$, each of which has two possible outcomes, and such that $M_i$ and $M_{i+1}$ (with the sum modulo $5$) are compatible. Therefore, in the KCBS contextuality scenario there are five contexts. For each contextuality scenario, a ``matrix of correlation,'' ``behavior,'' or simply ``correlation'' is a set of probabilities for all possible combination of outcomes in each of the contexts. One obtains one of these correlations using a specific initial state and measurements. Probabilities have to satisfy the corresponding normalization and nondisturbance (or nonsignaling) constraints. 

Like what happens in Bell scenarios \cite{Froissart:1981,Suppes:1981SYN,Fine:1982JMP,Fine:1982PRL,GM:1984,Pitowsky:1986JMP,Pitowsky:1989,Pitowsky:1991MP}, in any KS scenario the set of correlations satisfying outcome noncontextuality is a polytope. Here it is called the noncontextual polytope of the scenario. Correlations outside this set are contextual and violate one of the linear inequalities (in the probabilities) that define the facets of the noncontextuality polytope. Each of these facets corresponds to a inequality that is necessary for noncontextuality and is called a tight noncontextuality inequality. These inequalities were introduced long before quantum mechanics. 

In 1990, during a symposium in Jerusalem and Tel Aviv coincidentally entitled ``Einstein in context,'' Pitowsky distributed among the participants a draft [later published as \cite{Pitowsky:1994BJPS}] where he pointed out that \citet{Boole:1862PTRL}, one of the fathers of modern logic, had developed a set of equalities and inequalities he called ``conditions of possible experience'' \cite{Boole:1862PTRL} and that the Bell inequalities violated by quantum mechanics were a subset of them. 

This observation leads to the following questions: (i)~Which of Boole's inequalities can be violated? (ii)~What is the largest set of correlations possible for a given scenario? (iii)~How does this set compare to the one in quantum theory? Answering these questions would have helped to answer the central question Pitowsky asked: ``WHY is [it] that microphysical phenomena and classical phenomena differ in the way they do?'' \cite{Pitowsky:1994BJPS}. 

The answer to question~(i) was known in the 1960s. A theorem introduced by \citet{Vorobev1959,Vorobev1962,Vorobev1967coal} showed that a violation of Boole's inequalities can  occur only for scenarios in which the graph of compatibility contains an induced cyclic path with a size larger than 3 (i.e., following the path along some edges of the graph one obtains a square,  a pentagon, a hexagon, etc). The graph of compatibility is the one in which compatible measurements are represented by adjacent vertices. Otherwise, there is always a joint probability distribution, and therefore a noncontextual model. Bell inequalities violated by quantum mechanics correspond to scenarios with this property. 

The CHSH scenario \cite{Clauser:1970PRL}, with four dichotomic measurements whose graph of compatibility is a square, is therefore the one with the smallest number of ideal measurements that allow for contextual correlations. In fact, the CHSH inequality is the only nontrivial tight noncontextuality inequality for the CHSH scenario \cite{Fine:1982PRL,Fine:1982JMP}.

Both \citet{Bell:1966RMP} and \citet{Kochen:1967JMM} noticed that the statistics of ideal measurements on a two-dimensional quantum system (or qubit) can be reproduced with noncontextual models. Therefore, interesting questions are as follows: in which scenario does a three-dimensional quantum system (or qutrit) violate noncontextuality inequalities with ideal measurements? What are these inequalities? 
The answer to the first question is the KCBS scenario \cite{Klyachko:2008PRL}. The KCBS inquality is the only tight noncontextuality inequality for the KCBS scenario \cite{Araujo:2013PRA}. The KCBS scenario, which was previously considered in some papers on quantum logic \cite{GGM:1974,Wright:1978}, is also the scenario with the smallest number of ideal measurements whose relations of compatibility (and incompatibility) cannot occur in a Bell scenario. All these features made the KCBS inequality a key to the world of KS contextuality. 


\bibliographystyle{apsrmp4-1}
\bibliography{common}


\end{document}